\documentclass[12pt]{article}
\newcommand{\blind}{0}
\usepackage{easyReview}
\usepackage{amsmath,amsthm}
\usepackage{amsfonts}
\usepackage{amssymb}
\usepackage{bbm}
\usepackage{mathtools}

\usepackage{mathrsfs}

\usepackage{subcaption} 
\usepackage{graphicx}
\usepackage{pgfplots}
\usepackage[all]{nowidow}
\usepackage[utf8]{inputenc}
\usepackage{tikz}
\usepackage{multicol}
\usepackage{algpseudocode,algorithm,algorithmicx}
\usepackage{scalerel}
\usepackage{natbib}
\usepackage[normalem]{ulem}

\usepackage[english]{babel}

\usepackage{mathdots}
\usepackage{yhmath}
\usepackage{cancel}
\usepackage{xcolor}
\usepackage{array}
\usepackage{multirow}
\usepackage{gensymb}
\usepackage{tabularx}
\usepackage{booktabs}
\usetikzlibrary{fadings}
\usetikzlibrary{patterns}
\usetikzlibrary{shadows.blur}

\usepackage{enumerate}
\usepackage[algo2e]{algorithm2e}

\newtheorem{theorem}{Theorem}

\newtheorem{assumption}{Assumption}

\newtheorem{remark}{Remark}

\newtheorem{example}{Example}
\newtheorem*{excont}{Example \continuation}
\newcommand{\continuation}{??}
\newenvironment{continueexample}[1]
{\renewcommand{\continuation}{\ref{#1}}\excont[continued]}
{\endexcont}

\newcommand{\wt}{\widetilde}
\newcommand{\Z}{\mathbb{Z}}

\newcommand{\R}{\mathbb{R}}
\newcommand{\C}{\mathbb{C}}
\newcommand{\E}{\mathbb{E}}

\newcommand{\X}{\mathcal{X}}

\newcommand{\U}{\mathcal{U}}

\newcommand{\D}{\mathcal{D}}
\newcommand{\id}{\mathbbm{1}}

\newcommand{\fr}{{\lfloor nr\rfloor}}

\newcommand{\fa}{{\lfloor n\alpha \rfloor}}
\newcommand{\fb}{{\lfloor nb \rfloor}}

\newcommand{\ft}{{\lfloor nt\rfloor}}

\DeclareMathOperator*{\cov}{Cov}
\DeclareMathOperator*{\im}{Im}
\DeclareMathOperator{\sign}{sign}
\newcommand{\li}{\langle}
\newcommand{\ri}{\rangle}

\usepackage[nodisplayskipstretch]{setspace}

\let\svthefootnote\thefootnote
\newcommand\freefootnote[1]{%
	\let\thefootnote\relax%
	\footnotetext{#1}%
	\let\thefootnote\svthefootnote%
}

\begin{document}
	
	\freefootnote{We would like to thank the Editor, the Associate Editor, and the anonymous reviewers for their valuable
		comments and constructive suggestions, which lead to significant improvements in the paper. } 
	\def\spacingset#1{\renewcommand{\baselinestretch}%
		{#1}\small\normalsize} \spacingset{1}

	
	\if0\blind
	{
		\title{\bf Hypothesis Testing for a Functional Parameter via Self-normalization}
		\author{Yi Zhang\hspace{.2cm}
			and 
			Xiaofeng Shao\thanks{Yi Zhang is a PhD student at Department of Statistics, University of Illinois at Urbana-Champaign ({\tt yiz19@illinois.edu}); Xiaofeng Shao is a Professor at Department of Statistics and Data Science, and Department of Economics, Washington University in St Louis ({\tt shaox@wustl.edu}). The research is partially supported by NSF grants (DMS-2210002 and DMS-2412833).}\\
		}
		\maketitle
	} \fi
	
	\if1\blind
	{
		\bigskip
		\bigskip
		\bigskip
		\begin{center}
			{\LARGE\bf Title}
		\end{center}
		\medskip
	} \fi
	
	\bigskip
	\begin{abstract}
		Testing simple or composite hypothesis on a functional parameter has attracted considerable attention in time series analysis. To accommodate for the unknown  temporal dependence, classical nonparametric approaches such as block bootstrapping and subsampling all involve a bandwidth parameter, the choice of which can substantially affect the finite sample performance. The self normalization (SN) method is tuning parameter free when applied to the inference of a finite-dimensional parameter but its applicability to a functional parameter is unknown.

		In this paper, we propose a sample splitting based approach to generalize the SN method to hypothesis testing of a functional parameter.  
		Our SS-SN (sample splitting plus self-normalization) idea is broadly applicable to many testing problems for functional parameters, including testing for simple/composite hypothesis on marginal cumulative distribution function, testing for time-reversibility and testing for a change point on the spectral distribution of a multivariate time series.
		Specifically, we derive the pivotal limiting distributions of our SS-SN test statistics under the null for both simple and composite  null hypothesis, and derive the limiting power function under the local alternatives. Numerical simulations show that our new tests tend to yield accurate size with competitive power performance as compared to many existing ones. 
		
	\end{abstract}
	
	\noindent%
	\vfill
	
	\newpage
	\spacingset{1.8} 
	
	\section{Introduction}\label{ch1}

	In the analysis of time series, hypothesis testing plays an important role in understanding the 
	property of a time series. Prominent examples include whether the series contains a change point in the marginal distribution, the marginal distribution is 
	Gaussian,  serial dependence is present, or the time series is reversible, among others. All of these problems can be formulated as hypothesis testing of a functional parameter, which is of infinite dimension. 
	The literature of testing/inference for time series is huge, and 
	classical nonparametric approaches
	such as block bootstrapping \citep{kunsch1989,liu1992}, subsampling \citep{politisromano1994,politis1999} and blockwise empirical likelihood \citep{Yuichi1997} have been widely used  
	to accommodate temporal dependence via the idea of blocking; see \cite{politis1999} and \cite{lahiri2003} for book-length
	treatments of these topics.

	The self-normalized approach to inference for time series was first introduced in \cite{shao2010} as an extension of the early work of \cite{kiefer2000} and \cite{lobato2001}, who advocated the bandwidth-free inference in the context of time series regression and white noise testing, respectively. It can be viewed as a novel studentization technique which uses recursive subsample estimators in forming the self-normalizer,  and avoids or reduces the choice of bandwidth/block size,  the
	latter of which is common in existing inferential methods for time series (e.g., in the traditional HAC (Heteroscedasticity autocorrelation consistent)-based inference and afore-mentioned blocking-based resampling/subsampling methods).
	
	Self-normalization has been 
	further extended to many problems in time series analysis and we refer the reader to \cite{shao2015} for a review, which was exclusively on the inference of a low-/fixed-dimensional parameter.
	Recently, SN has been extended to do inference for a growing-dimensional parameter in \cite{zhang2023another} and high-dimensional parameter in \cite{wangshao2020}, \cite{wzvs2022}, \cite{zws2022}, among others. So far the literature has not provided a way of using SN for the inference of an infinite dimensional parameter. 
	Throughout this paper we call an infinite dimensional parameter a functional parameter to highlight the fact that the parameter itself is a function. There are ample examples for a functional parameter, such as marginal distribution function, marginal quantile function, and spectral distribution function, among others; see Section~\ref{sec:testing} for more examples.

	In this article, we shall fill this gap and propose a SN-based inference method for a functional parameter via the idea of sample splitting. The basic idea is to split the sample into two parts $\X_1=\{X_1,\dots,X_\fa\}$
	and $\X_2=\{X_{\lfloor n\alpha\rfloor +1},\dots,X_n\}$, where $\alpha\in (0,1)$ is the splitting ratio. Then on the basis of ${\cal X}_1$, we estimate the deviation of the functional parameter from the one specified under the null and project the deviation to the recursive subsample estimate of functional parameter in the second part. The resulting one-dimensional sequence is then used to build our SN test statistic.  Note that sample splitting and studentization were employed  by \cite{kim2024} as a way of achieving dimension-agnosticness in testing/inference for iid data recently. 
	\cite{gaowangshao2023} and \cite{zhang2023another} have extended the sample splitting and self-normalization idea to the time series setting, but their main focuses are on  a finite-dimensional parameter (e.g., mean), which can be fixed or growing as sample size goes to infinity.

	Our use of SS-SN in the testing of a functional parameter brings out some new and distinctive features: (a), Since our main focus is on a functional parameter, the projection direction is in a function space, and after projection, the dimension is reduced from $\infty$ to $1$. This is different from the naive extension by doing SN based test  of the functional parameter evaluated at a finite set of pre-specified points. The latter also reduces the intrinsic infinite dimension to finite dimension but the test is inconsistent and a simple aggregation may lead to non-pivotalness of the limiting null distribution; see Section~\ref{sec_simple_null} for more discussions. By contrast, our SS-SN test statistic has a pivotal limiting null distribution independent of the splitting ratio and its implementation is fairly straightforward. (b), In many practical situations, the null hypothesis is composite and contains an unknown finite-dimensional parameter, which  requires some nontrivial treatment in both methodological formulation and asymptotic theory. In a sense, our test for composite null can be viewed as an extension of \cite{kuan2006} from a finite dimensional parameter to a functional parameter. In particular,  \cite{kuan2006} addressed robust M testing for a finite-dimensional parameter  without consistent estimation of the
	asymptotic covariance matrix. It is an extension of the original SN method in 
	\cite{kiefer2000} and \cite{lobato2001} to robust M tests with estimation effects eliminated by using recursive estimators to compute the normalizing matrix. 
	In addition, we can extend SS-SN to change-point testing for a functional parameter and the new procedure is nearly tuning parameter free  except for the specification of trimming proportion which is common in change-point analysis \citep{andrews1993}; see Section~\ref{subsec:cp}.
	(c), The theoretical justification is very different from these early work, and it relies on the weak convergence of recursive subsample estimate of functional parameter; see Assumption \ref{assump_1null}. The Gaussian process limit and its independent increment  along the time coordinate are the key for us to derive the limiting distribution of our test statistic. Overall, the new SS-SN tests provide a new perspective to the testing/inference of a functional parameter for a weakly dependent time series as no block size/bandwidth is involved, and the finite sample performance is insensitive to the choice of splitting ratio $\alpha$, as we demonstrate in numerical studies.



	The remainder of the paper is organized as follows. Section~\ref{sec:testing} contains some common examples of functional parameter, presents our SS-SN test statistic in the case of simple null hypothesis in Section~\ref{sec_simple_null}, for composite null hypothesis in Section~\ref{sec_comp} and for single change-point testing in Section~\ref{subsec:cp}. Numerical results are presented in Section~\ref{sec:simulation} and Section~\ref{sec:conclusion} concludes. All the proofs, additional simulation results and two real data illustrations are relegated to supplemental material. 	
	
	Throughout the article, we use ``$\stackrel{d}{=}$", ``$\stackrel{\D}{\to}$" and ``$\stackrel{p}{\to}$" to denote equal in distribution, converge in distribution and converge in probability respectively. We use ``$\rightsquigarrow$" to denote process convergence in function space. We let $D^s[0,1]$ (when $s{=}1$, we omit the superscript and just use $D[0,1]$) denote the space of $\R^s$ valued functions on $[0,1]$ which are right continuous and have left limit, endowed with the topology induced by the multidimensional Skorokhod metric \citep{bill2013}. We use $\id(\cdot)$ to denote the indicator function. For positive integer $d$, we use $\mathbf{I}_d$ to denote the $d$-dimensional identity matrix and $\mathbf{1}_d\in \R^d$ is a vector with all components being 1.
	
	\section{Hypothesis testing for a functional parameter}
	\label{sec:testing}

	Throughout this paper, we shall assume $\{X_t\}_{ t\in \Z}$ is a strictly stationary and weakly dependent time series, unless specified otherwise.	Below we introduce some examples for a functional parameter. For  notational simplicity, we shall assume that $X_t$ is univariate, and multivariate extension is straightforward. 
	\begin{enumerate}
		\item \label{example_1}Marginal cdf (cumulative distribution function):
		$G(x)=P(X_1\le x), ~x\in \R$. 
		\item \label{example_2}Marginal quantile function:
		$Q(\tau)=\inf\{x|G(x)\ge \tau\},~\tau\in (0,1).$
		\item \label{example_3}Spectral distribution function: 
		$SDF(\lambda)=\int_{0}^{\lambda} f(\omega)d\omega,$ $\lambda\in [0,\pi]$, where $\{f(\omega),\omega\in [-\pi,\pi]\}$ is the spectral density function. As stated in many time series textbooks (see, e.g., \cite{fan2003}), the spectral distribution function fully characterizes the second order property of a time series, and plays an equally important role as the spectral density function. 
		
		
		\item \label{example_4}Characteristic function for $(X_t,X_{t-j})$: 
		$CF_j(u,v)=\E(e^{iX_t u} e^{iX_{t-j}v})-\E(e^{iX_t u}) \E(e^{iX_{t-j}v})$. Note that $X_t$ is independent of $X_{t-j}$ if and only if $CF_j(u,v)=0$ for all $(u,v)\in \R^2$. This function plays an important role in testing for pairwise independence of a time series; see \cite{hong1999}. 
		
		\item \label{example_5}Conditional mean dependence measure at lag $j$, $\gamma_j(x)=\E[(X_t-\E(X_t)) \id(X_{t-j}\le x)]$, $x\in \R$; see \cite{escanciano2006a}. Note that $\gamma_j(x)=0$ for any $x\in \R$ if and only if $\E(X_t|X_{t-j})=\E(X_t)$ almost surely. 
		
		\item \label{example_6}Generalized spectral distribution function: Let $H(\lambda,x)=2\int_0^{\lambda \pi} f(\omega,x) d\omega,~\lambda\in [0,1], x\in \R$, where $f(\omega,x)=(2\pi)^{-1}\sum_{j=-\infty}^{\infty}\wt\gamma_j(x) e^{-ij\omega}, \omega\in [-\pi,\pi]$ and $\wt \gamma_j(x) = \E[(X_t-\E(X_t)) e^{ixX_{t-j}}]$. This function is instrumental in the generalized spectral distribution based test for martingale difference in \cite{escanciano2006b}.
		
		\item \label{example_7}Copula spectral distribution function: 
		\[{\cal S}(\lambda;\tau_1,\tau_2)=\int_{0}^{\lambda} f(\omega;\tau_1,\tau_2)d\omega,~\lambda\in [0,\pi],~(\tau_1,\tau_2)\in (0,1)^2,\]
		where $f(\omega;\tau_1,\tau_2)=(2\pi)^{-1}\sum_{k\in Z}\gamma_k^{U}(\tau_1,\tau_2)e^{-i\omega k}$, $\omega\in [-\pi,\pi]$ and 
		$\gamma_k^{U}(\tau_1,\tau_2)=\cov(\id(G(X_k)\le \tau_1), \id(G(X_0)\le \tau_2))$.
		This function is the basis for time reversibility test developed in \cite{goto2021}. 
		
	\end{enumerate}
	
	Note that all the above-mentioned functions can be estimated at a $\sqrt{n}$ rate with their sample version weakly converges to their population counterpart at either pointwise or process level. We exclude some functional parameters, such as marginal density function $G'(\cdot)$, spectral density function $f(\cdot)$, as these functions can only be estimated at a slower than $\sqrt{n}$ rate nonparametrically. Fortunately, for many features of a time series, we can formulate it as a restriction on a functional parameter that is $\sqrt{n}$-estimable. For example, if we are interested in knowing whether the marginal distribution is Gaussian, then we can write $G(x)=\Phi((x-\mu)/\sigma)$ for some $\mu\in R$
	and $\sigma\in (0,\infty)$, where $\Phi$ is the cdf of a $N(0,1)$ random variable. Alternatively, one can specify the density function $G'(x)=(1/\sigma)\phi((x-\mu)/\sigma)$, where $\phi(\cdot)$ is the density function of $N(0,1)$ random variable. However, the latter formulation involves nonparametric density estimation, which yields a rate slower than $\sqrt{n}$, so we shall put our testing problem into the former framework (i.e. testing cdf). 

	\subsection{Hypothesis testing for simple null}\label{sec_simple_null}
	
	Suppose we are interested in the functional parameter $F:\Omega \to \U^d$, $\U\in\{\R,\C\}$, $d\in \Z_+$, which depends on the joint distribution of a stationary time series $\{X_t\}_{t\in \Z}$. Here $\Omega$ is $\R^s$ or a compact subset of $\R^s$, where $s\in\Z_+$. We want to test the null hypothesis $H_0:F(x)=F_0(x)$ for any $x\in \Omega$ against the alternative $H_A:F(x)\neq F_0(x)$ for some $x\in \Omega$, where $F_0$ is a pre-specified function. It is worth noting that we aim to develop an omnibus test, but if there is prior knowledge on the type of alternatives, our test can be modified to account for the prior knowledge; see Remark \ref{rmk_weight}.

	For illustration purpose, we assume $\U = \R$ and $d=1$. As shown in \cite{shao2010}, traditional SN methods can be used to test hypothesis on finitely many predetermined points of $F$. To be specific, if we want to test the null hypothesis $H_0': F(x_0)= F_0(x_0)$ for a given $x_0\in\Omega$, then the SN statistic is defined as: 
	$$SN_n(x_0)=\frac{n \big\{F_n(x_0)- F_0(x_0)\big\}^2}{ n^{-2}\sum_{k=1}^{n} k^2\big\{ F_k(x_0)- F_n(x_0)\big\}^2},$$
	where $F_k(x_0)$ is an estimator of $F(x_0)$ based on the subsample $\{X_1,X_{2},\dots,X_k\}$. If we assume that $n^{-1/2}\lfloor nr\rfloor [F_{\fr}(x_0)-F(x_0)]\rightsquigarrow \sigma B(r)$ holds on $D[0,1]$ for some $\sigma>0$ where $ B(r)$ is a standard Brownian motion, then under $H_0'$, 
	$$SN_n(x_0) \stackrel{\D}{\to}  \Lambda_1=\frac{B(1)^2}{\int_{0}^{1}\{B(r)-rB(1)\}^2dr},$$ which is pivotal and its critical values have been tabulated in \cite{lobato2001}. In a similar fashion, one can test $H_0^{''}:F(x_j)=F_0(x_j)$, for $j=0,1,\cdots,M{-}1$, where $\{x_0,\dots,x_{M-1}\}$ are pre-selected points in $\Omega$. A multivariate version of $SN_n(x_0)$ can be formed and the limiting null is expected to be $\Lambda_M$, which is an extension of $\Lambda_1$ to the multivariate setting; see Lobato (2001) and Shao (2010) for the exact expression.

	The above approach only applies to testing at a finite number of points in $\Omega$, but since we are testing the whole functional form of $F$,
	the finite dimensional approach may not be consistent. Recently \cite{zhang2023another} extended the SN method to hypothesis testing on the mean of a multi-dimensional time series where the dimension is allowed to grow with $n$. As pointed out by a referee, their method can  be applied to test $H''_0$ (with $M$ allowed to diverge to infinity) for certain functional parameters. However, this approach is not applicable to general functional parameters, such as (copula) spectral distribution function and can not handle composite null hypothesis. In addition, applying \cite{zhang2023another} involves discretizing $\Omega$, which is highly nontrivial in practice and can significantly influence the finite sample performance; see Appendix \ref{app_snCP} of the supplement for more discussion and results.
	
	So far the literature does not  offer a viable solution to this problem and it seems difficult to apply the SN approach directly. To illustrate the difficulty, a natural approach is to aggregate the test for a single point, e.g.,  $\sup_{x \in\Omega}SN_n(x)$ or $\int_\Omega SN_n(x)d F_0(x)$. Following the same idea as above, under $H_0$ and assuming $n^{-1/2}\fr [F_{\fr}(x)-F(x)]\rightsquigarrow H(r,x)$, where $H(r,x)$ is a mean zero Gaussian process on $[0,1]\times \Omega$, we have
	\begin{align}
		\sup_{x \in\Omega}SN_n(x)\stackrel{\D}{\to}&\sup_{x \in\Omega} \frac{H(1,x)^2}{\int_{0}^{1}\{H(r,x)-rH(1,x)\}^2dr},\label{sn_infeas1}\\
		\int_\Omega SN_n(x)d F_0(x) \stackrel{\D}{\to}&\int_\Omega \frac{H(1,x)^2}{\int_{0}^{1}\{H(r,x)-rH(1,x)\}^2dr} dF_0(x).\label{sn_infeas2}
	\end{align}
	Unfortunately neither limiting null distribution in the above display is pivotal because the covariance structure of $H(r,x)$ typically depends on the joint distribution of $\{X_t\}_{t\in \Z}$ and is unknown.

	To tackle this problem, we propose to do sample splitting before constructing the SN statistic. We split the sample into two parts: $\X_1 =\{X_1,\dots,X_\fa\}$ and $\X_2 = \{X_{\fa+1},\dots,X_n\}$, where $\alpha\in (0,1)$ is the splitting ratio. Let $F_{a:b}(x)$ be an estimator of $F(x)$ based on subsample $\{X_a,X_{a+1},\dots,X_b\}$ and define 
	\begin{align}\label{eq_11}
		P_\alpha(x) = \frac{\fa}{\sqrt{n}}\big[  F_{1:\fa}(x)  - F_0(x) \big],
	\end{align}
	that is, $P_\alpha(x)$ estimates the deviation of $F(x)$  from $F_0(x)$ at $x$ (up to some scaling factor). For $k=\fa{+}1,\dots,n$, let $S_k = \int_\Omega  P_\alpha(x)  
	\{F_{\fa+1:k}(x){-}F_0(x)\} dx=\li P_{\alpha},F_{\fa+1:k}{-}F_0\ri$, where we define $\li g_1,g_2\ri=\int_{\Omega} g_1(x) g_2(x) dx$. Hence $\{S_k\}_{k=\fa+1,\dots,n}$ can be viewed as the projected one-dimensional sequence (if $d>1$, we have $\mathbf{P}_\alpha(x) = (P_\alpha^1(x),\dots,P_\alpha^d(x))^\top$ and similar superscripts are used for $\mathbf{F}_{\fa+1:k}(x)$ and $\mathbf{F}_0(x)$. We define $S_k$ = $\sum_{i=1}^{d}\li P^i_{\alpha},F^i_{\fa+1:k}-F^i_0\ri$). In this way, we manage to reduce the dimension from $\infty$ to $1$. 
	
	After projection and dimension reduction, we can apply SN to the one-dimensional sequence $S_k$ and define the test statistic as
	\begin{align}
		\label{eq:Tn}
		T_n=\frac{\sqrt{n-\fa} S_n}{ (n-\fa)^{-1}\sqrt{\sum_{k=\fa+1}^{n} (k-\fa)^2( S_k- S_n)^2}}.
	\end{align} 
	
	\begin{remark}
		In the i.i.d. data setting, \cite{kim2024} proposed to test hypothesis on a functional parameter via cross U-statistics (see Equation (31) therein). For linear estimators $F_{a:b}(x) = \frac{1}{b-a+1}\sum_{t=a}^bg(X_t,x)$, our $S_n$ is proportional to their cross U-statistic with kernel $h(X_i,X_j) = \int_{\Omega}[g(X_i,x)-F_0(x)][g(X_j,x)-F_0(x)]dx$. Under the alternative, $S_n$ is proportional to the $L_2$ norm of the signal $\sqrt{\int_{\Omega}|F(x)-F_0(x)|^2dx}$. So in a sense our test statistic can be regarded as an extension of the method in \cite{kim2024} from i.i.d. to time series setting. With that being said, our formulation also covers the case when the estimator $F_{a:b}(x)$ is not  linear and the composite null case (see Section~\ref{sec_comp}), so is more broadly applicable. In addition, our projection is more explicit and our theoretical treatment is considerably different from that in \cite{kim2024}. 
		
	\end{remark}
	
	Let $\Gamma=\{(r,t)|r\leq t; r,t \in[0,1]\}$, we use $\ell_\infty[\Gamma\times \Omega]$ to denote the space of bounded real valued functions on $\Gamma \times \Omega$ (similarly, $\ell^p_\infty[\Gamma\times \Omega]$ denote the space of bounded $\R^p$ valued functions on $\Gamma \times \Omega$). The following weak convergence assumption is needed in deriving the asymptotic properties of $T_n$.
	\begin{assumption}[FCLT]\label{assump_1null}
		we assume that:
		\begin{eqnarray}
			\frac{\ft{-}\fr {+}1}{\sqrt{n}}\big[F_{\fr :\ft}(x){-}F(x)\big] \rightsquigarrow H(t,x){-}H(r,x) \mbox{ on } \ell_\infty[\Gamma\times \Omega],
		\end{eqnarray}
		where $H(r,x): [0,1]\times \Omega \to \R$ is a mean zero Gaussian process with $\cov[H(r_1,x_1),H(r_2,x_2)]=\min\{r_1,r_2\} C(x_1,x_2)$ for any $r_1,r_2\in [0,1]$ and $x_1,x_2 \in \Omega$. We further assume that the covariance function  $C(x,y):\Omega\times \Omega\to\R$ of the mean zero Gaussian process $H(1,x):\Omega \to \R$ satisfy the condition that $\int_\Omega C(x,x)dx<\infty$.
	\end{assumption}
	
	By the continuous mapping theorem, Assumption \ref{assump_1null} is implied by the uniform FCLT: $\frac{\ft}{\sqrt{n}}\big[F_{1:\ft}(x)-F(x)\big] \rightsquigarrow H(t,x) \mbox{ on } \ell_\infty[[0,1]\times \Omega]$ (see \cite{mohr2020}, \cite{volgushev2014}, \cite{andrews1994}) if the estimator $F_{\fr:\ft}(x)$ is approximately linear, that is, for any $0\leq r<t\leq 1$,
	\begin{equation}\label{eq_linear}
		F_{1:\ft}(x)=\frac{\fr{-}1}{\ft}F_{1:\fr{-}1}(x){+}\frac{\ft{-}\fr+1}{\ft}F_{\fr:\ft}(x)+R'(r,t,x), 
	\end{equation}
	where $\frac{\ft}{\sqrt{n}}R'(r,t,x)=o_p(1)$ uniformly in $r,t$ and $x$. This provides a simple way to verify Assumption \ref{assump_1null} and we now demonstrate how it can be verified for Examples \ref{example_1}-\ref{example_3} stated at the beginning of Section \ref{sec:testing}. For more complex functionals, verifying Assumption~\ref{assump_1null} can be nontrivial and very challenging; see Appendix \ref{app_verify} of the supplement for more detailed discussion.
		
		\begin{example}\label{example1}
			If $F(x){=}G(x)$ is the marginal cdf of $\{X_t\}_{t\in \Z}$ and $F_{a:b}(x)=\frac{1}{b-a+1}\sum_{t=a}^b\mathbf{1}(X_t\leq x) $, then $R'(r,t,x)\equiv 0$ and the uniform FCLT for $F_{1:\ft}(x)$ is proved in \cite{berkes2009} under some suitable  weak dependence condition. To show the limiting distributions in Equations (\ref{sn_infeas1}) and (\ref{sn_infeas2}) are not pivotal, suppose we want to test $H_0:G(x) = \Phi(x)$, then we have $\sup_{x \in\Omega}SN_n(x)\stackrel{\D}{\to}\sup_{x \in\Omega} \frac{H(1,x)^2}{\int_{0}^{1}\{H(r,x)-rH(1,x)\}^2dr}$ where $\cov(H(r_1,x_1),H(r_2,x_2)) = \min\{r_1,r_2\}\sum_{t=-\infty}^\infty\E\{[\id(X_0\leq x_1)-\Phi(x_1)][\id(X_t\leq x_2)-\Phi(x_2)]\}$, which depends on the long run cross-covariance between $\{\id(X_t\leq x_1)\}$ and $\{\id(X_t\leq x_2)\}$. Hence the limiting distribution $\sup_{x \in\Omega} \frac{H(1,x)^2}{\int_{0}^{1}\{H(r,x)-rH(1,x)\}^2dr}$ is not pivotal. The non-pivotalness of the distribution in Equation (\ref{sn_infeas2}) can be shown in a similar way.
		\end{example}
		\begin{example}\label{example2}
			If $F(x){=}Q(\tau)$ is the quantile function of $X_1$ with $x{=}\tau\in\Omega{=}(p_1,p_2)\subset (0,1)$, then we let $F_{a:b}(\tau)$ be the sample quantile function calculated on $\{X_a,X_{a+1},\dots,X_b\}$. A uniform FCLT for $F_{1:\ft}(\tau)$ and $\frac{\ft}{\sqrt{n}}R'(r,t,\tau)=o_p(1)$ uniformly in $(r,t,\tau)$ can be shown by following the arguments in deriving the Bahadur Representation for sample quantile of weakly dependent time series (see Theorems 1-3 and their proofs in \cite{wu2005bahadur}).
		\end{example}
		
		\begin{example}\label{example3}
			If $F(x){=}SDF(\lambda)$ is the spectral distribution function (sdf) of $\{X_t\}_{t\in \Z}$ with $x{=}\lambda\in\Omega{=}[0,\pi]$, then we let $F_{a:b}(\lambda) = \int_0^\lambda I_{a:b}(s)ds$ where 
			$I_{a:b}(s) = (2\pi (b-a))^{-1}|\sum_{t=a}^b(X_t-\bar X)e^{its}|^2$ is the periodogram calculated on subsample $\{X_a,X_{a+1},\dots,X_b\}$ and $\bar X$ is the sample mean of $\{X_t\}_{t=1}^n$. The uniform FCLT for $F_{1:\ft}(\lambda)$ and Equation (\ref{eq_linear}) are proved in Theorem 3.2 and Proposition 4.1 in \cite{giraitis1992testing} under mild assumption on the temporal dependence and spectral density function. 
		\end{example}

	As shown below, the limiting null distribution of $T_n$ is $U_1$, which is symmetric around zero and the upper critical values of $U_1^2 = \Lambda_1$ have been tabulated in \cite{lobato2001}. Given the level $\gamma \in (0,1)$ (say, $0.05$), our test is $\boldsymbol{1}(T_n>U_{1,\gamma})$, where $U_{1,\gamma}$ is the $100(1-\gamma)$th upper percentile of $U_1$ (which is also the $100(1-2\gamma)$th upper percentile of $\Lambda_1$). The following theorem shows the asymptotic properties of $T_n$, which is proved in Appendix \ref{appen_th_1null2}.
	\begin{theorem}\label{th_1null2}
		Suppose Assumption \ref{assump_1null} holds. Then (i) under $H_0$, we have 
		\begin{equation}
			T_{n} \stackrel{\D}{\to}  U_1 = \frac{B(1)}{\sqrt{\int_{0}^{1}\{B(r)-rB(1)\}^2dr}},
		\end{equation}
		(ii) under $H_A$, assume the true value of $F(x)$ is $F_1(x)=F_{1n}(x)$ and denote $\|F_1{-}F_0\|_2=\big[\int_\Omega |F_1(x){-}F_0(x)|^2dx\big]^{1/2}$. We have 
		\begin{enumerate}[1.]
			\item If $\sqrt{n}\|F_{1n}{-}F_0\|_2\to\infty$, then $T_{n} \stackrel{p}{\to} \infty$, thus the limiting power of our test is 1.
			\item If $\sqrt{n}[F_{1n}(x){-}F_0(x)]\to c(x)$ for some function $c(x)$ such that $\int_\Omega |c(x)|^2dx<\infty$, then 
			$$	P_\alpha(x) = \frac{\fa}{\sqrt{n}}[F_{1:\fa}(x){-}F_0(x)] \rightsquigarrow H(\alpha,x)+\alpha c(x)=\wt c(x) \mbox{ and }T_n\stackrel{\D}{\to}\wt U_1,$$
			where the conditional distribution of $\wt U_1$ given $\wt c(x)$ is 
			$$\wt U_1\Big| \wt c(x) \stackrel{d}{=}\frac{B(1)+\frac{\sqrt{1-\alpha}\int_\Omega\wt c(x)c(x)dx}{\sqrt{C^A}}}{\sqrt{\int_0^{1}\big[B(r)-rB(1)\big]^2 dr}},$$
			where $C^A=\int_\Omega\int_\Omega \wt c(x)C(x,y)\wt c(y)dxdy$ and $\wt c(x)$ is independent of $\{B(r)\}_{r\in[0,1]}$.
			\item If $\sqrt{n}\|F_{1n}{-}F_0\|_2\to 0$, then $T_{n}\stackrel{\D}{\to} U_1,$ so our test has trivial power asymptotically.
		\end{enumerate}
	\end{theorem}
	\begin{remark}
		The results in Theorem \ref{th_1null2} still hold for fixed $d>1$, with $\|F_1{-}F_0\|_2=\sqrt{\sum_{i=1}^d\|F_1^i-F_0^i\|_2}$, $\int_\Omega |c(x)|^2dx=\int_\Omega \sum_{i=1}^{d}|c^i(x)|^2dx, \int_\Omega |c(x)|\sqrt{C(x,x)}dx = \sum_{i=1}^{d}\int_\Omega |c^i(x)|\sqrt{C^{ii}(x,x)}dx$ and $C^A =\int_\Omega\int_\Omega \sum_{i,j=1}^{d}\wt c^i(x)C^{ij}(x,y)\wt c^j(y)dxdy$. Here we use the $i$th superscript to indicate the $i$th component of an $\R^d$-valued function and $C^{ij}(x,y)$ is the covariance between $H^i(1,x)$ and $H^j(1,y)$. 
		
		When $d>1$, the same limiting null distribution $U_1$ can be derived under a multivariate version of Assumption~\ref{assump_1null}. The key is the independent increment along the time coordinate for the limiting Gaussian process.   Under $H_A$, after projection the signal $F_1(x){-}F_0(x)$ contained in $\X_1$ and $\X_2$ are reflected in the leading term of the numerator of $T_n$, which is proportional to  $\int_\Omega |F_1(x)-F_0(x)|^2dx$ (see Equation \ref{eq_app2_1} in the proof of Theorem \ref{th_1null2}). The non-central drift term is always positive and  adding up projected data from different components of an $\R^d$-valued function does not get the signal canceled out, and we only reject $H_0$ if $T_n$ is large.

	\end{remark}
	\begin{remark}
		Note that the projection $P_\alpha(x)$ defined in Equation (\ref{eq_11}) may not be the optimal direction of projection in terms of power maximization. To see that, assume $\Omega$ is compact, $C(x,y)$ is continuous and let $L^2(\Omega;m)$ be the space of square integrable functions on $\Omega$ with respect to the Lebesgue measure $m$ on $\R^s$, then by Mercer's Theorem \citep[Theorem 12.20,][]{wainwright2019high}, there exist a sequence of eigenfunctions $\{\phi_i(x)\}_{i=1}^\infty$ that form an orthonormal basis of $L^2(\Omega;m)$ and non-negative eigenvalues $\{\lambda_i\}_{i=1}^\infty$ such that $C(x,y)=\sum_{i=1}^\infty\lambda_i\phi_i(x)\phi_i(y)$. Assume $F_1(x){-}F_0(x)=c_n(x) = \sum_{i=1}^{\infty}b_i\phi_i(x)\in L^2(\Omega;m)$ under the alternative. Fix $P(x) =  \sum_{i=1}^{\infty}a_i\phi_i(x)\in L^2(\Omega;m)$ and we project the data in the second subsample along $P(x)$ and construct a one dimensional SN statistic. Then similar to part (ii).2 of Theorem \ref{th_1null2}, the statistic approximately follows the same distribution as 
		$$U_n{=}\frac{B(1)+\frac{\sqrt{n(1-\alpha)}\int_\Omega P(x)c_n(x)dx}{\sqrt{C^n}}}{\sqrt{\int_0^{1}\big[B(r)-rB(1)\big]^2 dr}},$$
		where $C^n=\int_\Omega\int_\Omega P(x)C(x,y)P(y)dxdy$
		for large enough $n$. Note that the numerator and denominator of $U_n$ are independent and conditioning on $\sqrt{\int_0^{1}\big\{B(r)-rB(1)\big\}^2 dr}$, $B(1)+\frac{\sqrt{n(1-\alpha)}\int_\Omega P(x)c_n(x)dx}{\sqrt{C^n}}$ follows normal distribution with variance one and mean $\frac{\sqrt{n(1-\alpha)}\int_\Omega P(x)c_n(x)dx}{\sqrt{C^n}}$. It is clear that the optimal direction of projection which maximizes  $P(U_n\geq t)$ for all $t>0$ is the one that maximizes  the mean. By letting $\sign(b_i) = \sign(a_i)$, we can instead maximize
		$$\frac{n(1-\alpha)\big[\int_\Omega P(x)c_n(x)dx\big]^2}{C^n} = \frac{n(1-\alpha)\big[\sum_{i=1}^{\infty}a_ib_i\big]^2}{\sum_{i=1}^{\infty}\lambda_ia_i^2}.$$
		
		If we assume $c_n(x) = \int_\Omega C(x,y)f(y)dy$ for some $f(y)\in L^2(\Omega;m)$, that is, $c_n(x)$ belongs to the range of the Hilbert–Schmidt operator $C(x,y)$, by Cauchy–Schwarz inequality the optimal direction of projection is proportional to $P_n^\ast(x) = \sum_{i=1}^{\infty}\frac{b_i}{\lambda_i}\phi_i(x)$ (here we use the convention $\frac{0}{0}{=}0$), while $P_\alpha(x)$ is an estimator of $c_n(x) = \sum_{i=1}^{\infty}b_i\phi_i(x)$ (up to a scaling constant). In practice, the estimation of optimal projection involves the consistent estimation of $C(x,y)$ and its eigenvalues and eigenfunctions, which seems a difficult task. As shown in \cite{zhang2023another}, the pursuit of optimal direction via consistent long run variance (LRV) estimation in the finite-dimensional parameter case may not be worthwhile due to the estimation error involved in LRV estimation, and we expect the same issue would arise here and our LRV operator/function $C(x,y)$ is even more difficult to estimate due to its infinite dimensional nature. 
		
	\end{remark}
	\begin{remark}\label{rmk_weight}
			If a practitioner has prior knowledge on the type of alternative hypothesis, one way to account for this prior knowledge and potentially increase the power of SS-SN test is to incorporate a weighting function $w(\cdot)$ when defining the projected data, i.e., 
			$$S_k = \int_\Omega  P_\alpha(x)  
			\{F_{\fa+1:k}(x){-}F_0(x)\}w(x) dx,$$
			and then build the SS-SN test statistic according to (\ref{eq:Tn}).  
			For example, if $F(x)$ is the marginal cdf and the practitioner is targeting an alternative under which $F(x)$ differs from $F_0(x)$ only in the right tail, then we can set $w(x) = \id(x\geq \tau)$ for some $\tau\in \R$ and the theoretical results derived in Theorem \ref{th_1null2} still holds.
	\end{remark}
	Under the null, the limiting distribution $U_1$ is pivotal and does not depend on the splitting ratio $\alpha$. Under the local alternative $\sqrt{n}[F_1(x)-F_0(x)]\to c(x)$, the limiting distribution of our test statistic depends on $\alpha$, $c(x)$, $H(\alpha,x)$ and $C(x,x)$. It appears difficult to discuss the optimal $\alpha$ theoretically and we shall address this empirically later in simulations.


	\subsection{Hypothesis testing for composite null}\label{sec_comp}

	In practice, one often can only specify the null up to a finite dimensional parameter. For example, suppose we test the null hypothesis that the marginal distribution of a time series is Gaussian with some unknown mean and variance, then the null is not completely specified, but depends on the unknown parameters. In this section, we focus on testing the null hypothesis $H_0: F(x)=F_0(x;\boldsymbol{\theta}_0)$ for any $x\in\Omega$ for a known function $F_0$ indexed by an unknown $s$-dimensional parameter $\boldsymbol{\theta}_0\in \boldsymbol{\Theta}\subset \R^s$ against the alternative $H_A:F(x)\neq F_0(x;\boldsymbol{\theta})$ for any $\boldsymbol{\theta} \in \boldsymbol{\Theta}$. In the rest of this section, we shall extend the SS-SN test statistic to this more realistic scenario. 
	
	As in the previous section, let $\{X_t\}_{t\in \Z}$ be a stationary time series and $\boldsymbol{\hat\theta}_{a:b}$ be an estimator of $\boldsymbol{\theta}$ based on the subsample $\{X_a,X_{a+1},\dots,X_b\}$. The following assumption is needed to derive the asymptotic properties of our test statistic. 
	
	\begin{assumption}\label{assump_nnull}
		For some $\boldsymbol{\theta}^\ast \in \boldsymbol{\Theta}$, (i) the derivative of $F_0(x;\boldsymbol{\theta})$ with respect to $\boldsymbol{\theta}$, $\nabla F_0(x;\boldsymbol{\theta})$, exists in a neighborhood of $\boldsymbol{\theta}^\ast$ for any $x\in \Omega$.
		
		(ii) the following Taylor expansion holds: for any $r<t \in [0,1]$ and $x \in \Omega$:
		\begin{align}
			F_0(x;\boldsymbol{\hat\theta}_{\fr :\ft})  = F_0(x;\boldsymbol{\theta}^\ast) + \nabla F_0(x;\boldsymbol{\theta}^\ast)^\top(\boldsymbol{\hat\theta}_{\fr :\ft}{-}\boldsymbol{\theta}^\ast) + R_{\fr :\ft }(x)
		\end{align}
		where $R_{\fr :\ft }(x)$ satisfies $\big\{\int_{\Omega}|R_{1 :n }(x)|^2dx\big\}^{1/2}{=}o_p(n^{-1/2})$ and $n^{-2}\sum_{t=1}^{n}\big\{t^2\int_{\Omega}|R_{1 :t }(x)|^2dx\big\} {=} o_p(1)$.
		
		(iii) let $F'_{\fr :\ft}(x)=F_{\fr :\ft}(x){-}\nabla F_0(x;\boldsymbol{\theta}^\ast)^\top\boldsymbol{\hat\theta}_{\fr :\ft}$ and $F'_0(x;\boldsymbol{\theta}^\ast)=F(x){-}\nabla F_0(x;\boldsymbol{\theta}^\ast)^\top\boldsymbol{\theta}^\ast$,
		\begin{equation}
			\frac{\ft{-}\fr { +}1}{\sqrt{n}}\big[F'_{\fr :\ft}(x){-}F'_0(x;\boldsymbol{\theta}^\ast)\big] \rightsquigarrow H'(t,x){-}H'(r,x) \mbox{ on } \ell_\infty[\Gamma\times \Omega],\nonumber
		\end{equation}
		where $H'(r,x): [0,1]\times \Omega \to \R$ is a mean zero Gaussian process with $\cov(H'(r_1,x_1),H'(r_2,x_2))=\min\{r_1,r_2\} C'(x_1,x_2)$ for any $r_1,r_2\in [0,1]$ and $x_1,x_2 \in \Omega$. We further assume that the covariance function  $C'(x,y):\Omega\times \Omega\to\R$ of the mean zero Gaussian process $H'(1,x):\Omega \to \R$ satisfy the condition that $\int_\Omega C'(x,x)dx<\infty$.
	\end{assumption}
	Assumption \ref{assump_nnull} is implied by the joint uniform FCLT of $F_{1:\ft}(x)$ and $\nabla F_0(x;\boldsymbol{\theta}^\ast)^\top\boldsymbol{\hat\theta}_{1:\ft}$ if both $F_{\fr:\ft}(x)$ and $\boldsymbol{\hat\theta}_{\fr:\ft}$ are approximately linear in the sense of Equation (\ref{eq_linear}) and $F_0(x;\boldsymbol{\theta})$ satisfies certain smoothness conditions. We now demonstrate how to verify Assumption \ref{assump_nnull} for Examples \ref{example1} and \ref{example3} discussed in Section \ref{sec_simple_null}.

		\begin{continueexample}{example1}
			Suppose we test the null hypothesis $H_0:F(x)=\Phi(\frac{x-\mu_0}{\sigma_0})$ for some $\boldsymbol{\theta}_0=(\mu_0,\sigma_0^2)^\top\in\boldsymbol{\Theta}=\R\times(0,\infty)$, where $\Phi(x)$ is the cdf of the standard normal distribution. We use the estimator $\boldsymbol{\hat \theta}_{a:b}=(\hat \mu_{a:b},\hat \sigma^2_{a:b})^\top$ where $\hat \mu_{a:b}=1/(b{-}a{+}1)\sum_{t=a}^{b}X_t$ and $\hat \sigma^2_{a:b}=1/(b{-}a{+}1)\sum_{t=a}^{b}(X_t{-}\hat \mu_{1:n})^2$. Let $F_{\fr:\ft}(x)$ be the empirical cdf based on the subsample $\{X_\fr,X_{\fr+1},\dots,X_\ft\}$. Clearly $F_{\fr:\ft}(x)$, $\hat \mu_{a:b}$ and $\hat \sigma^2_{a:b}$ satisfies Equation (\ref{eq_linear}). Since each coordinate of $\nabla F_0(x;\boldsymbol{\theta}_0)=(-\frac{1}{\sigma_0}\phi(\frac{x-\mu_0}{\sigma_0}),-\frac{x-\mu_0}{2\sigma_0^3}\phi(\frac{x-\mu_0}{\sigma_0}))^\top$ is continuous and uniformly bounded in $x$, it follows from  a multivariate generalization of Corollary 2.7 in \cite{mohr2020}, the following Uniform FCLT holds. That is,  
			\begin{equation}
				\frac{\ft}{\sqrt{n}} \begin{bmatrix}
					F_{1:\ft}(x) {-}F_0(x)  \\
					-\frac{1}{\sigma_0}\phi(\frac{x-\mu_0}{\sigma_0})\big[\hat\mu_{1:\ft}{-}\mu_0\big] \\
					{-}\frac{x-\mu_0}{2\sigma_0^3}\phi(\frac{x-\mu_0}{\sigma_0})\big[\frac{1}{\ft}\sum_{i=1}^{\ft}(X_i-\mu_0)^2{-}\sigma^2_0\big] 
				\end{bmatrix} \rightsquigarrow \boldsymbol{G}'(t,x) \text{ on }  \ell^3_\infty[[0,1]\times \Omega],
			\end{equation}
			for some $\R^3$ valued Gaussian process $\boldsymbol{G}'(t,x)$. By continuous mapping theorem, Assumption \ref{assump_nnull} holds under the null.
		\end{continueexample}
		\begin{continueexample}{example3}
			Suppose we test the null hypothesis $H_0:F(x) = SDF(\lambda)=\int_0^\lambda f(s)ds$ where $f(s)=\frac{\sigma_0^2}{2\pi(1 {-}\rho_0 \cos(s){+}\rho_0^2)}$ is the spectral density function of the AR(1) process $X_t=\rho_0X_{t {-}1}{+}\epsilon_t$ with $\epsilon_t\stackrel{i.i.d}{\sim}N(0,\sigma^2)$. That is, we are assessing the goodness-of-fit of AR(1) process to the time series using spectral distribution function. 
			Here we have $\boldsymbol{\theta}_0=(\rho_0,\sigma_0^2)^\top\in\boldsymbol{\Theta}=(0,1)\times(0,\infty)$ and each component of 
			$$\nabla SDF(\lambda;\boldsymbol{\theta}_0)=\left(\int_0^\lambda\frac{\sigma_0^2(\cos(s) {-}2\rho_0)}{2\pi(1 {-}\rho_0 \cos(s){+}\rho_0^2)^2}ds, \int_0^\lambda \frac{1}{2\pi(1 {-}\rho_0 \cos(s){+}\rho_0^2)}  ds  \right)^\top$$ is continuous and uniformly bounded in $\lambda$. We use the estimator $\boldsymbol{\hat \theta}_{a:b}=(\hat \gamma_{a:b}(1)/\hat \gamma_{a:b}(0),\hat \gamma_{a:b}(0)-\hat \gamma_{a:b}(1)^2/\hat \gamma_{a:b}(0))^\top$ where $\hat \gamma_{a:b}(1)=1/(b {-}a{+}1)\sum_{t=a}^{b {-}1}(X_t {-}\hat \mu_{1:n})(X_{t{{+}}1}{ {-}}\hat \mu_{1:n})$ and $\hat \gamma_{a:b}(0)=1/(b {-}a{+}1)\sum_{t=a}^{b}(X_t {-}\hat \mu_{1:n})^2$. Note that $\hat \gamma_{a:b}(1)$, $\hat \gamma_{a:b}(0)$ and $F_{a:b}(\lambda)=\int_{0}^{\lambda}I_{a:b}(s)ds$ can all be written in the form $\int_0^\pi b(s)I_{a:b}(s)ds$ (for $b_1(s)=\cos(s)$, $b_2(s)=1$ and $b_3(s)=\mathbf{1}_{[0,\lambda]}(s)$), for which the FCLT and approximate linearity can be shown in a similar way as Theorem 3.2 and Proposition 4.1 in \cite{giraitis1992testing}. Then by continuous mapping theorem, Assumption \ref{assump_nnull} holds under the null.
		\end{continueexample}

	Define $P_\alpha'(x)=\frac{\fa}{\sqrt{n}} [F_{1:\fa}(x){-}F_0(x;\boldsymbol{\hat\theta}_{1:\fa})]$ and for $k=\fa{+}1,\dots,n$, let $S'_k = \int_\Omega  P_\alpha'(x)  [F_{\fa{+}1:k}(x){-}F_0(x;\boldsymbol{\hat\theta}_{\fa{+}1:k})] dx$. Then the SS-SN test statistic is defined as
	\small
	\begin{equation*}
		T'_{n}=\frac{\sqrt{n{-}\fa} S'_n}{ (n{-}\fa)^{{-}1}\sqrt{\sum_{k=\fa{+}1}^{n} (k{-}\fa)^2\big\{S'_k-S'_n\big\}^2}}.
	\end{equation*}
	\normalsize
	The following theorem shows the asymptotic properties of $T'_{n}$ under the null and alternative, which is proved in Appendix \ref{appen_th_nnull}.
	
	\begin{theorem}\label{th_nnull}
		(i) Under $H_0$, suppose Assumption \ref{assump_nnull} holds with $\boldsymbol{\theta}^\ast=\boldsymbol{\theta}_0$, we have 
		\begin{equation}
			T'_{n} \stackrel{\D}{\to}       U_1,
		\end{equation}
		(ii) under $H_A$, let the true functional parameter be $F_1=F_{1n}$ and Assumption \ref{assump_nnull} holds with $\boldsymbol{\theta}^\ast=\boldsymbol{\theta}_{1n}$, we have 
		\begin{enumerate}[1.]
			\item If $\sqrt{n}\|F_1(x){-}F_0(x;\boldsymbol{\theta}_{1n})\|_2\to\infty$, then $T'_{n} \stackrel{p}{\to} \infty$, thus the limiting power of the test $\boldsymbol{1}(T'_n>U_{1,\gamma})$ is 1.
			\item If $\sqrt{n}[F_1(x){-}F_0(x;\boldsymbol{\theta}_{1n})]\to c'(x)$ for some function $c'(x)$ such that $\int_\Omega |c'(x)|^2dx<\infty$, then 
			$$P_\alpha'(x) \rightsquigarrow  H'(\alpha,x)+\alpha c'(x)=\wt c'(x) \mbox{ and }T'_n\stackrel{\D}{\to}\wt U_1',$$
			where the conditional distribution of $\wt U_1'$ given $\wt c'(x)$ is 
			$$\wt U_1'\Big| \wt c'(x) \stackrel{d}{=}\frac{B(1)+\frac{\sqrt{1-\alpha}\int_\Omega\wt c'(x)c'(x)dx}{\sqrt{\wt C^A}}}{\sqrt{\int_0^{1}\big[B(r)-rB(1)\big]^2 dr}},$$
			where $\wt C^A=\int_\Omega\int_\Omega \wt c'(x)C'(x,y)\wt c'(y)dxdy$ and $\wt c'(x)$ is independent of $\{B(r)\}_{r\in[0,1]}$.
			\item If $\sqrt{n}\|F_1(x){-}F_0(x;\boldsymbol{\theta}_{1n})\|_2\to 0$, then $T'_{n}\stackrel{\D}{\to} U_1$, our test has trivial power asymptotically.
		\end{enumerate}
	\end{theorem}

	\begin{remark}
		
		For the $s$-dimensional parameter $\boldsymbol{\theta}$, typically we need at least $s$ sample points to calculate $\boldsymbol{\hat\theta}_{a:b}$. Let $\wt k \geq \fa{+}1$ be the smallest integer such that $\boldsymbol{\hat\theta}_{\fa+1:\wt k}$ can be calculated. We can set $\boldsymbol{\hat\theta}_{\fa+1: k}=\boldsymbol{\hat\theta}_{\fa+1:\wt k}$ for any $\fa{+}1\leq k\leq \wt k$ and it will not affect the result in the above theorem.
	\end{remark}
	
	\begin{remark}
		
		Our treatment of the estimation effect is similar to the one proposed in \cite{kuan2006}, where they used SN method to test composite null hypothesis on a finite dimensional parameter. Note that we used the recursive estimator $\boldsymbol{\hat\theta}_{\fa{+}1:k}$, instead of $\boldsymbol{\hat\theta}_{\fa{+}1:n}$, in the denominator of $T'_n$ because if we use $\boldsymbol{\hat\theta}_{\fa{+}1:n}$, the $\pm F_0(x;\boldsymbol{\hat\theta}_{\fa{+}1:n})$ terms in $S'_k{-}S'_n$ cancel out but the estimation effect in the numerator remains, which leads to a non-pivotal limiting distribution under the null. 
	\end{remark}

	\subsection{Change point testing}
	\label{subsec:cp}
	
	Let $\{X_t\}_{t=1}^n$ be an observed time series and $\{Y_t\}_{t\in \Z}$, $\{Z_t\}_{t\in \Z}$ be two stationary time series with unknown functional parameter $F^{(0)}\neq F^{(1)}$, which depends on the joint distributions of $\{Y_t\}_{t\in \Z}$ and $\{Z_t\}_{t\in \Z}$ respectively. Under the null hypothesis, there is no change point and $X_t=Y_t$ is a stationary time series. Under the single change-point alternative, we follow the framework of \cite{dette2020} and assume that $X_t=Y_t{\id}(1\le t\le k^*)+Z_t{\id}(k^*+1\le t\le n)$, where $k^\ast=\lfloor  n \tau_0 \rfloor$ for some unknown $\tau_0 \in (b,1-b)$. Here $b$ is usually called a trimming parameter \citep{andrews1993}. 
	Let $F_t=F^{(0)}$ if $X_t=Y_t$ and $F^{(1)}$ otherwise. Then it is equivalent to test the null hypothesis $H_0:F_1=F_2=\cdots=F_n$ against $H_A: F_1=F_2\cdots= F_{k^\ast}\neq F_{k^\ast+1}=F_{k^\ast+2}\cdots=F_n$. Without loss of generality, assume $F_t=F^{(0)}$ under $H_0$ and $F_t=F^{(0)}\id(t\leq k^\ast)+F^{(1)}\id(t> k^\ast)$ under $H_A$.

	We now extend our SS-SN method to this change-point testing problem. 
	Define $P_{b}(x)=\frac{\fb}{\sqrt{n}}[ F_{1:\fb}(x){-} F_{n-\fb+1:n}(x)]$, that is, $P_{b}(x)$ estimates the deviation of $F_0$ from $F_1$ at $x$ up to some scaling factor. For $a_1,a_2=\fb{+}1,\dots,n{-}\fb{-}1$, let $Q_{a_1:a_2} = \int_\Omega P_{b}(x)F_{a_1:a_2}(x)dx=\li P_b,F_{a_1:a_2}\ri $ be the projected sequence. Let 
	$$T_n(k)=\frac{(k-\fb)(n-\fb-k)}{(n-2\fb)^{3/2}}\big\{Q_{\fb+1:k}- Q_{k+1:n-\fb}\big\}~\mbox{and}$$
	\begin{align}
		V_n(k) =& (n-2\fb)^{-1}\Big\{\sum_{t=\fb+1}^{k-1}\frac{(t-\fb)^2(k-t)^2}{(k-\fb)^2}\{ Q_{\fb+1:t}- Q_{t+1:k}\}^2+\nonumber \\
		&\qquad\qquad\qquad\qquad  \sum_{t=k+2}^{n-\fb}\frac{(n-\fb-t+1)^2(t-k-1)^2}{(n-\fb-k)^2}\{ Q_{t:n-\fb}- Q_{k+1:t-1} \}^2 \Big\}^{1/2}.\nonumber
	\end{align}
	The SS-SN test statistic is defined as
	$$G_n = \sup_{k=\fb+2,\fb+3,\dots,n-\fb-2}\frac{T_n(k)}{V_n(k)}.$$ 
	Note that in the self-normalizer, we used the contrast statistic form as advocated in \cite{lavitas2018}. For mean testing problems, CUSUM and contrast statistics are equivalent but not otherwise. Below we shall derive the asymptotic distribution of $G_n$ under $H_0$ under Assumption \ref{assump_1null} and  derive the asymptotic properties of $G_n$ under local alternatives based on the following assumption.
	
	\begin{assumption}\label{assump_cp}
		Let $\hat F_{a_1:a_2}$, $\wt F_{a_1:a_2}$ be estimators of the functional parameter based on $\{Y_{a_1},Y_{a_1+1},\dots,Y_{a_2}\}$ and $\{Z_{a_1},Z_{a_1+1},\dots,Z_{a_2}\}$. We assume that:
		\begin{enumerate}[(a)]
			\item \label{assump_cp_a} $\fb {+}1 < \lfloor n\tau_0 \rfloor  < n{-}\fb{-}1$ for some $b \in (0,0.5)$.
			\item \label{assump_cp_b}$F_{\fr:\ft}$ is approximately linear in the sense of Equation (\ref{eq_linear}).
			\item \label{assump_cp_c}		\begin{equation}
				\frac{\ft}{\sqrt{n}} \begin{pmatrix}
					\hat F_{1:\ft}(x) {-}F^{(0)}(x)  \\
					\wt F_{1:\ft}(x) {-}F^{(1)}(x)
				\end{pmatrix} \rightsquigarrow 
				\begin{pmatrix}
					H_1(t,x) \\
					H_2(t,x)
				\end{pmatrix} \text{ on }  \ell^2_\infty[[0,1]\times \Omega],
			\end{equation}
			where $H_i(r,x):[0,1]\times\Omega\to \R$ is mean zero Gaussian processes with $\cov[H_i(r_1,x_1),H_i(r_2,x_2)]=\min\{r_1,r_2\} C_i(x_1,x_2)$ and the covariance functions $C_i(x,y)$ satisfy the condition that $\int_\Omega C_i(x,x)dx<\infty$ for $i=1,2$.
			\item \label{assump_cp_d} $\cov[H_1(r_1,x_1),H_2(r_2,x_2)]=\min\{r_1,r_2\} C_3(x_1,x_2)$, where $|\min\{r_1,r_2\} C_3(x_1,x_2)|\leq \sqrt{r_1r_2C_1(x_1,x_1)C_2(x_2,x_2)}$.
			\item \label{assump_cp_e}$C_1(x,y) = C_2(x,y)$ for any $x,y \in \Omega$.
		\end{enumerate}
	\end{assumption}
	
	Under Assumption~\ref{assump_cp}(a), the change point can not be in the first and last $\fb$ sample points. We have the following theorem on the asymptotic properties of $G_n$, which is proved in Appendix \ref{appen_th_cpn}.
		\newcounter{cccp11}
		\begin{theorem}\label{th_cpn}
			(i) Under $H_0$, suppose Assumption \ref{assump_1null} holds, then we have
			\begin{equation}
				G_{n} \stackrel{D}{\to} G_1 \stackrel{d}{=} \sup_{r \in [0,1]}  \frac{B(r)- rB(1)}  {\Big\{\int_{0}^{r} \big[B(s)-\frac{s}{r}B(r)\big]^2 ds + \int_{r}^{1}\big[ B(1)-B(s)-\frac{1-s}{1-r}(B(1)-B(r)) \big]^2 ds \Big\}^{1/2}}.\nonumber
			\end{equation}
			(ii) Under $H_A$, suppose part (\ref{assump_cp_a}), (\ref{assump_cp_b}) and (\ref{assump_cp_c}) of Assumption \ref{assump_cp} hold and denote ${\Delta}(x)=\Delta_n(x)=F^{(1)}(x){-}F^{(0)}(x)$, then we have 
			\begin{enumerate}[1.]
				\item  If $\sqrt{n}\|{\Delta}\|_2\to\infty$, then $G_{n} \stackrel{p}{\to} \infty$, thus the limiting power of the test $\boldsymbol{1}(G_n>G_{1,\gamma})$ is 1, where $G_{1,\gamma}$ is the $100(1{-}\gamma)$th upper percentile of $G_1$. 
				\item  If $\sqrt{n}\Delta(x)\to \delta(x)$ for some continuous function $\delta(x)$ such that $\int_\Omega |\delta(x)|^2dx<\infty$, then we have $P_b(x) \rightsquigarrow H_1(b,x){-}H_2(1,x){+}H_2(1{-}b,x){-}b\delta(x)= \wt c''(x) \mbox{ and } G_{n} \stackrel{\D}{\to}  \wt G''$ where $\wt G''$ is defined in Equations (\ref{eqg1}) and (\ref{eqg2}).
				\setcounter{cccp11}{\value{enumi}} 
			\end{enumerate}
			Below we further assume part (\ref{assump_cp_d}) and (\ref{assump_cp_e}) of Assumption \ref{assump_cp} hold.
			\begin{enumerate}[1.]
				\setcounter{enumi}{\value{cccp11}}
				\item If $\delta(x)$ is non-zero continuous function such that $\int_\Omega |\delta(x)|^2dx<\infty$, then the conditional distribution of $\wt G''$ given $\wt c''(x)$ is
				\begin{equation}
					\wt G''\big|\wt c''(x) {\stackrel{d}{=}} \sup_{r \in [0,1]}  \frac{B''(r){-} rB''(1)}  {\Big\{\int_{0}^{r} \big[B''(s){-}\frac{s}{r}B''(r)\big]^2 ds {+} \int_{r}^{1}\big[ B''(1){-}B''(s){-}\frac{1{-}s}{1{-}r}(B''(1){-}B''(r)) \big]^2 ds\Big\}^{1/2} },\nonumber
				\end{equation}
				where the process $B''(r)$ is defined as $B''(r)=B(r){+} \frac{1}{\sqrt{1-2b}}J((1{-}2b)r{+}b)$ with $J(r)=\frac{1}{\sqrt{\int_\Omega\int_\Omega \wt c''(x) C_1(x,y)\wt c''(y)dxdy}}\int_\Omega\wt c''(x){J}(r,x)dx$, ${J}(r,x)=(r-\tau_0)\delta(x)\id(r{\geq} \tau_0)$ and $\wt c''(x)$ is independent of $\{B(r)\}_{r\in[0,1]}$.
				\item If $\sqrt{n}\|{\Delta}\|_2\to 0$, then $G_n \stackrel{D}{\to} G_1,$ so our test has trivial power asymptotically.
			\end{enumerate}
	\end{theorem}
	
	It is interesting to note that the above limiting null distribution $G_1$ is identical to the one in \cite{gaowangshao2023}, it does not depend on $b$, and its critical values have been simulated in \cite{gaowangshao2023}.

	\section{Simulation studies}
	\label{sec:simulation}
	
	In this section, we shall examine the finite sample performance of our proposed SS-SN tests and compare with some existing methods. Specifically, we present simulation results for time reversibility tests in Section~\ref{sim:1}, testing composite null of Gaussian marginal distribution in Section~\ref{sim:2} and testing for a change point in spectral distribution function in Section~\ref{sim:3}. We set the nominal level to be $0.05$ in all testing results reported here. 
	
	\subsection{Testing for time-reversibility}
	\label{sim:1}

	Time reversibility is an important feature of time series data.   To be specific, let $F_k(x,y)$ denote the joint distribution of $(X_t,X_{t+k})$ for a stationary time series $\{X_t\}_{t\in\Z}$, we say that $\{X_t\}_{t\in\Z}$ is pairwise time-reversible if $F_k=F_{-k}$ for all $k\in\Z$. Due to the fact that $\cov(X_t,X_{t-k})=\cov(X_t,X_{t+k})$, classical spectral analysis is not equipped to  capture this feature since only the second order property is preserved in spectral density/distribution function  \citep{fan2003}. 
	
	\cite{goto2021} proposed to use copula spectral distribution function (csdf) to test time-reversibility, which is defined as $F_{csdf}(\lambda,\tau_1,\tau_2)=\int_0^\lambda f(\omega,\tau_1,\tau_2)d\omega$ where $f(\omega,\tau_1,\tau_2)=\frac{1}{2\pi}\sum_{k\in\Z}e^{-i\omega k}\cov(\id(F_{cdf}(X_t)\leq \tau_1),\id(F_{cdf}(X_{t-k})\leq \tau_2))$ and $F_{cdf}$ is the marginal cdf of $X_t$. As stated in Proposition 4.1 of \cite{goto2021}, $\{X_t\}_{t\in\Z}$ is pairwise time-reversible if and only if $\im(F_{csdf}(\lambda,\tau_1,\tau_2))=0$ for all $(\lambda,\tau_1,\tau_2)\in[0,\pi]\times(0,1)^2$, where $\im(\cdot)$ denotes the imaginary part of a complex number. Let $I_{a:b}(\omega,\tau_1,\tau_2)$ be the cross-periodogram calculated on subsample $\{ (\id(\hat F_{cdf}(X_t)\leq \tau_1),\id(\hat F_{cdf}(X_t)\leq \tau_2))\}_{t=a}^b$ where $\hat F_{cdf}$ is the empirical cdf of $\{X_t\}_{t=1}^n$, then our SS-SN method can be used to test time-reversibility by letting $F(x)=\im(F_{csdf}(\lambda,\tau_1,\tau_2))$ and $F_{\fr,\ft}(x)=\im(\int_{0}^{\lambda}I_{\fr,\ft}(\omega,\tau_1,\tau_2)d\omega)$, where $x=(\lambda, \tau_1,\tau_2)$. To avoid complication on the boundary of $(\tau_1,\tau_2)$, we set $\Omega=[0,\pi]\times[\eta,1-\eta]^2$ for some small $\eta\in(0,1/2)$, and the null hypothesis of time-reversibility implies the following hypothesis
	$H_0: \sup_{(\lambda,\tau_1,\tau_2)\in\Omega}|\im(F_{csdf}(\lambda,\tau_1,\tau_2))|=0.$
	
	To examine the finite sample size of SS-SN test statistic, under $H_0$, we assume the data comes from the AR(1) model: ${X}_t{ =} \rho{X}_{t-1}{+}{\epsilon}_t,$ where ${\epsilon}_t \stackrel{iid}{\sim} N(0,1)$. 
	The experiment is repeated 2000 times with the length of the time series $n\in \{128,256,512\}$ and $\rho\in\{-0.7,-0.5,0.2,0.5,0.7\}$. As comparison, we also include the test statistics used in \cite{goto2021} (both with and without correction term, denoted as $Goto^c$ and $Goto$ respectively). Since their test requires subsampling, we set the subsampling block length $b\in\{n/8,n/4,n/2\}$. 
	In actual calculation, $(\lambda,\tau_1,\tau_2)\in[0,\pi]\times(0,1)^2$ needs to be discretized. In Table~\ref{tab_triv_size}, we first take $\lambda$ from the set $\{\frac{\pi i}{40}:i=0,1,\dots,40\}$ and $\tau_1,\tau_2\in\{k/40;k=1,2,\dots,39\}$. We call this discretization scheme $D_1$. Then we employ a coarser grid, which is the same as the one used in \cite{goto2021} and in this setting $\lambda\in\{\frac{\pi i}{16}:i=0,1,\dots,16\}$ and $\tau_1,\tau_2\in\{k/8:k=1,2,\dots,7\}$. This latter scheme is called $D_2$. 
	
	As shown in Table \ref{tab_triv_size}, SS-SN has accurate size across all combinations of $(n,\rho,\alpha)$, and the size is not influenced by the resolution of discretization. By contrast, the sizes for $Goto$ and $Goto^c$ depend heavily on the block length and the discretization resolution. As shown in $D_1$ of Table \ref{tab_triv_size}, when we take a finer grid for $(\lambda,\tau_1,\tau_2)$, $Goto$ is severely oversized for $n=128,256$ and $Goto^c$ goes from oversized to undersized as $n$ increases when $b=n/8$. For $b=n/4,n/2$, $Goto^c$ becomes more undersized as $n$ increase. When we take a coarser grid for $(\lambda,\tau_1,\tau_2)$, as shown in $D_2$ of Table \ref{tab_triv_size}, $Goto$ is still oversized for $n=128,256$. For $Goto^c$, the size accuracy deteriorates as $n$ increases when $b=n/8$ and for $b=n/4,n/2$ it is still severely undersized.

	\begin{footnotesize}
		
		\begin{table}[H]
			\centering
			\setlength{\tabcolsep}{8pt}
			\scalebox{0.84}{
				\begin{tabular}{ccc|ccc|ccc|ccc}
					\toprule
					\midrule
					\multirow{2}{*}{ }   &	\multirow{2}{*}{$n$}       &\multirow{2}{*}{$\rho$}&\multicolumn{3}{c|}{SS-SN}&\multicolumn{3}{c|}{$Goto$} &\multicolumn{3}{c}{$Goto^c$}    \\
					&	&&$\alpha=0.15$&$\alpha=0.3$&$\alpha=0.5$&$b_1$& $b_2$&$b_3$ &$b_1$& $b_2$&$b_3$   \\ \hline
					\multirow{15}{*}{$D_1$}& \multirow{5}{*}{128}                        &-0.7 &    5.20   &3.95   &5.10   &24.90 &12.35&10.85  &16.65 &2.80 &0.00          \\ \cline{3-12} 
					&&-0.5 &    5.85   &5.75   &6.05   &23.30 &10.95&10.70  &14.80 &1.85 &0.00     \\ \cline{3-12} 
					&&0.2  &    4.55   &5.45   &5.30   &24.35 &11.15&10.60  &15.30 &1.60 &0.00          \\ \cline{3-12} 
					&&0.5  &    5.85   &5.85   &5.85   &27.85 &12.25&10.30  &17.45 &2.00 &0.00       \\ \cline{3-12} 
					&&0.7  &    5.30   &4.85   &5.40   &32.55 &14.40&11.00  &20.50 &2.85 &0.00         \\ \cline{2-12} 
					&\multirow{5}{*}{256}                        &-0.7 &    5.35   &5.10   &5.45   &12.70 &8.75 &9.10   &6.70  &1.20 &0.00          \\ \cline{3-12} 
					&&-0.5 &    5.45   &5.95   &5.35   &10.65 &7.80 &8.95   &5.80  &0.75 &0.00     \\ \cline{3-12} 
					&&0.2  &    5.10   &4.70   &5.10   &10.60 &7.40 &8.10   &5.45  &0.65 &0.00          \\ \cline{3-12} 
					&&0.5  &    5.55   &4.60   &5.35   &11.90 &8.20 &7.75   &6.45  &0.90 &0.00       \\ \cline{3-12} 
					&&0.7  &    5.15   &4.70   &4.90   &12.95 &7.85 &8.15   &6.25  &0.80 &0.00         \\ \cline{2-12} 							
					&\multirow{5}{*}{512}                        &-0.7 &    5.55   &5.15   &5.00   &7.30  &7.00 &8.40   &3.55  &0.90 &0.00        \\ \cline{3-12} 
					&&-0.5 &    5.75   &4.95   &5.50   &6.45  &5.60 &7.35   &3.15  &0.65 &0.00     \\ \cline{3-12} 
					&&0.2  &    5.60   &5.40   &4.95   &5.90  &5.40 &7.95   &2.75  &0.85 &0.00         \\ \cline{3-12} 
					&&0.5  &    5.75   &5.85   &5.30   &7.75  &6.65 &7.65   &4.05  &1.00 &0.00        \\ \cline{3-12} 
					&&0.7  &    4.75   &4.70   &5.10   &6.95  &5.90 &7.55   &2.75  &0.65 &0.00          \\\hline	
					\multirow{15}{*}{$D_2$} 			&\multirow{5}{*}{128}                        &-0.7 &    4.65   &4.75   &5.45   &10.50 &7.75 &8.25   &6.75 &2.45 &0.05          \\ \cline{3-12} 
					&&-0.5 &    4.95   &5.45   &6.05   &8.30  &6.55 &8.15   &5.60 &1.45 &0.10     \\ \cline{3-12} 
					&&0.2  &    4.65   &4.65   &5.35   &8.40  &6.70 &8.30   &4.55 &0.70 &0.00          \\ \cline{3-12} 
					&&0.5  &    6.00   &5.50   &5.75   &10.10 &7.45 &8.05   &6.20 &2.15 &0.00       \\ \cline{3-12} 
					&&0.7  &    4.80   &4.85   &6.00   &12.55 &7.70 &8.50   &7.20 &1.95 &0.05           \\ \cline{2-12}                   
					&\multirow{5}{*}{256}                        &-0.7 &    5.65   &5.55   &5.85   &8.30  &6.95 &9.50   &5.10  &1.95 &0.20          \\ \cline{3-12} 
					&&-0.5 &    5.15   &6.25   &5.75   &7.20  &6.15 &7.85   &4.15  &1.50 &0.10     \\ \cline{3-12} 
					&&0.2  &    4.75   &5.30   &5.45   &7.45  &6.90 &8.75   &4.20  &1.30 &0.10          \\ \cline{3-12} 
					&&0.5  &    4.45   &5.15   &5.55   &6.70  &6.10 &7.90   &3.75  &1.40 &0.00       \\ \cline{3-12} 
					&&0.7  &    5.10   &5.25   &5.55   &7.45  &7.05 &7.80   &4.30  &1.40 &0.00         \\ \cline{2-12} 		
					&\multirow{5}{*}{512}                        &-0.7 &    6.15   &4.85   &5.85   &6.35  &5.90 &7.75   &2.95  &1.60 &0.05        \\ \cline{3-12} 
					&&-0.5 &    5.60   &4.35   &5.20   &5.75  &5.85 &8.15   &3.35  &1.05 &0.05     \\ \cline{3-12} 
					&&0.2  &    5.10   &4.90   &5.05   &5.80  &5.95 &6.55   &2.65  &1.30 &0.00         \\ \cline{3-12} 
					&&0.5  &    5.65   &5.35   &5.20   &6.05  &6.85 &7.65   &3.50  &1.35 &0.15        \\ \cline{3-12} 
					&&0.7  &    5.15   &5.20   &5.10   &5.55  &5.60 &7.80   &3.05  &1.35 &0.10          \\\bottomrule										
			\end{tabular}}
			\vspace{0.3em}
			\caption{Empirical rejection rate (in percentage) under the null when testing for time-reversibility, $(b_1,b_2,b_3)=(n/8,n/4,n/2)$.}\label{tab_triv_size} 
		\end{table}
		
	\end{footnotesize}

\begin{figure}[H]
		\begin{subfigure}{0.5\textwidth}
			\includegraphics[width=0.83\textwidth]{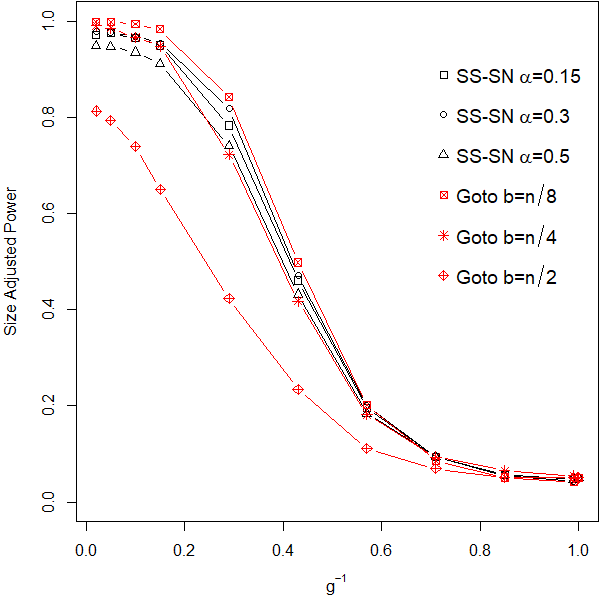}
			\caption{}
			\label{goto256notc}
		\end{subfigure}
		\hfill
		\begin{subfigure}{0.5\textwidth}
			\includegraphics[width=0.83\textwidth]{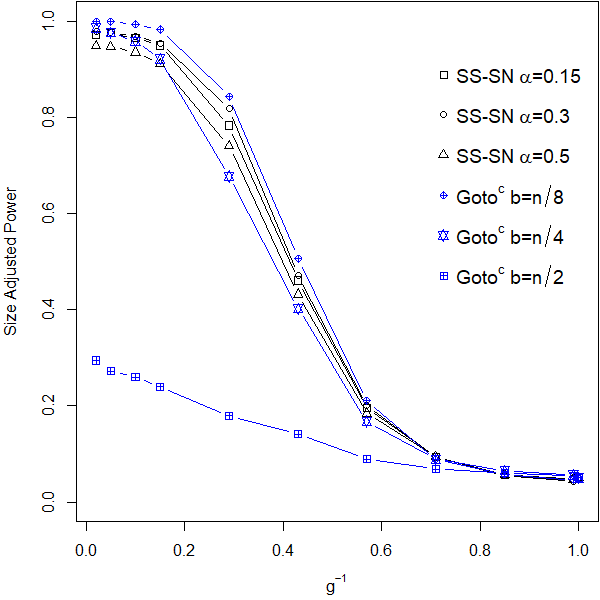}
			\caption{}
			\label{goto256c}
		\end{subfigure}

		\begin{subfigure}{0.5\textwidth}
			\includegraphics[width=0.83\textwidth]{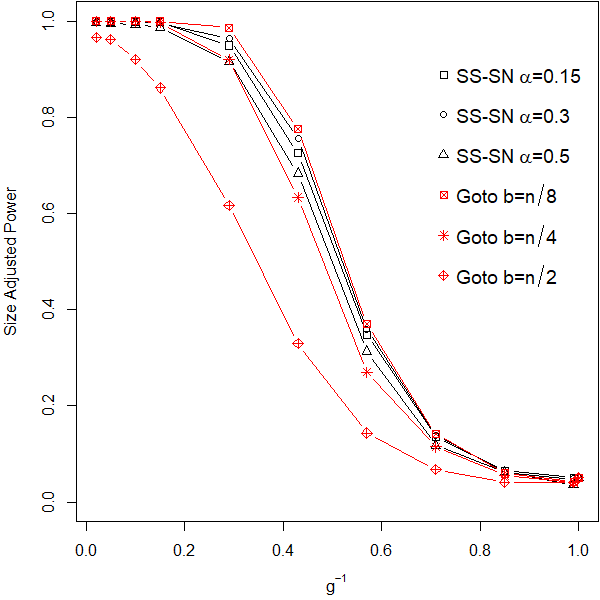}
			\caption{}
			\label{goto512notc}
		\end{subfigure}
		\hfill
		\begin{subfigure}{0.5\textwidth}
			\includegraphics[width=0.83\textwidth]{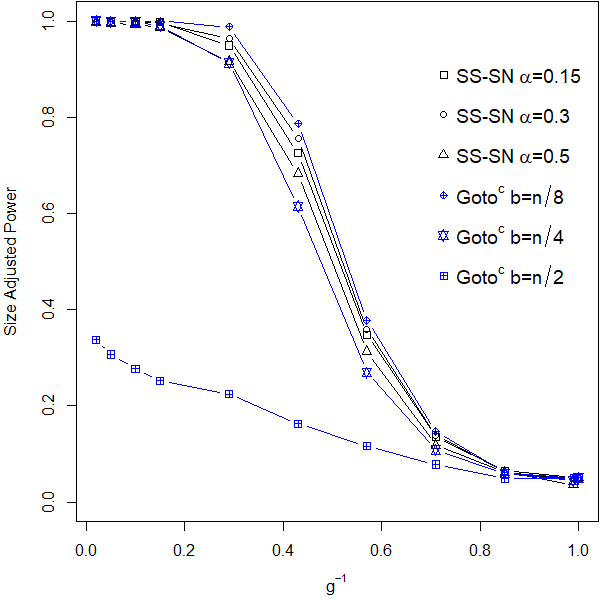}
			\caption{}
			\label{goto512c}
		\end{subfigure}

		\caption{Size adjusted power of SS-SN for testing time-reversibility when compared with $Goto$ (left) and $Goto^c$ (right) with sample size 256 (first row) and 512 (second row).}	\label{fig_time_rev_power}
	\end{figure}

	For the size adjusted power, we assume the data comes from the process: ${X}_t =C_g^{-1}(U_t|X_{t-1}),$ where $U_t \stackrel{iid}{\sim} Uniform(0,1)$ and $C_g(u,v): [0,1]^2\to[0,1]$ is the asymmetric Gumbel copula defined as 
	$C_g(u,v) = u^{1-a}v^{1-b}e^{-\{ (-a\log u)^g+(-b\log v)^g    \}^{1/g}}$ 
	with $(a,b)=(1,0.5)$ and $g\geq 1$. Note that $\{{X}_t\}$ is a stationary time series with $Uniform(0,1)$ marginal distribution and the joint cdf of $(X_t,X_{t-1})$ is $C_g(u,v)$. 
	For $Goto$ and $Goto^c$, the size adjusted powers are calculated using the same method proposed in \cite{dominguez2000}, who  developed a way of calculating size adjusted power for  bootstrap-based tests. For $n\in\{256,512\}$, we plot the size adjusted power against $g^{-1}\in\{0.02,0.05,0.1,0.15,0.29,\allowbreak 0.43,0.57,0.71,0.85,0.99,1\}$ and the experiment is repeated 2000 times. We use the same discretization method, i.e., $D_2$ as used in \cite{goto2021}. As shown in Figure \ref{fig_time_rev_power}, the size adjusted power for SS-SN with $\alpha\in\{0.15,0.3\}$ is close to $Goto$ and $Goto^c$ with $b=n/8$, which is the block length corresponding to the largest power. SS-SN with $\alpha=0.5$ has mild power loss compared with $\alpha=0.15,0.3$, which is reasonable since a larger portion of data are used to estimate the direction of projection, but it still outperforms $Goto$ and $Goto^c$ with $b=n/4$ for most values of $g^{-1}$. For $Goto$ and $Goto^c$ with $b=n/2$, the power loss is much larger compared with other block lengths. Overall, our SS-SN test has very stable and accurate size, and the power performance is close to $Goto$ and $Goto^c$ with the best block length, which is hard to choose in practice. Relatively speaking, the choice of sample splitting ratio $\alpha$ is less important as compared to the choice of subsampling block length, as the power is much less sensitive to $\alpha$. We also tried larger $\alpha$, such as $\alpha=0.7$, but its performance is always inferior to $\alpha=0.5$, so the result is not  reported. The same phenomenon is also observed in \cite{zhang2023another}. 

	\subsection{Testing for composite hypothesis on marginal cumulative distribution}
	\label{sim:2}

	In this subsection, we focus on testing the null hypothesis $H_0:F(x)=F_0(x;\boldsymbol{\theta}_0)=\Phi(\frac{x-\mu_0}{\sigma_0})$ for some $\boldsymbol{\theta}_0=(\mu_0,\sigma_0^2)^\top\in\boldsymbol{\Theta}=\R\times(0,\infty)$. As stated in Section \ref{sec_comp}, our SS-SN method can be applied  to testing this composite null hypothesis by setting $F_{a:b}(x)=\frac{1}{b-a+1}\sum_{t=a}^{b}\boldsymbol{1}(X_t\leq x)$ and $\boldsymbol{\hat \theta}_{a:b}{=}(\hat \mu_{a:b},\hat \sigma^2_{a:b})^\top$ where $\hat \mu_{a:b}{=}1/(b{-}a{+}1)\sum_{t=a}^{b}X_t$ and $\hat \sigma^2_{a:b}{=}1/(b{-}a{+}1)\sum_{t=a}^{b}(X_t{-}\hat \mu_{1:n})^2$.

	To examine the finite sample performance of our method, we have compared with the block bootstrap and subsampling methods, and as expected, our test has much more accurate size and the block bootstrap/subsampling methods are sensitive to the block size choice (results not shown). To capture the impact of block length, \cite{shao2013} have developed a fixed-b approach to calibrate the finite sample distribution of the p-value after applying block bootstrap or subsampling. Here  we shall extend the fixed-b subsampling method 
	proposed in \cite{shao2013} from testing a  simple null hypothesis to a composite null. To be specific, the p-value based on fixed-b subsampling is defined as:
	$$pval^\ast_{n,l}=\frac{1}{N} \sum_{t=1}^N\boldsymbol{1}\big\{\sqrt{l}\| F_{t:t+l-1}(x)- F_{1:n}(x) \|_\infty \geq \sqrt{n}\|  F_{1:n}(x)- F_0(x;\boldsymbol{\hat\theta}_{1:n})\|_\infty    \big\},$$
	where $\|\cdot\|_{\infty}$ is the infinity norm, $l=\lfloor nb \rfloor$, $N=n{-}l{+}1$ and $b\in (0,1)$ is the block-length parameter. The distribution of $pval^\ast_{n,l}$ is not pivotal and depends on $b$, but can be approximated by a second-level subsampling.  Let $n'=30$ be the second-level  subsampling length, $l' = \max(\lceil n'b\rceil, 2)$ and $N'=n'{-}l'{+}1$. For $t=1,2,\dots,n{-}n'{+}1$, define
	$$h^\ast_{n',t}=\frac{1}{N'}\sum_{j=t}^{t+N'-1}\boldsymbol{1}\big\{   \sqrt{l'}\| F_{j:j+l'-1}(x)- F_{t:t+n'-1}(x) \|_\infty \geq \sqrt{n'}\|  F_{t:t+n'-1}(x)- F_0(x;\boldsymbol{\hat\theta}_{t:t+n'-1})\|_\infty       \big\}.$$
	Then the final p-value is defined as 
	$pval^\ast = \frac{1}{n-n'+1} \sum_{t=1}^{n-n'+1}  \boldsymbol{1}\{h^\ast_{n',t} < pval^\ast_{n,l}\}.$
	The null hypothesis is rejected at level $\gamma$ if $pval^*\leq \gamma$.  This approach captures the choice of $b$ and is shown to be more accurate in size for simple null hypothesis in \cite{shao2013} and the same has been observed for composite null in our simulations (not shown). 
	
\begin{figure}[H]
		
		\begin{subfigure}{0.5\textwidth}
			\includegraphics[width=0.83\textwidth]{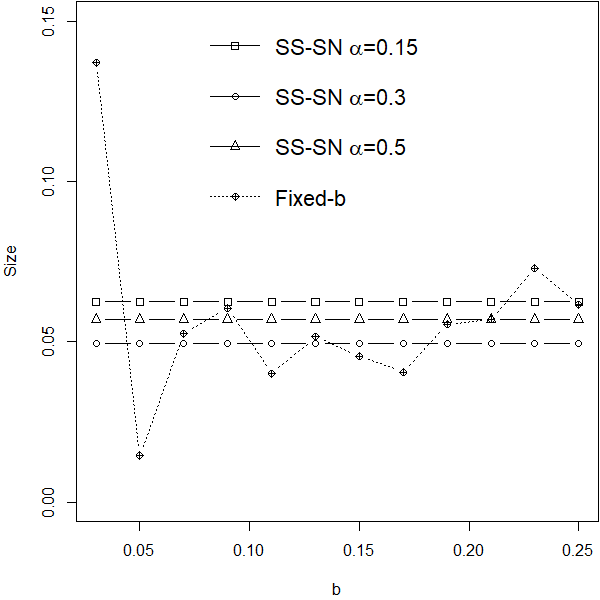}
			\caption{}
			\label{distfamily_n200rho-0.7}
		\end{subfigure}
		\hfill
		\begin{subfigure}{0.5\textwidth}
			\includegraphics[width=0.83\textwidth]{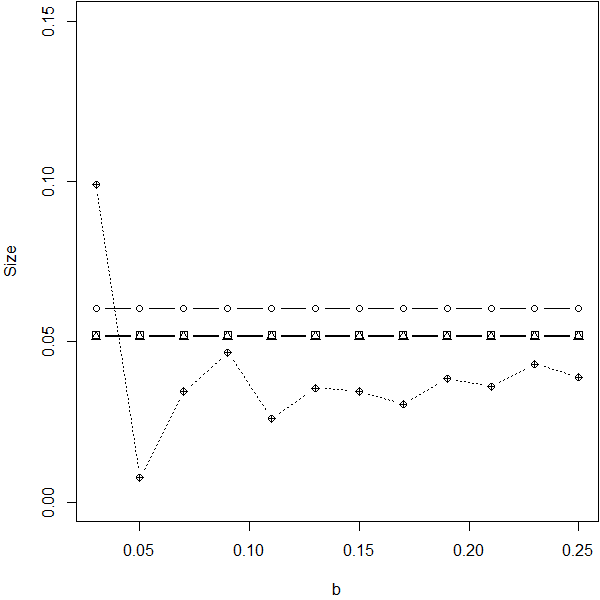}
			\caption{}
			\label{distfamily_n400rho-0.7}
		\end{subfigure}

		\begin{subfigure}{0.5\textwidth}
			\includegraphics[width=0.83\textwidth]{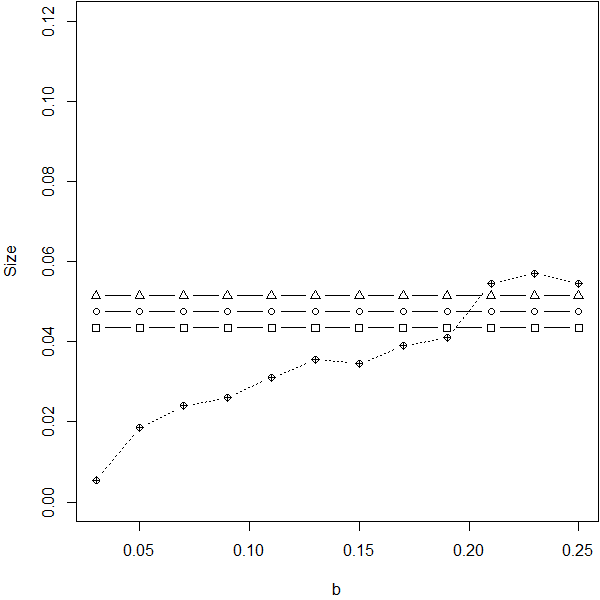}
			\caption{}
			\label{distfamily_n200rho0.7}
		\end{subfigure}
		\hfill
		\begin{subfigure}{0.5\textwidth}
			\includegraphics[width=0.83\textwidth]{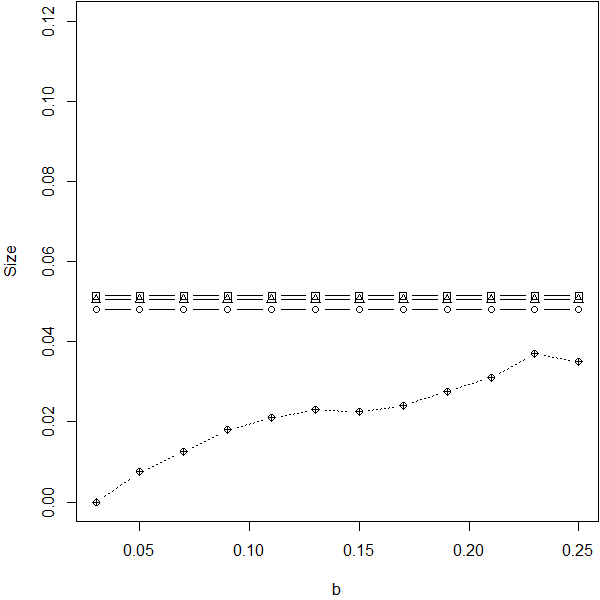}
			\caption{}
			\label{distfamily_n400rho0.7}
		\end{subfigure}
		
		\caption{Empirical size for testing composite hypothesis on marginal distribution for $n=100$ (left column), $n=200$ (right column) and $\rho=-0.7$ (first row), $\rho=0.7$ (last row).}
		\label{fig_distfamily_size}
	\end{figure}

	To examine the finite sample size, we generate the data from the AR(1) process $X_t=\rho X_{t-1}+\epsilon_t$ with ${\epsilon}_t \stackrel{iid}{\sim} N(0,1)$. We let $n\in\{100,200\}$, $\rho\in\{-0.7,0.7\}$, $b\in\{\frac{2k+1}{100}:k=1,2,\dots,12\}$ and the experiment is repeated 2000 times. For all combinations of $(n,\rho)$, we plot the empirical size for SS-SN statistic with $\alpha\in\{0.15,0.3,0.5\}$ and fixed-b subsampling based statistic (denoted as Fixed-b) against $b$. As shown in Figure \ref{fig_distfamily_size}, the size for SS-SN is close to the nominal level for all $\rho$ and the size accuracy improves as $n$ increases. The fixed-b subsampling method becomes more under-sized as $\rho$ gets closer to 1 and smaller value of $b$ corresponds to larger size distortion. Also, the size accuracy does not improve as $n$ increases.

\begin{figure}[H]
		\scalebox{0.9}{
			
			\begin{subfigure}{0.5\textwidth}
				\includegraphics[width=\textwidth]{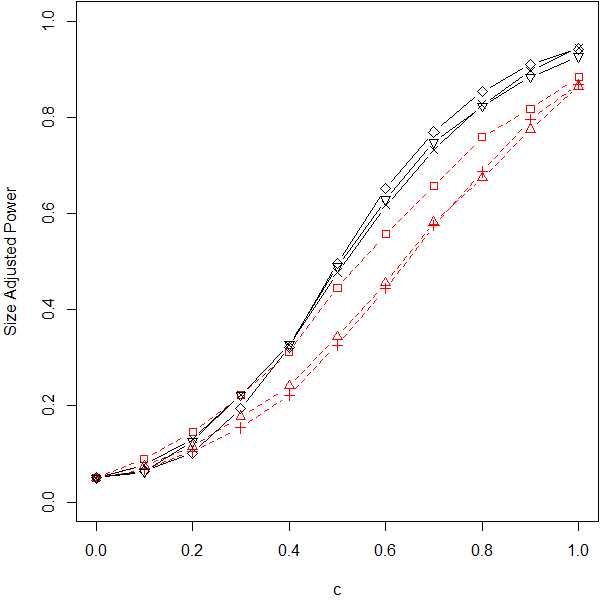}
				\caption{}
				\label{fig_dist_family_n100}
			\end{subfigure}
			\hfill
			\begin{subfigure}{0.5\textwidth}
				\includegraphics[width=\textwidth]{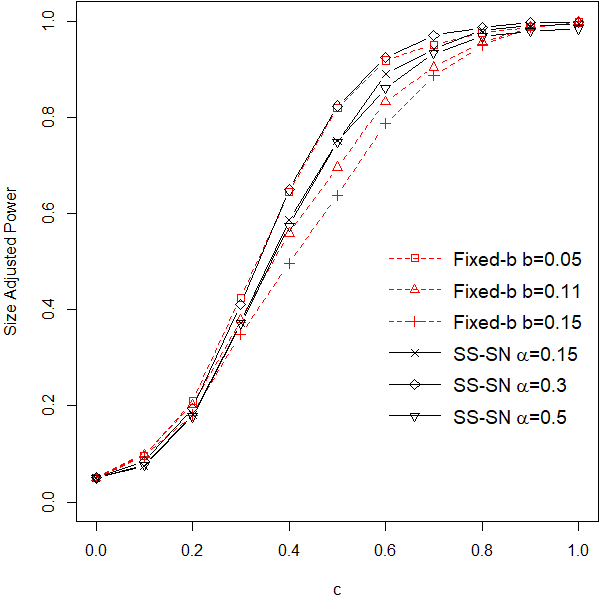}
				\caption{}
				\label{fig_dist_family_n200}
		\end{subfigure}}

		\caption{Size adjusted power for testing composite hypothesis on marginal cumulative distribution function for $n=100$ (left) and $n=200$ (right).}
		\label{fig_dist_family_power}
	\end{figure}

	To examine the size adjusted power, under the alternative we assume the data comes from the process $Z_t=(1{-}\delta_t)X_t{+}\delta_tY_t$, where $X_t=0.4X_{t-1}+\epsilon_t$ with ${\epsilon}_t \stackrel{iid}{\sim} N(0,1)$, $Y_t\stackrel{i.i.d}{\sim}Exp(1){-}1$ and $\delta_t\stackrel{i.i.d}{\sim}Bernoulli(c)$. We also assume ${X_t}$, $Y_t$ and $\delta_t$ are independent. The experiment is run 2000 times for $n\in\{100,200\}$ and the size adjusted powers are plotted against $c\in[0,1]$. When $n=100$, as shown in Figure \ref{fig_dist_family_n100}, SS-SN with all three values of $\alpha$ outperform Fixed-b with all three values of $b$ when $c\geq 0.4$ and they have similar power when $c\in[0,0.4)$. When $n=200$, SS-SN with $\alpha=0.3$ has similar power performance as Fixed-b with $b=0.05$, which corresponds to the best power performance for both methods respectively. For $\alpha\in\{0.15,0.5\}$, SS-SN outperforms Fixed-b with $b\in \{0.11,0.15\}$.
	
	\subsection{Testing for a change point in the spectral distribution}
	\label{sim:3}
	
	In this subsection we examine the finite sample size and size adjusted power of SS-SN statistic in testing for a change point in the spectral distribution. Following the theoretical results in Section \ref{subsec:cp}, let $F_t(x) = SDF^{(0)}(\lambda) = \int_{0}^{\lambda} f^{(0)}(\omega)d\omega$ if $t\leq \lfloor  n \tau_0 \rfloor$ and $F_t(x) = SDF^{(1)}(\lambda) = \int_{0}^{\lambda} f^{(1)}(\omega)d\omega$ otherwise with $x{=}\lambda\in\Omega{=}[0,\pi]$. Let $F_{a:b}(\lambda) = \int_0^\lambda I_{a:b}(s)ds$, then the statistic $G_n$ defined in Section \ref{subsec:cp} can be used to test for a change point in the spectral distribution function.
	
	In the literature, testing for a change in spectrum
	has been considered by \cite{picard1985},
	\cite{giraitis1992testing}, \cite{lavielle2000}, \cite{shaozhang2010} and \cite{preuss2015}, among others.
	In particular, \cite{picard1985} developed a Kolmogorov–Smirnov test for spectrum change under the Gaussian assumption, which was later relaxed by \cite{giraitis1992testing}. Since the limiting null distribution of the test statistic of \cite{giraitis1992testing} depends on the
	fourth order cumulants of the process, the implementation of their test typically requires block bootstrap or subsampling. \cite{lavielle2000} 
	allowed multiple change points but assumed a parametric form for the spectral density in each segment. \cite{shaozhang2010} applied the SN-based test to detect the changes in spectral distribution function evaluated at a pre-specified number of frequencies. More recently, 
	\cite{preuss2015} proposed a nonparametric test (denoted as PR) which compares the difference between $F_{\fr-N+1:\fr}(\lambda)$ and $F_{\fr+1:\fr+N}(\lambda)$ around a potential change point $\fr$, where $N$ is a pre-specified block length. Specifically, their test statistic is defined as
	\begin{equation} \label{eqpr}
		\hat D_n = \sup_{(r,\lambda)\in[0,1]\times[0,\pi]}|F_{\fr-N+1:\fr}(\lambda)-F_{\fr+1:\fr+N}(\lambda)|.
	\end{equation}
	For $\R^d$-valued data, they replace $|F_{\fr-N+1:\fr}(\lambda){-}F_{\fr+1:\fr+N}(\lambda)|$ in Equation (\ref{eqpr}) by $\max_{i,j=1,\dots,d}\big|F^{ij}_{\fr-N+1:\fr}(\lambda)-F^{ij}_{\fr+1:\fr+N}(\lambda)\big|,$
	where $F^{ij}_{\fr-N+1:\fr}(\lambda)$ is the $(i,j)$th component of the matrix valued estimator $\mathbf{F}_{\fr-N+1:\fr}(\lambda)$.
	Since the asymptotic distribution of $\hat D_n$ under the null is not pivotal, they proposed a bootstrap procedure where the bootstrap replicates are calculated using data generated from an AR(p) process with standard normal innovation and parameters estimated from the original data.
	
\begin{figure}[H]
		
		\scalebox{0.9}{
			\begin{subfigure}{0.5\textwidth}
				\includegraphics[width=\textwidth]{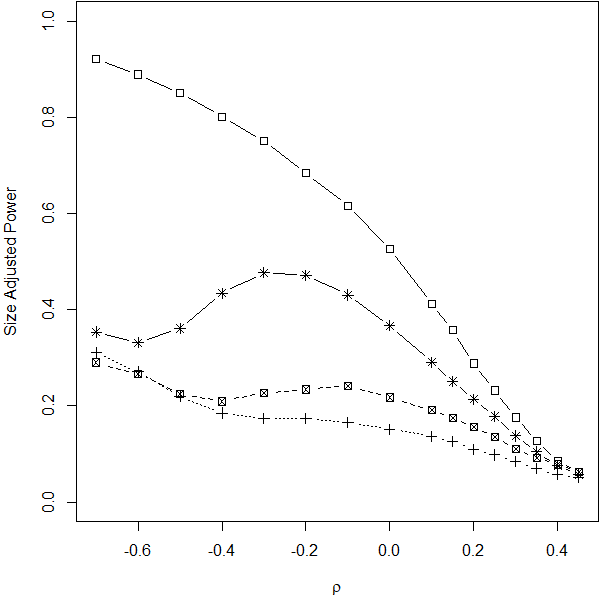}
				\caption{}
				\label{fig_cp_var1}
			\end{subfigure}
			\hfill
			\begin{subfigure}{0.5\textwidth}
				\includegraphics[width=\textwidth]{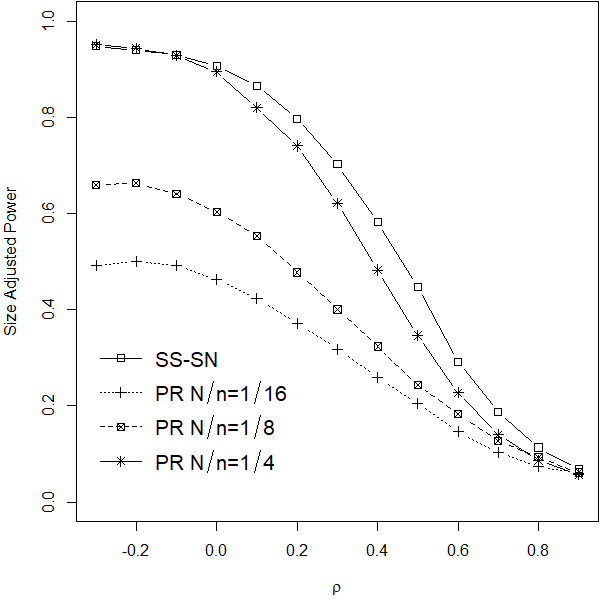}
				\caption{}
				\label{fig_cp_vma1}
		\end{subfigure}}

		\caption{Size adjusted power when testing for a change point in spectral distribution function for Alt-VAR(1) (left) and Alt-VMA(1) (right) model.}
		\label{fig_cp2}
	\end{figure}

	Under the null, we consider 2 DGPs used in \cite{preuss2015}: (a),
	$\text{VAR(1)}: X_t=\big\{(\rho{-}0.2)\mathbf{I}_2+0.2\mathbf{1}_2\mathbf{1}_2^\top \big\}\mathbf{X}_{t-1}+\boldsymbol{\epsilon}_t
	$ and (b), 
	$\text{VMA(1)}: X_t=\big\{(\rho{-}0.2)\mathbf{I}_2+0.2\mathbf{1}_2\mathbf{1}_2^\top \big\} \boldsymbol{\epsilon}_{t-1}+\boldsymbol{\epsilon}_t
	$, 
	where $\boldsymbol{\epsilon}_t\stackrel{i.i.d}{\sim}{N}(0,I_2)$. We set $\rho\in\{-0.5,0.5\}$, $n\in\{256,512\}$, $N\in\{n/16,n/8,n/4\}$ and the experiment is repeated 2000 times. The SS-SN statistic is calculated with $b=0.15$. 
	\begin{table}[H]
		\centering
		\vspace{1em}
		\setlength{\tabcolsep}{4pt}
		\begin{tabular}{cc|cccc|cccc}
			\toprule
			\midrule
			\multirow{3}{*}{$n$}       &\multirow{3}{*}{$\rho$}&\multicolumn{4}{c|}{VAR(1)}&\multicolumn{4}{c}{VMA(1)}    \\
			&& \multirow{2}{*}{SS-SN} &\multicolumn{3}{c|}{PR} & \multirow{2}{*}{SS-SN} &\multicolumn{3}{c}{PR} \\ \cline{4-6}\cline{8-10} 
			&&&$\frac{N}{n}=\frac{1}{16}$&$\frac{N}{n}=\frac{1}{8}$&$\frac{N}{n}=\frac{1}{4}$&&$\frac{N}{n}=\frac{1}{16}$&$\frac{N}{n}=\frac{1}{8}$&$\frac{N}{n}=\frac{1}{4}$\\ \hline
			\multirow{2}{*}{256}                        &-0.5 &    5.45      &1.65   &1.75 &2.15    &4.60   &1.60 &1.60  &1.60        \\ \cline{2-10} 
			&0.5  &    5.25      &2.95   &2.05 &1.60    &5.65   &2.30 &1.95  &2.15   \\ \cline{1-10}  	
			\multirow{2}{*}{512}                        &-0.5 &    6.70      &2.10   &2.60 &3.00    &6.55   &3.05 &2.55  &2.95        \\ \cline{2-10} 
			&0.5  &    5.80      &3.40   &3.15 &2.95    &5.70   &2.75 &3.00  &3.35      \\ \bottomrule		
		\end{tabular}
		\vspace{0.3em}
		\caption{Empirical rejection rate (in percentage) under the null when testing for a change point in spectral distribution function.}\label{tab_CP_size2} 
	\end{table}
	For the empirical size, as shown in Table \ref{tab_CP_size2}, under both VAR(1) and VMA(1) models, SS-SN has relative accurate size for all combinations of $n$ and $\rho$ except $(n,\rho)=(512,0.5)$, and is slightly oversized under this setting. In contrast, PR is undersized in general and the level of size distortion depends on both $n$ and $N$. For the size adjusted power, we consider the following 2 DGPs:
	\small
	$$\text{Alt-VAR(1)}: X_t=
	\begin{cases}
		\big\{0.4\mathbf{I}_2+0.1\mathbf{1}_2\mathbf{1}_2^\top \big\} \mathbf{X}_{t-1}+\boldsymbol{\epsilon}_t,  & t\leq \lfloor \frac{n}{2}\rfloor \\
		\big\{(\rho_1{-}0.1)\mathbf{I}_2+0.1\mathbf{1}_2\mathbf{1}_2^\top \big\}\mathbf{X}_{t-1}+\boldsymbol{\epsilon}_t, & t> \lfloor \frac{n}{2}\rfloor
	\end{cases}$$
	$$\text{Alt-VMA(1)}: X_t=
	\begin{cases}
		\big\{0.9\mathbf{I}_2+0.1\mathbf{1}_2\mathbf{1}_2^\top \big\}  \boldsymbol{\epsilon}_{t-1}+\boldsymbol{\epsilon}_t,  & t\leq \lfloor \frac{n}{2}\rfloor \\
		\big\{(\rho_2{-}0.1)\mathbf{I}_2+0.1\mathbf{1}_2\mathbf{1}_2^\top \big\}\boldsymbol{\epsilon}_{t-1}+\boldsymbol{\epsilon}_t, & t> \lfloor \frac{n}{2}\rfloor.
	\end{cases}$$
	\normalsize
	We set $n=256$ and plot the power curve for $\rho_1\in[-0.7,0.5]$ and $\rho_2\in[-0.3,1]$. The experiment is repeated 2000 times. As shown in Fig.~\ref{fig_cp2}, SS-SN outperforms PR under both Alt-VAR(1) and Alt-VMA(1). The power for PR depends on the block length and is nonmonotonic. That is, the power can decrease when the alternative gets farther away from the null, and it is unclear what is the root cause of this non-monotonic power phenomenon. This example shows that our SS-SN test statistic can outperform some existing methods in power besides its size accuracy.

	\section{Summary and conclusion}
	\label{sec:conclusion}
	In this paper, we propose a sample splitting approach to the inference of a functional parameter in the setting of weakly dependent time series. Specifically we develop tests in the case of simple null, composite null and for a change-point, and study the limiting distributions under both the null and alternatives. The proposed sample splitting plus self-normalization tests have broad applicability, and can assess properties of a time series, such as whether the marginal distribution is Gaussian, whether there is a change in second order properties (in terms of a change point in spectral distribution distribution), and whether the time series is reversible. From numerical experiments, we find that these new tests are more accurate in size with competitive power performance, when comparing to bandwidth-dependent procedures. The relative insensitivity to the splitting ratio is also notable in simulations, so the optimal choice of $\alpha$ is less critical in practice, as long as it is in the range $[0.3,0.5]$.
	
	As potential future research topics, it would be interesting to look into extensions to the inference of a functional parameter in the setting of locally stationary time series, unit root nonstationary time series, and functional time series. Further research along these directions are well
	underway.

	{\fontsize{11.15}{11.15}\selectfont 
		\setlength{\bibsep}{0pt plus 0.24ex}
		
		\bibliographystyle{chicago}
		
		\bibliography{Bib_major_rev}
	}

	\newpage
	\bigskip
	\begin{center}
		{\large\bf SUPPLEMENT to ``Hypothesis Testing for a Functional Parameter via Self-normalization''}
	\end{center}
	
	\begin{center}
		{BY Yi Zhang AND Xiaofeng Shao}
	\end{center}
	
	The supplement is organized as follows. In Appendix \ref{app_ep}, we use two real data examples to illustrate the finite sample performance of our SS-SN method. Appendix \ref{app_verify} discusses how to verify Assumption \ref{assump_1null} for Examples 4-7 stated at the beginning of Section \ref{sec:testing}. Appendix \ref{app_snCP} discusses how to apply the SN method proposed in \cite{zhang2023another} in testing simple null hypothesis on a functional parameter. Appendix \ref{app_proof} contains the proof for all the theoretical results in Section \ref{sec:testing}. In particular, Appendixes~\ref{appen_th_1null2}-\ref{appen_th_cpn}  contain the proofs for Theorems \ref{th_1null2}-\ref{th_cpn}.
	
	\appendix
	
	\renewcommand{\theequation}{A-\arabic{equation}}
	\setcounter{equation}{0}  
	
	\section{Empirical examples}\label{app_ep}
	
	\subsection{Testing for with noise for weekly stock indices}
	
	In this section, we consider a real data example where we test if a given time series is a white noise sequence. This is an important step in ARMA model building. 
	To be specific, for the real valued stationary time series $\{X_t\}_{t\in\Z}$ we want to test the null hypothesis $H_0: \gamma(j)=0,j=1,2,\dots$ where $\gamma(j)=\E[(X_t-\mu)(X_{t-j}-\mu)]$ for $j=0,1,2,\dots$ and $\mu=\E(X_t)$. \cite{shao2011boot} proposed a  spectral distribution  based test which can accommodate high-order serial dependence under the null. The limiting null distribution of the test statistic in \cite{shao2011boot} is not pivotal and its critical values are approximated by a  blockwise wild bootstrap method which involves a block length parameter $b_n$. Here we can treat the functional parameter as $F(\lambda)=\sum_{j=1}^\infty\gamma(j)\frac{\sin j\lambda}{j\pi}$. Under the null $F_0(\lambda)\equiv0$, so it is a simple null and our SS-SN test can be used for this problem. 

\begin{figure}[H]
		\centering
		\includegraphics[width=1\textwidth]{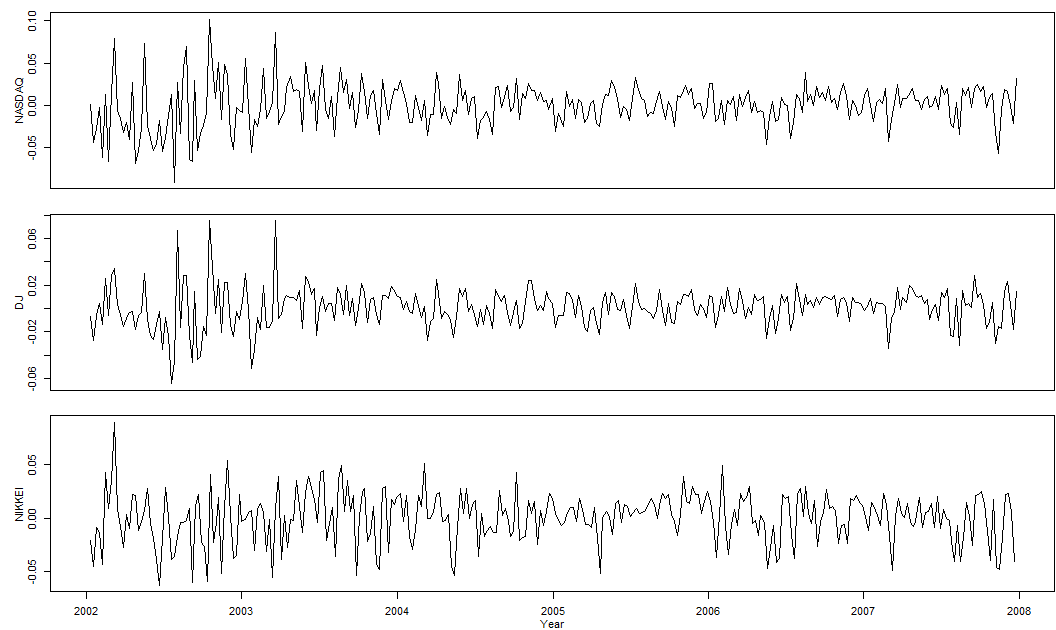}
		\caption{Stock indices growth rate}
		\label{fig_cpi}
	\end{figure}

	We use the weekly average price data for three stock indices and present the result for the growth rate of NASDAQ 100 (NASDAQ), Dow Jones (DJ) and Nikkei Stock Average (NIKKEI) indices, which cover the period from January 2002 to December 2007. There are 313 observations in total and Figure \ref{fig_cpi} plots three indices' growth rate series. The data is contained in the datasets \textbf{NASDAQ}, \textbf{DJ} and \textbf{NIKKEI} in the R package \textbf{qrmdata}.
	
\begin{figure}[H]
		\centering
		\includegraphics[width=1\textwidth]{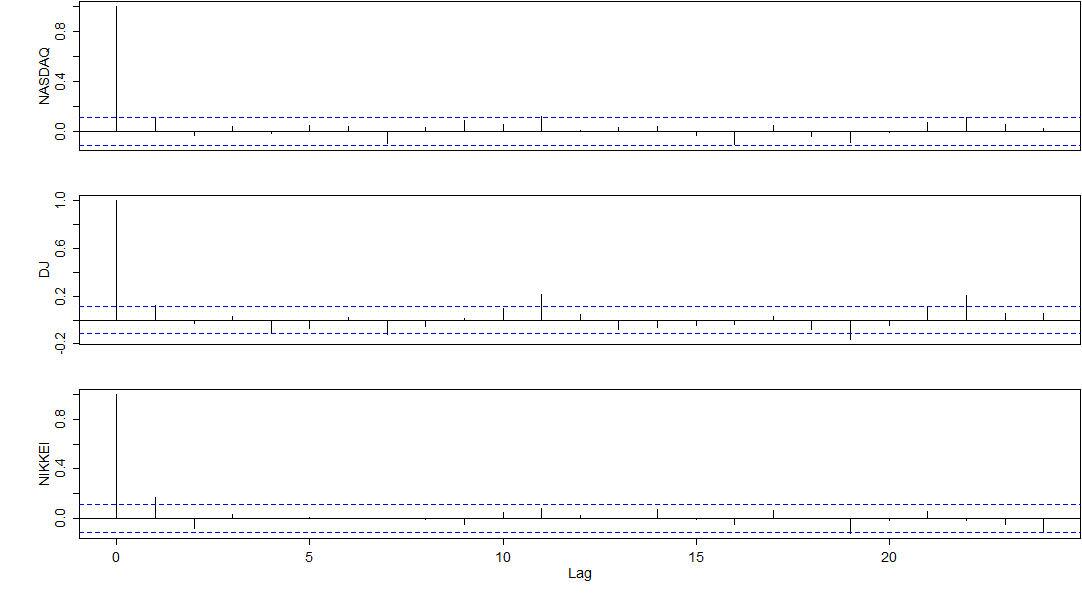}
		\caption{Sample ACF of the stock indices growth rate series}
		\label{fig_sample_corr}
	\end{figure}

\begin{figure}[H]
		\centering
		\includegraphics[width=1\textwidth]{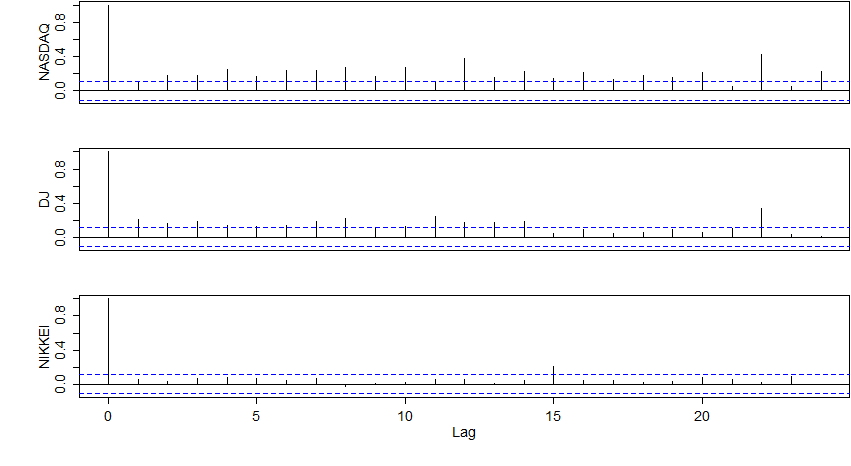}
		\caption{Sample ACF of the square of the demeaned growth rate series}
		\label{fig_sample2_corr}
	\end{figure}

	The sample autocorrelation at lag $p$, $\hat r_p = \hat{\gamma}_p/\hat{\gamma}_0$, where $ \hat{\gamma}_p=\frac{1}{n}\sum_{t=1}^{n-p}(X_t-\bar X_n)(X_{t+p}-\bar X_n)$ are shown in the ACF plot in Figure \ref{fig_sample_corr}. For NASDAQ, $\hat r_p$s at all lags $p\leq 25$ are within the $95\%$ confidence interval around zero, as indicated by the dashed line, with some $\hat r_p$ close to the dashed line. For DJ and NIKKEI, some $\hat r_p$s are outside the $95\%$ confidence interval. Note that the two blue dashed lines are constructed on the basis of iid assumption under the null, and may be misleading when there are high-order serial dependence in the data, such as conditional heterocedasticity; see \cite{romano1996}. The ACF plot for the square of the demeaned growth rate series are shown in Figure \ref{fig_sample2_corr}, which shows the existence of high-order serial dependence.
	

	The p-values for the SS-SN test are shown in Table \ref{tab_emp_example}, which also includes the p-values for the test proposed in \cite{shao2011boot} (denoted as Shao) with different values of $b_n$, and the Ljung-Box test (denoted as Ljung-Box) at different lags $p$. For all three indices, SS-SN rejects the null at $5\%$ level for all values of $\alpha$. For the Ljung-Box test, the null hypothesis is not rejected at $5\%$ level for NASDAQ and the p-values get larger as $p$ increase for NIKKEI with the null being rejected for $p=1,5$. However, for DJ the Ljung-Box test rejects the null at $5\%$ level for all values of $p$ except $p=5$, and the smallest p-value appears at $p=20$. Note that the Ljung-Box test employs the $\chi^2$ reference distribution, which is no longer valid when there is high-order serial  dependence \citep{romano1996}, so it is hard to interpret these results. 
	
	In comparison with Ljung-Box test, the test in \cite{shao2011boot} and our SS-SN tests are both robust to high-order serial dependence in theory. However, the choice of block lengh $b_n$ in Shao is not addressed in the literature yet.
	As shown in Table~\ref{tab_emp_example}, Shao's test is quite sensitive to the choice of $b_n$, making the interpretation of the results difficult. Based on our SS-SN tests, we feel confident to assert that there are significance autocorrelations in all three series. 
	

	
	\begin{table}[H]
		
		\centering
		\setlength{\tabcolsep}{7pt}
		\scalebox{0.73}{
			\begin{tabular}{c|ccc|ccccc|cccc}
				\toprule
				\midrule
				\multirow{2}{*}{Indices}  &\multicolumn{3}{c|}{SS-SN}&\multicolumn{5}{c|}{Shao}&\multicolumn{4}{c}{Ljung-Box}\\
				&$\alpha{=}0.15$ & $\alpha{=}0.3$&  $\alpha{=}0.5$&$b_n{=}1$ & $b_n{=}{\sqrt{n}}/{2}$& $b_n{=}\sqrt{n}$& $b_n{=}2\sqrt{n}$& $b_n{=}4\sqrt{n}$ &$p$=1 &$p$=5&$p$=10&$p$=20   \\ \hline
				NASDAQ&    3.6\% &0.0\% &0.4\%  &11.4\% &3.8\% &3.0\% &3.2\%    & 9.0\%  &5.2\%&35.9	\%& 25.5\%  &16.4\%    \\ \cline{2-13} 
				DJ&    0.4\% &2.2\% &0.4\%  &12.4\% &2.0\% &0.2\% &2.8\%     & 9.0\%  &3.2\%&7.3	\%& 3.9\%  &0.0\%   \\ \cline{2-13} 
				NIKKEI& 1.1\% &0.1\% &0.2\%  &0.2\% &0.4\% &1.6\% &3.8\%   & 2.0\%  &0.3\%&4.9	\%& 24.3\%  &25.3\%  \\  \bottomrule                        
		\end{tabular}}
		\caption{P-values of SS-SN, Shao and Ljung-Box.}\label{tab_emp_example}
	\end{table}

			\subsection{Testing for time reversibility for weekly stock indices}
			
			According to \citep[][Definition 1]{ramsey1996time}, a time series $\{X_t\}_{t\in\Z}$ is time reversible if $(X_{t_1},X_{t_2},\dots,X_{t_m})$ and $(X_{M-t_1},X_{M-t_2},\dots,X_{M-t_m})$ have the same joint distribution for any integers $t_1,\dots,t_m,M,m$. Testing for time reversibility can provide important insights for subsequent modeling. For example, \cite{weiss1975time} showed that a stationary Gaussian process is time reversible. On the other hand, a linear, non
			Gaussian process is time irreversible in general, except when its coefficients satisfy certain constraints \citep{chen2000testing}. For nonlinear process, time reversibility appears to be the exception rather than the rule \citep{tong1990non}. As such, a test of time reversibility can be viewed as a diagnostic tool to check if the linearity and Gaussianity assumptions in model building are appropriate. 
			
			In this section, we apply our SS-SN test to examine if the weekly returns of eight major stock indices are time reversible. The data, which was also analyzed in \cite{psaradakis2008assessing}, consists of weekly price indices for the stock exchanges in: Amsterdam (AMSTEOE), Frankfurt (DAXINDX), Hong Kong (HNGKNGI), London (FTSE100), New York (SPCOMP), Paris (FRCAC40), Singapore (SNGALLS) and Tokyo (JAPDOWA). The data covers the period from 6 January 1986 to 31 December 1997 (626 observations) except for Paris, which has 547 observations covering from 9 July 1987 to 31 December 1997. The data can be downloaded from https://sites.google.com/view/dickvandijk/nltsmef. 
			The price indices $p_t$ are transformed into weekly returns $X_t$ using the formula $X_t=100\ln (p_t/p_{t-1})$.  
			
			The p-values for the SS-SN test under the two discretization schemes ($D_1$ and $D_2$) defined in Section \ref{sim:1} are shown in Table \ref{tab_emp_rev}, which also includes the p-values for the test proposed in \cite{goto2021} (both with and without correction term, denoted as $Goto^c$ and $Goto$ respectively; see Section \ref{sim:1}) with different values of block length $b$. For SS-SN, we can not reject the null hypothesis of time reversibility for Amsterdam, Hong Kong, London, New York and Singapore at the 10\% level under both discretization schemes. For Paris and Tokyo, the null is rejected at 5\% level under both discretization schemes (except for Tokyo under scheme $D_2$ with $\alpha=0.5$, for which the p-value is very close to 5\%). For Frankfurt, we can not reject the null at 10\% level under both $D_1$ and $D_2$ for $\alpha=0.3$ and for $\alpha=0.5$, the null is rejected at 10\% level under $D_1$ but the p-value is slightly larger than 10\% under $D_2$.
			
			For $Goto$ and $Goto^c$, The results are basically in line with SS-SN for Amsterdam, Hong Kong, London, Singapore and Frankfurt. For New York, Paris and Tokyo, the p-values vary drastically under different discretization schemes and for different block length, which makes it hard to decide whether to reject the null or not. The above result again demonstrates that the p-values for our SS-SN test is relatively stable with respect to the choice of $\alpha$ and disretization scheme, as compared to the ones in \cite{goto2021}. The SS-SN test  is convenient to implement and its result is  easy to interpret, which are important in practical applications.

			\begin{table}[H]
				\centering
				\setlength{\tabcolsep}{8pt}
				\scalebox{0.92}{
					\begin{tabular}{cc|cc|ccc|ccc}
						\toprule
						\midrule
						\multirow{2}{*}{ }   &	\multirow{2}{*}{City}   &\multicolumn{2}{c|}{SS-SN}&\multicolumn{3}{c|}{$Goto$} &\multicolumn{3}{c}{$Goto^c$}    \\
						&	&$\alpha=0.3$&$\alpha=0.5$&$b_1$& $b_2$&$b_3$ &$b_1$& $b_2$&$b_3$   \\ \hline
						\multirow{8}{*}{$D_1$}& Amsterdam &64.93\%   &88.51\%   &45.54\%   &68.51\%  &62.23\%   &99.68\%   &91.28\%  &72.26\%          \\ \cline{2-10} 
						& Frankfurt &12.61\%   &8.99\%    &44.90\%   &37.23\%  &20.44\%   &100.00\%  &64.68\%  &34.31\%       \\ \cline{2-10} 
						& Hong Kong &22.18\%   &32.55\%   &28.98\%   &47.66\%  &43.07\%   &99.04\%   &72.98\%  &58.76\%     \\ \cline{2-10} 
						& London    &93.90\%   &91.89\%   &77.07\%   &87.02\%  &81.57\%   &100.00\%  &99.36\%  &91.24\%     \\ \cline{2-10} 
						& New York  &20.21\%   &13.10\%   &5.73\%    &2.34\%   &5.11\%    &87.90\%   &31.28\%  &8.21\%     \\ \cline{2-10} 		
						& Paris     &1.30\%    &0.87\%    &18.25\%   &0.73\%   &1.25\%    &100.00\%  &17.27\%  &4.18\%   \\ \cline{2-10} 
						& Singapore &24.96\%   &21.43\%   &9.55\%    &19.36\%  &26.28\%   &93.31\%   &42.13\%  &41.24\%     \\ \cline{2-10} 
						& Tokyo     &0.04\%    &2.10\%    &11.78\%   &0.21\%   &0.55\%    &82.80\%   &4.04\%   &2.19\%     \\ \hline	         
						\multirow{8}{*}{$D_2$}          & Amsterdam &57.32\%   &69.40\%   &38.22\%   &48.09\%  &52.19\%   &97.13\%   &69.94\%  &63.14\%           \\ \cline{2-10} 
						& Frankfurt &17.23\%   &11.13\%   &16.88\%   &18.29\%  &17.88\%   &88.54\%   &34.25\%  &21.53\%       \\ \cline{2-10} 
						& Hong Kong &88.17\%   &42.39\%   &55.46\%   &71.49\%  &61.50\%   &99.36\%   &88.72\%  &70.44\%      \\ \cline{2-10} 
						& London    &93.38\%   &99.79\%   &91.08\%   &95.74\%  &92.15\%   &99.36\%   &100.00\% &94.89\%     \\ \cline{2-10} 
						& New York  &18.24\%   &16.13\%   &0.32\%    &2.13\%   &2.37\%    &56.05\%   &14.26\%  &3.65\%      \\ \cline{2-10} 		
						& Paris     &0.78\%    &1.57\%    &0.36\%    &0.49\%   &0.21\%    &77.37\%   &5.11\%   &1.04\%     \\ \cline{2-10} 
						& Singapore &45.65\%   &21.65\%   &40.13\%   &40.64\%  &49.64\%   &96.18\%   &65.11\%  &59.31\%     \\ \cline{2-10} 
						& Tokyo     &0.24\%    &5.68\%    &28.34\%   &3.62\%   &5.11\%    &77.07\%   &16.17\%  &9.49\%     \\    \bottomrule 
				\end{tabular}}
				\vspace{0.3em}
				\caption{P-values for SS-SN, $Goto$ and $Goto^c$. $(b_1,b_2,b_3)=(n/2,n/4,n/8)$.}\label{tab_emp_rev} 
			\end{table}

	\section{Verification of Assumption \ref{assump_1null} for Examples \ref{example_4}-\ref{example_7}}\label{app_verify}
		
		\begin{example}
			Suppose we want to test if $X_t$ and $X_{t-1}$ are independent for a stationary time series $\{X_t\}_{t=0}^n$, then $F(x) = CF_1(u,v)$ and $F_0\equiv0$. For $t=1,2,\dots,n$, denote $Y_t =Y_t(u)= e^{iuX_t}$, $Z_t = Z_t(v) = e^{ivX_{t-1}}$, which can be viewed as random elements taking values in the Hilbert space $L_2(\C,\mu) = \{f:\R\to \C|\,\int|f(x)|^2\mu(dx)<\infty \}$ where $\mu(\cdot)$ is some finite measure. Then the null hypothesis is equivalent to the fact that the cross covariance operator between $Y_t$ and $Z_t$ is zero. Define $F_{\fr:\ft} = F_{\fr:\ft}(u,v) = \frac{1}{\ft-\fr+1}\sum_{i=\fr}^{\ft}(Y_i-\bar Y)(Z_i-\bar Z)$ where $\bar Y=\frac{1}{n}\sum_{t=1}^{n}Y_t$ and $\bar Z=\frac{1}{n}\sum_{t=1}^{n}Z_t$. Clearly $F_{\fr:\ft}$ is approximately linear and a uniform FCLT for $F_{1:\ft}(u,v)$ is proven in theorem 2.2 of \cite{sharipov2020bootstrapping} under reasonable moment and mixing conditions on $\{X_t\}_{t=0}^{n}$. So Assumption 1 is verified for Example 4.
		\end{example}
		
		\begin{example}
			For this example, Assumption \ref{assump_1null} can be verified in the same way as for Example \ref{example_4}.
		\end{example}
		
		\begin{example}
			For this example, $F(\lambda,x)=H(\lambda,x)$ is the generalized spectral distribution. The estimator $F_{\fr:\ft}(\lambda,x)$ can be defined as $F_{\fr:\ft}(\lambda,x) = 2\int_{0}^{\lambda\pi}\hat f_{\fr:\ft}(s,x)ds$, where $\hat f_{\fr:\ft}(s,x) = \frac{1}{2\pi}\sum_{j=-(\ft-\fr)}^{\ft-\fr}\hat \gamma_{\fr:\ft}(j,x)e^{-ijs} $, $\hat \gamma_{\fr:\ft}(-j,x) = \hat \gamma_{\fr:\ft}(j,x) = \frac{1}{\ft-\fr+1-j}\sum_{t=\fr+j}^{\ft-\fr+1}(X_t-\bar X)e^{ixX_{t-j}}$ for $j=0,1,\dots,\ft-\fr$ and $\bar X = \frac{1}{n}\sum_{t=1}^{n}X_t$. An FCLT for $F_{1:n}(\lambda,x)$ is proved in Theorem 1 of \cite{escanciano2006b}. We conjure that a uniform FCLT for $F_{\fr:\ft}(\lambda,x)$ can be shown based on the FCLT for $F_{1:n}(\lambda,x)$ and some additional assumptions, in view of Theorem 4.10 in \cite{volgushev2014}. However, the proof is highly non-trivial and is beyond the scope of this paper. 
			
		\end{example}
		
		\begin{example}
			For this example, The functional parameter $F(x)$ and the estimator $F_{\fr:\ft}(x)$ is defined in Section \ref{sim:1} when testing for time-reversibility and an FCLT for $F_{1:n}(x)$ is proved in Theorem 3.1 of \cite{goto2021}. It seems possible to extend the same method as in Example \ref{example_6} to show a uniform FCLT for $F_{\fr:\ft}(x)$, but we leave this for future research.
		\end{example}

		\section{Comparison with \cite{zhang2023another}}\label{app_snCP}
		
		Theoretically, for simple null hypothesis testing, the SS-SN method in \cite{zhang2023another} developed for a growing dimensional parameter can be applied if the functional parameter of interest can be written in the expectation form: $F(x) = \E f(X_1,x)$ for any $x\in \Omega$. Then we can transform the original one dimensional time series $\{X_t\}_{t=1}^n$ into a $p=p_n$ dimensional time series $\{Z_t=(Z_t^1,\dots,Z_t^p)^\top\}_{t=1}^n$, where $Z_t^i = f(X_t,x^i)$ and $x^1<x^2<\cdots<x^p$ are discrete points in $\Omega$. The original null hypothesis is then transformed into testing if the mean of $Z_t$ equals $(F_0(x^1),\dots,F_0(x^p)))^\top$. However, it seems not directly applicable if $F(x)$ can not be written in the expectation form, or if we want to test composite null hypothesis or perform change point testing. For example, if $F(x)$ is the spectral distribution function, it can not be written in the expectation form because it depends on the joint distribution of $\{X_t\}_{t\in\Z}$.
		
\begin{figure}[H]
			
			\begin{subfigure}{0.5\textwidth}
				\includegraphics[width=\textwidth]{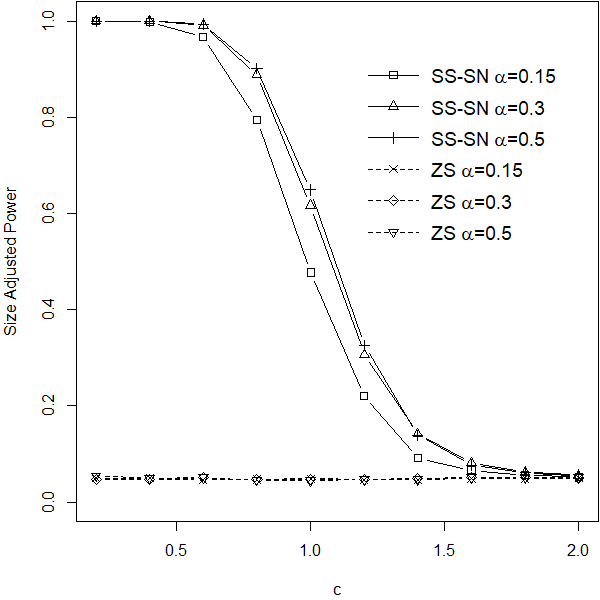}
				\caption{}
				\label{fig_alt2_n100_50}
			\end{subfigure}
			\hfill
			\begin{subfigure}{0.5\textwidth}
				\includegraphics[width=\textwidth]{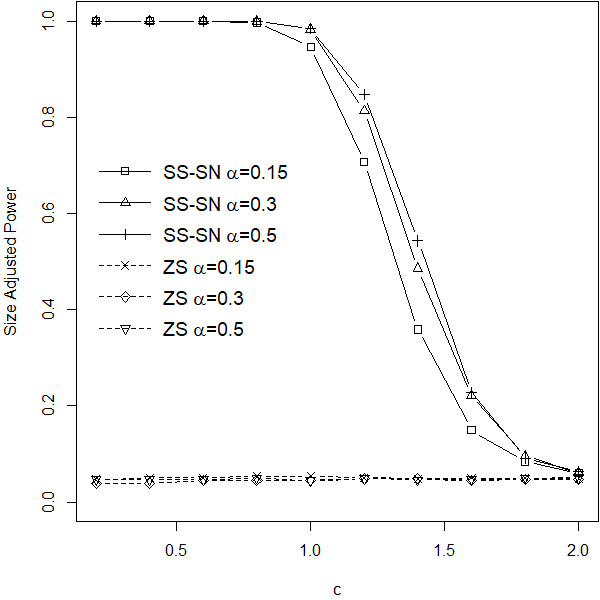}
				\caption{}
				\label{fig_alt2_n400_50}
			\end{subfigure}
			
			\caption{Size adjusted power for $p=50$, $n=100$ (left) and $n=400$ (right).}
			\label{fig_alt2}
		\end{figure}

		Even if $F(x)$ can be written in the expectation form,  the fastest rate at which $p$ can grow is $n^{1/4}$ (see Section 2.4 of \cite{zhang2023another}), which may be too small (for a moderate sample size) when $\Omega$ is not bounded (e.g. $\Omega=\R$ or $\R^d$ for some positive integer $d>1$). In addition, when determining the projection direction, the sample variance of each component of $Z_t$ calculated on the first subsample can not be zero, which means for fixed $i=1,\dots,p$, $\{Z^i_1,\dots,Z_\fa^i\}$ can not take the same value. This means oftentimes we need to choose the discretization points $\{x^1,\dots,x^p\}$ in a data-dependent way, which may lead to power loss, or  trivial power for some data generating processes (DGP). To show this, suppose we are interested in testing if the marginal cdf of a stationary time series is $\Phi(x)$. In this case we have $f(X_1,x) = \id(X_1\leq x)$. Let $X_{(1)}<X_{(2)}<\cdots<X_{(\fa)}$ be the order statistics of $\{X_t\}_{t=1}^\fa$, in order to make the sample variance of each component of $Z_t$ positive, we set $\{x^1,\dots,x^p\}$ to be equally spaced and
        \begin{equation}\label{eqdist}
            		X_{(1)}{+}0.1=x^1<x^2<\cdots<x^p = X_{(\fa)}{-}0.1.
        \end{equation}
		Consider the alternative hypothesis: $X_t=\id(X'_t\leq c)X'_t +c\id(X'_t>c)$ where $X'_t=\rho X'_{t-1}+\epsilon_t$ with ${\epsilon}_t \stackrel{iid}{\sim} N(0,1-\rho^2)$. The experiment is run 2000 times for $\alpha\in\{0.15,0.3,0.5\}$, $p\in\{5,10,50,200\}$, $\rho=-0.7$, $n\in\{100,400\}$ and the size adjusted powers are plotted against $c$. Denote the $L_2$-type SN statistic in Section 2.4 of \cite{zhang2023another} as ZS. We plot the size adjusted power curve for ZS with $p=50$ since the power for other values of $p$ are similar. As shown in Figure \ref{fig_alt2}, ZS have trivial power when $n\in\{100,400\}$ while SS-SN can successfully detect this alternative. This is because under the alternative, $x^p<c$, $F(x) =\Phi(x)$ for any $x< c$ and $F(x) = 1>\Phi(x)$ for any $x\geq c$.
		
        As pointed out by one reviewer, we can also define un-studentized versions of ZS without dividing each coordinate of the projection direction by the sample variance (equivalently, we just replace the sample variances of each coordinate by one). In this case, we can also consider discretization schemes other than Equation (\ref{eqdist}) that does not depend on the data. Specifically, for $L\in\{3,6,15\}$, we set $\{x^1,\dots,x^p\}$ to be equally spaced in the interval $[-L,L]$ and define the un-studentized statistic ZS$_L$ in the same way as ZS (without dividing each coordinate of the projection direction by the sample variance).
       
\begin{figure}[H]
		
		\begin{subfigure}{0.5\textwidth}
			\includegraphics[width=\textwidth]{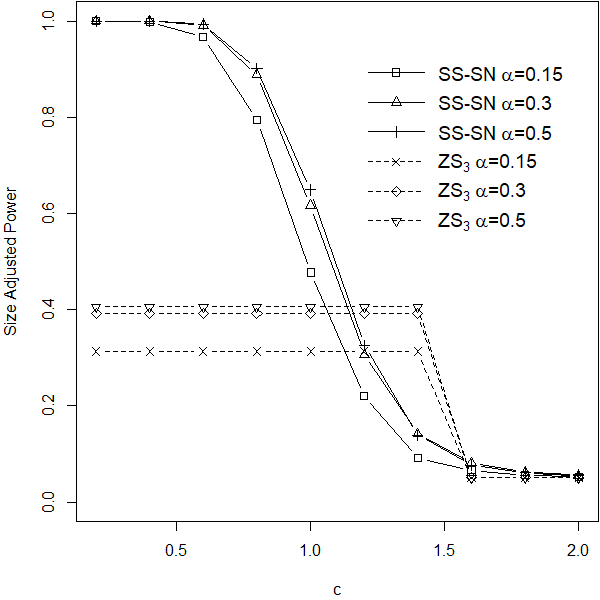}
			\caption{}
			\label{fig_3_alt2_n100_5}
		\end{subfigure}
		\hfill
		\begin{subfigure}{0.5\textwidth}
			\includegraphics[width=\textwidth]{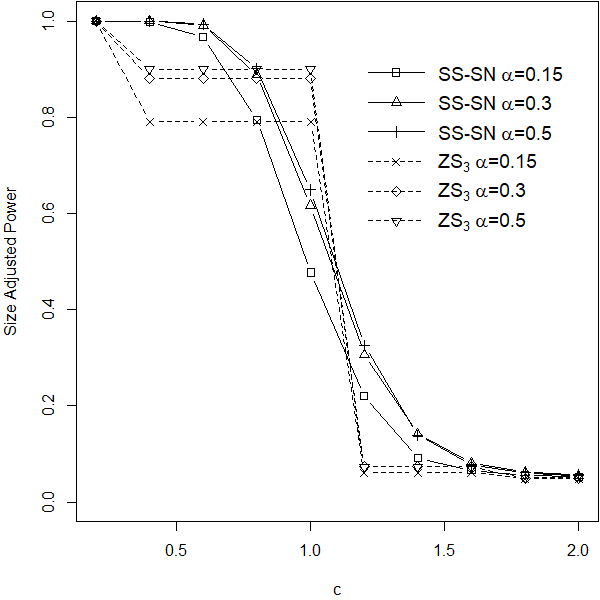}
			\caption{}
			\label{fig_3_alt2_n100_10}
		\end{subfigure}

		\begin{subfigure}{0.5\textwidth}
			\includegraphics[width=\textwidth]{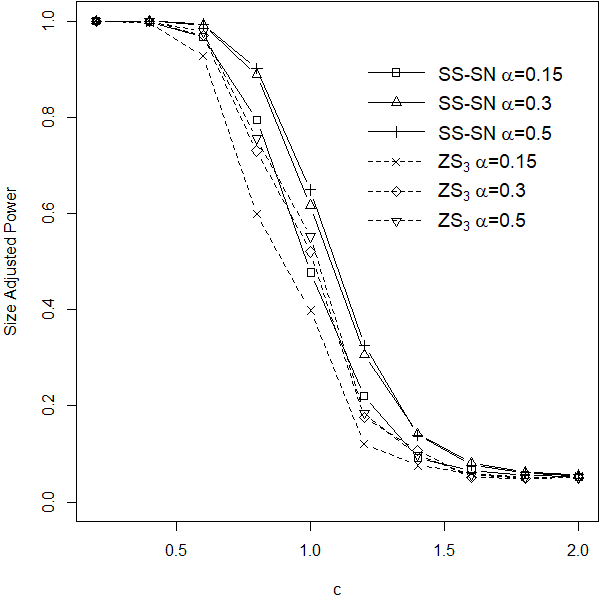}
			\caption{}
			\label{fig_3_alt2_n100_50}
		\end{subfigure}
		\hfill
		\begin{subfigure}{0.5\textwidth}
			\includegraphics[width=\textwidth]{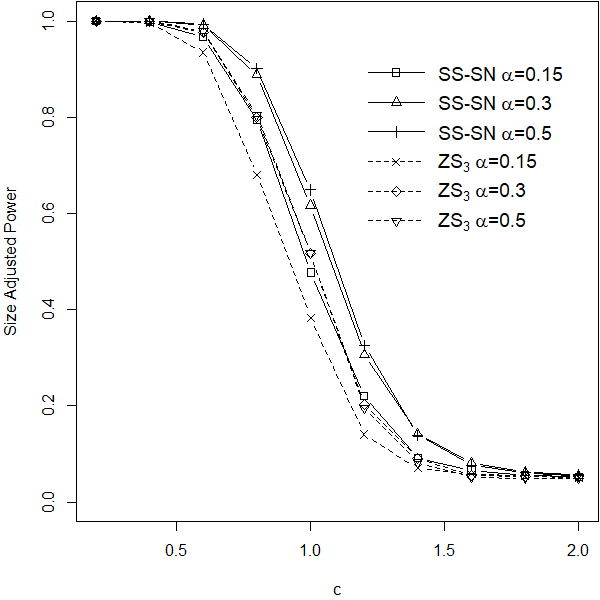}
			\caption{}
			\label{fig_3_alt2_n100_200}
		\end{subfigure}
		
		\caption{Size adjusted power for SS-SN and ZS$_3$ with $n=100$: (a), $p=5$; (b), $p=10$; (c), $p=50$; (d), $p=200$.}
		\label{fig_3_alt2_n100}
	\end{figure}

\begin{figure}[H]
		
		\begin{subfigure}{0.5\textwidth}
			\includegraphics[width=\textwidth]{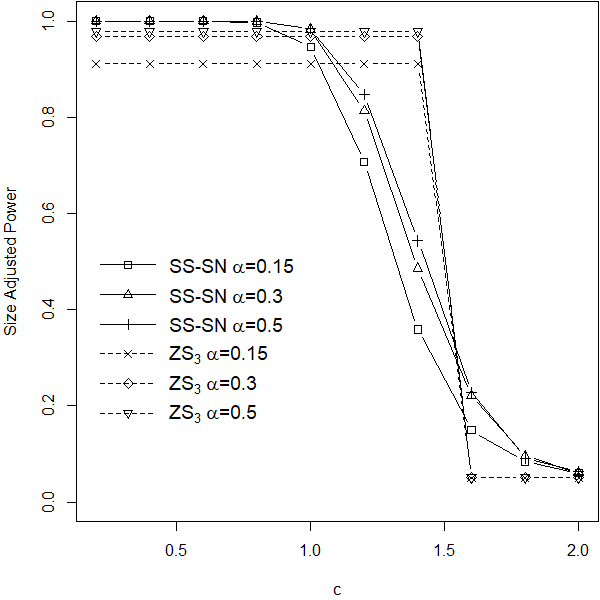}
			\caption{}
			\label{fig_3_alt2_n400_5}
		\end{subfigure}
		\hfill
		\begin{subfigure}{0.5\textwidth}
			\includegraphics[width=\textwidth]{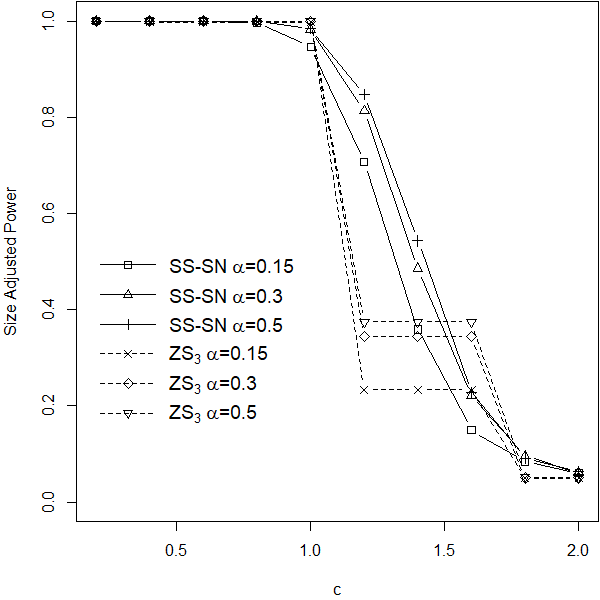}
			\caption{}
			\label{fig_3_alt2_n400_10}
		\end{subfigure}

		\begin{subfigure}{0.5\textwidth}
			\includegraphics[width=\textwidth]{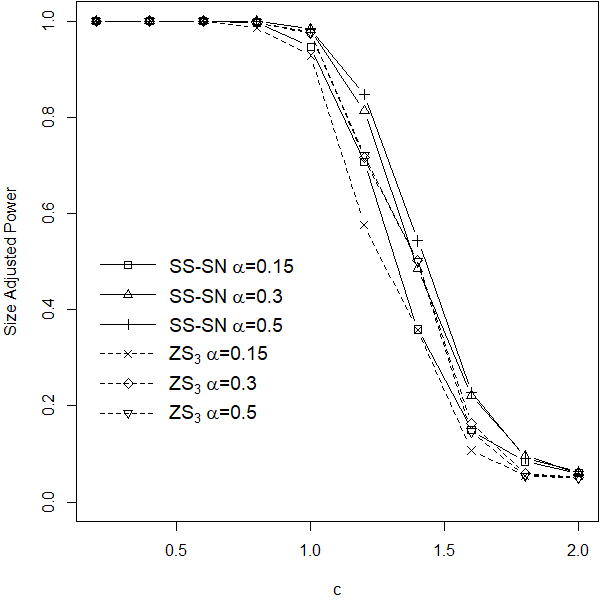}
			\caption{}
			\label{fig_3_alt2_n400_50}
		\end{subfigure}
		\hfill
		\begin{subfigure}{0.5\textwidth}
			\includegraphics[width=\textwidth]{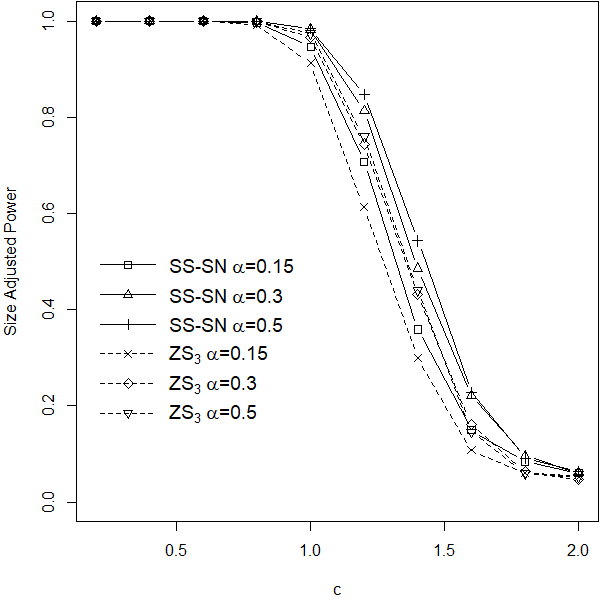}
			\caption{}
			\label{fig_3_alt2_n400_200}
		\end{subfigure}
		
		\caption{Size adjusted power for SS-SN and ZS$_3$ with $n=400$: (a), $p=5$; (b), $p=10$; (c), $p=50$; (d), $p=200$.}
		\label{fig_3_alt2_n400}
	\end{figure}

\begin{figure}[H]
		
		\begin{subfigure}{0.43\textwidth}
			\includegraphics[width=\textwidth]{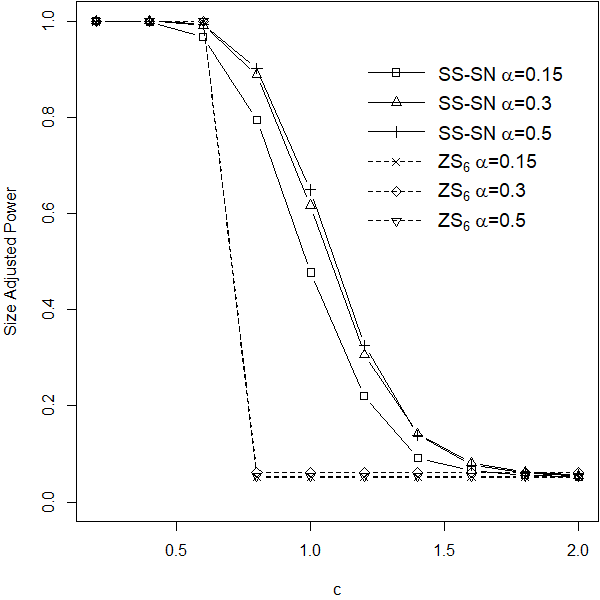}
			\caption{}
			\label{fig_6_alt2_n100_10}
		\end{subfigure}
		\hfill
		\begin{subfigure}{0.43\textwidth}
			\includegraphics[width=\textwidth]{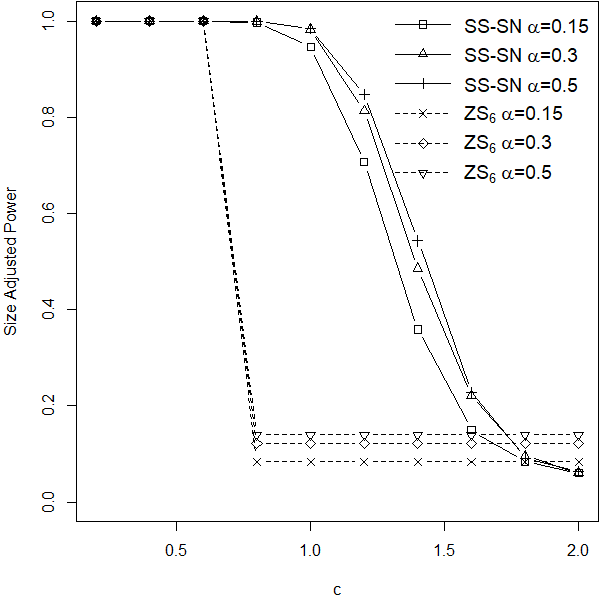}
			\caption{}
			\label{fig_6_alt2_n400_10}
		\end{subfigure}

		\begin{subfigure}{0.43\textwidth}
			\includegraphics[width=\textwidth]{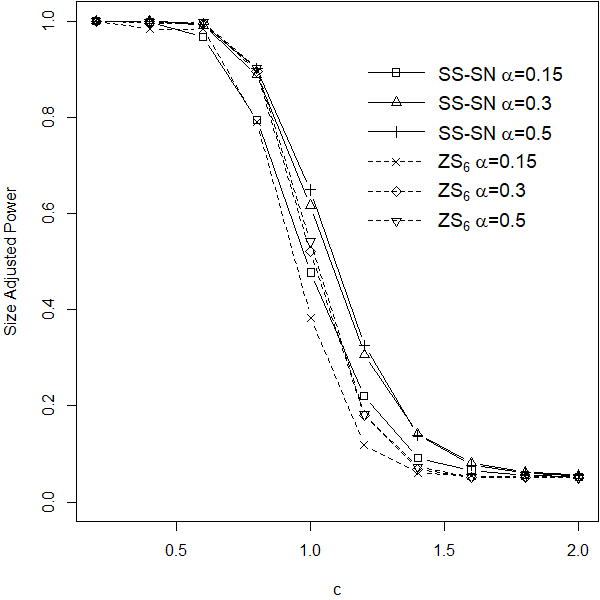}
			\caption{}
			\label{fig_6_alt2_n100_50}
		\end{subfigure}
		\hfill
		\begin{subfigure}{0.43\textwidth}
			\includegraphics[width=\textwidth]{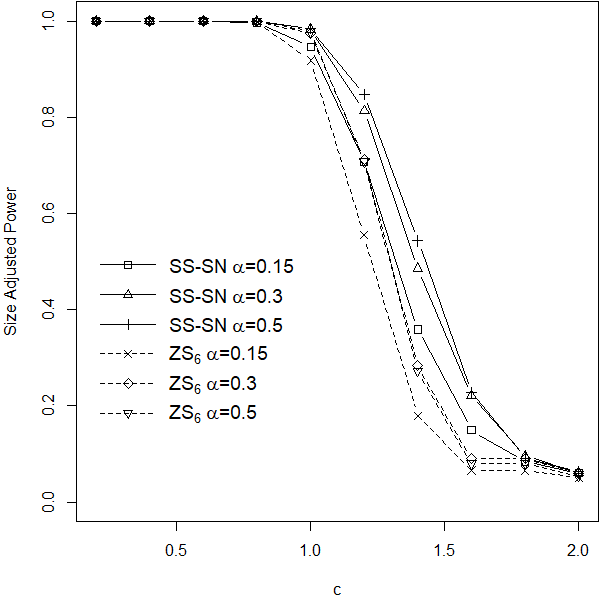}
			\caption{}
			\label{fig_6_alt2_n400_50}
		\end{subfigure}
		
		\begin{subfigure}{0.43\textwidth}
			\includegraphics[width=\textwidth]{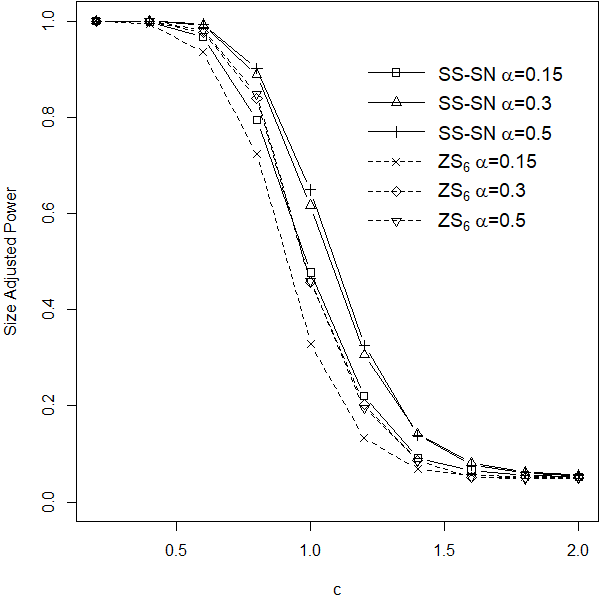}
			\caption{}
			\label{fig_6_alt2_n100_200}
		\end{subfigure}
		\hfill
		\begin{subfigure}{0.43\textwidth}
			\includegraphics[width=\textwidth]{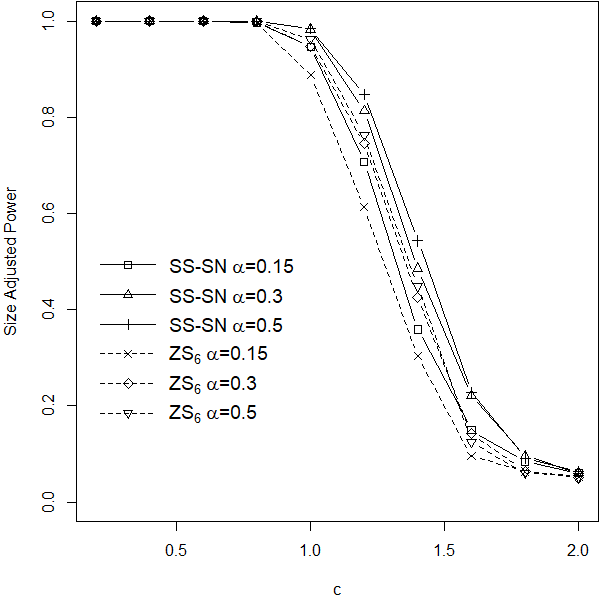}
			\caption{}
			\label{fig_6_alt2_n400_200}
		\end{subfigure}
		
		\caption{Size adjusted power for SS-SN and ZS$_6$ with $n=100$ (left column) and $n=400$ (right column) for $p=10$  (first row), $p=50$  (second row) and $p=200$ (third row).}
		\label{fig_6_alt2}
	\end{figure}

\begin{figure}[H]
		
		\begin{subfigure}{0.43\textwidth}
			\includegraphics[width=\textwidth]{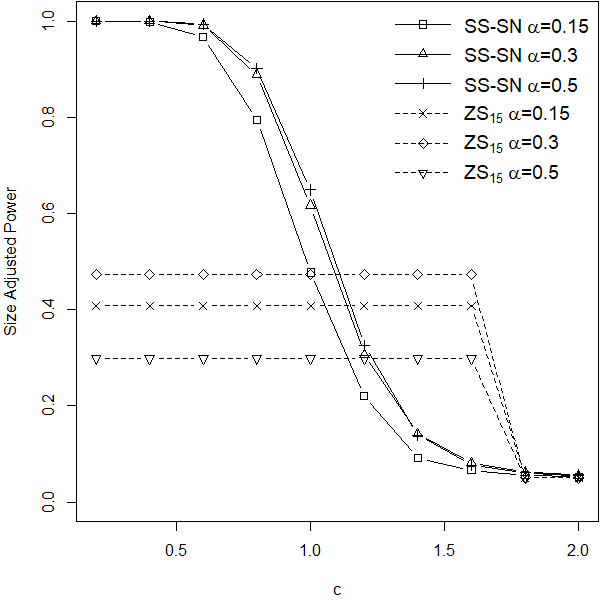}
			\caption{}
			\label{fig_15_alt2_n100_10}
		\end{subfigure}
		\hfill
		\begin{subfigure}{0.43\textwidth}
			\includegraphics[width=\textwidth]{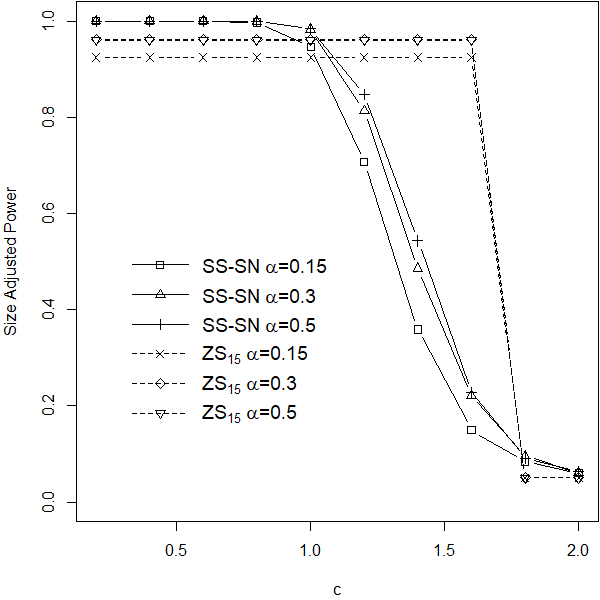}
			\caption{}
			\label{fig_15_alt2_n400_10}
		\end{subfigure}

		\begin{subfigure}{0.43\textwidth}
			\includegraphics[width=\textwidth]{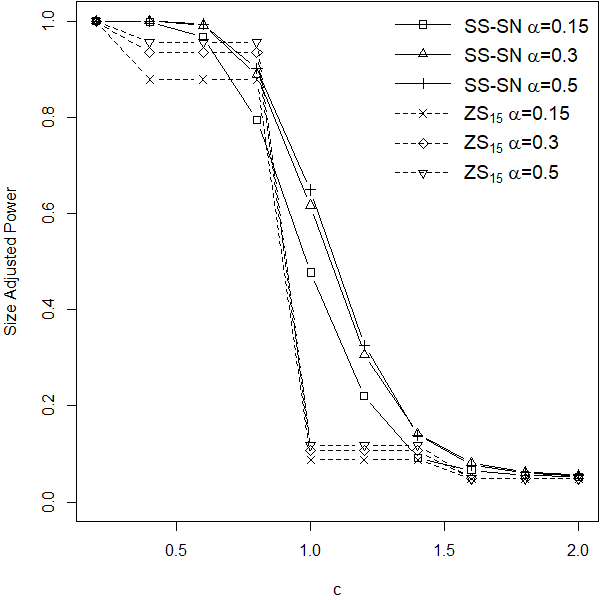}
			\caption{}
			\label{fig_15_alt2_n100_50}
		\end{subfigure}
		\hfill
		\begin{subfigure}{0.43\textwidth}
			\includegraphics[width=\textwidth]{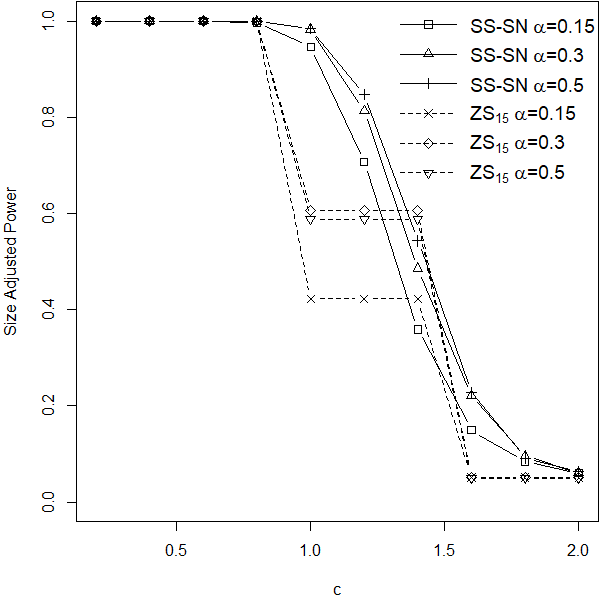}
			\caption{}
			\label{fig_15_alt2_n400_50}
		\end{subfigure}
		
		\begin{subfigure}{0.43\textwidth}
			\includegraphics[width=\textwidth]{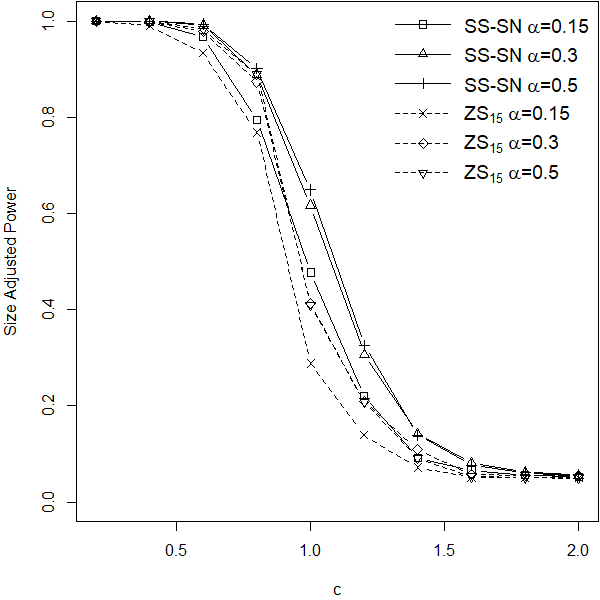}
			\caption{}
			\label{fig_15_alt2_n100_200}
		\end{subfigure}
		\hfill
		\begin{subfigure}{0.43\textwidth}
			\includegraphics[width=\textwidth]{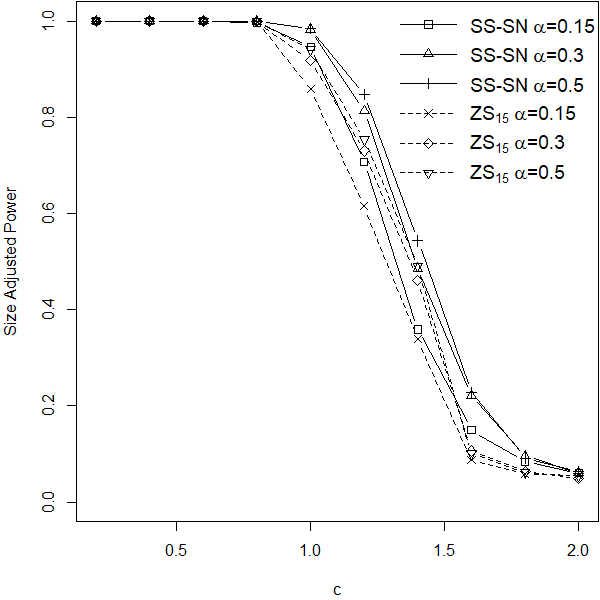}
			\caption{}
			\label{fig_15_alt2_n400_200}
		\end{subfigure}
		
		\caption{Size adjusted power for SS-SN and ZS$_{15}$ with $n=100$ (left column) and $n=400$ (right column) for $p=10$  (first row), $p=50$  (second row) and $p=200$ (third row).}
		\label{fig_15_alt2}
	\end{figure}

       The size adjusted powers for ZS$_{3}$, ZS$_{6}$ and ZS$_{15}$ with $n\in\{100,400\}$ are plotted in Figures \ref{fig_3_alt2_n100}-\ref{fig_15_alt2} and their power performances are similar to SS-SN when $p\in\{50,200\}$ (except for ZS$_{15}$ when $p=50$; see Figures \ref{fig_15_alt2_n100_50} and \ref{fig_15_alt2_n400_50}). The power performances of ZS$_3$ for $p\in\{5,10\}$, as well as the power performances of ZS$_{6}$ for $p=10$ and ZS$_{15}$ for $p\in\{10,50\}$, become very unstable. Although the power curves are still monotone, they may have sudden jumps when the signal increased/decreased beyond certain level; see Figures \ref{fig_6_alt2}-\ref{fig_15_alt2}. This is because under the alternative, if we assmue $c\in(x^i,x^{i+1})$ for some fixed $i\in\{1,2,\dots,{p-1}\}$, then $\{(Z^{i+1}_t,Z^{i+2}_t,\dots,Z^{p}_t)^\top\}_{t=1}^n$ are all $(p-i)$ dimensional vectors of ones and $\{(Z^{1}_t,Z^{2}_t,\dots,Z^{i}_t)^\top\}_{t=1}^n$ remains unchanged as c varies inside $(x^i,x^{i+1})$. So, the data $\{Z_t\}_{t=1}^n$ used to construct ZS$_{3}$, ZS$_{6}$ and ZS$_{15}$ are the same for different local alternatives $c\in(x^i,x^{i+1})$, resulting in the stepwise power curves for ZS$_{3}$, ZS$_{6}$ and ZS$_{15}$. Hence for this particular DGP, this artifact is solely  due to the discretization as the  power curve for our SS-SN approach (i.e., projection via integral) is not stepwise. 
       
		Additional simulation results (not shown) suggest that 
		the test proposed in \cite{zhang2023another} is more suitable for the inference of a moderate dimensional parameter, and its applicability to the functional parameter is rather restricted, and may lead to size distortion and power loss as compared to the SS-SN test we develop here due to the suboptimal choice of discretization scheme.
        

	\section{Proofs of main results}\label{app_proof}
	
	\subsection{Proofs of Theorem \ref{th_1null2}}\label{appen_th_1null2}
	
	First we prove part (i) of Theorem \ref{th_1null2}. By continuous mapping theorem, we have 
	\begin{align}
		P_\alpha(x) &\rightsquigarrow H(\alpha,x)\mbox{ on } \ell_\infty[ \Omega], \nonumber\\
		T_n &\stackrel{\D}{\to} \frac{\int_\Omega H(\alpha,x)[H(1,x)-H(\alpha,x)]dx}{\Big\{\frac{1}{1-\alpha}\int_\alpha^1\big\{\int_\Omega H(\alpha,x)[H(r,x)-H(\alpha,x)]dx-\frac{r-\alpha}{1-\alpha}\int_\Omega H(\alpha,x)[H(1,x)-H(\alpha,x)]dx\big\}^2 dr\Big\}^{1/2}}\nonumber \\
		&=T'. \nonumber
	\end{align}
	Since $\{H(r,x)-H(\alpha,x)\}_{r\in (\alpha,1],x\in\Omega}$ and $\{H(t,y)\}_{t\in[0,\alpha],y\in\Omega}$ are independent and $\{H(r,x)-H(\alpha,x)\}_{r\in (\alpha,1],x\in\Omega}$ is a Gaussian process with covariance function $\cov[H(r_1,x_1)-H(\alpha,x_1),H(r_2,x_2)-H(\alpha,x_2)]=\min\{r_1-\alpha,r_2-\alpha\}C(x_1,x_2)$, we have that conditioning on $\{H(\alpha,x)\}_{x\in\Omega}$, $\{\int_\Omega H(\alpha,x)[H(r,x)-H(\alpha,x)]dx\}_{r\in (\alpha,1]} \coloneqq \{\wt B(r)\}_{r\in (\alpha,1]}$ is a Gaussian process with covariance function $\cov[\wt B(r_1),\wt B(r_2)]=\min\{r_1-\alpha,r_2-\alpha\}\wt C$ where $\wt C = \int_\Omega\int_\Omega H(\alpha,x)C(x,y)H(\alpha,y)dxdy$. So the conditional distribution of $T'$ given $\{H(\alpha,x)\}_{x\in\Omega}$ is 
	\begin{align}
		T'\Big|\{H(\alpha,x)\}_{x\in\Omega} &\stackrel{d}{=}\frac{\wt B(1)}{\big\{\frac{1}{1-\alpha}\int_\alpha^1\big\{\wt B(r)-\frac{r-\alpha}{1-\alpha}\wt B(1)\big\}^2 dr\big\}^{1/2}}\Big|\{H(\alpha,x)\}_{x\in\Omega}\nonumber\\
		&\stackrel{d}{=}\frac{\frac{\wt B(1)}{\sqrt{\wt C}}}{\big\{\frac{1}{1-\alpha}\int_\alpha^1\big\{\frac{\wt B(r)}{\sqrt{\wt C}}-\frac{r-\alpha}{1-\alpha}\frac{\wt B(1)}{\sqrt{\wt C}}\big\}^2 dr\big\}^{1/2}}\Big|\{H(\alpha,x)\}_{x\in\Omega}\nonumber \\
		&\stackrel{d}{=}\frac{B(1)-B(\alpha)}{\big\{\frac{1}{1-\alpha}\int_\alpha^1\big\{B(r)-B(\alpha)-\frac{r-\alpha}{1-\alpha}(B(1)-B(\alpha))\big\}^2 dr\big\}^{1/2}}\label{eeq1}\\
		&\stackrel{d}{=}\frac{B(1)-B(\alpha)}{\big\{\frac{1}{1-\alpha}\int_0^{1-\alpha}\big\{B(r+\alpha)-B(\alpha)-\frac{r+\alpha-\alpha}{1-\alpha}(B(1)-B(\alpha))\big\}^2 dr\big\}^{1/2}}\label{eeq5}\\
		&\stackrel{d}{=}\frac{B(1-\alpha)}{\big\{\frac{1}{1-\alpha}\int_0^{1-\alpha}\big\{B(r)-\frac{r}{1-\alpha}B(1-\alpha)\big\}^2 dr\big\}^{1/2}} \label{eeq2}\\
		&\stackrel{d}{=}\frac{B(1-\alpha)}{\big\{\int_0^{1}\big\{B((1-\alpha)r)-rB(1-\alpha)\big\}^2 dr\big\}^{1/2}} \label{eeq4}\\
		&\stackrel{d}{=}\frac{B(1)}{\big\{\int_0^{1}\big\{B(r)-rB(1)\big\}^2 dr\big\}^{1/2}}. \label{eeq3}
	\end{align}
	For Equation (\ref{eeq1}), we used the fact that conditioning on $\{H(\alpha,x)\}_{x\in\Omega}$, $\{\wt B(r)/\sqrt{\wt C}\}_{r\in (\alpha,1]}\stackrel{d}{=}\{B(r)-B(\alpha)\}_{r\in (\alpha,1]}$. For Equation (\ref{eeq2}), we used the property $\{B(r+\alpha)-B(\alpha):r\in[0,1-\alpha]\}\stackrel{d}{=}\{B(r):r\in[0,1-\alpha]\}$. For Equations (\ref{eeq5}) and (\ref{eeq4}), we used the change of variable property for integration. For Equation (\ref{eeq3}), we used the property $\{B((1-\alpha)r):r\in[0,1]\} \stackrel{d}{=}\{\sqrt{1-\alpha}B(r):r\in[0,1]\}$.
	
	So the limiting null distribution is
	\begin{equation*}
		T'=\frac{B(1)}{\big\{\int_0^{1}\big\{B(r)-rB(1)\big\}^2 dr\big\}^{1/2}}=U_1, 
	\end{equation*}
	and part (i) is proved.

	To prove part (ii).1 of Theorem \ref{th_1null2}, note that the numerator of $T_n$ can be written as 
	\begin{align}
		&\frac{\sqrt{n}}{\sqrt{n{-}\fa}}\Big\{\int_\Omega   \frac{\fa}{\sqrt{n}}\big[F_{1:\fa}(x){-}F_1(x){+}F_1(x){-}F_0(x)\big]\frac{n{-}\fa}{\sqrt{n}}\big[F_{\fa{+}1:n}(x){-}F_1(x){+}F_1(x){-}F_0(x)\big] dx \Big\}\nonumber\\
		=&\frac{\sqrt{n}}{\sqrt{n{-}\fa}}\Big\{ N_1+N_2+N_3+ \frac{\fa(n{-}\fa)  }{n} \|F_1-F_0\|_2^2     \Big\},\label{eq_app2_1}
	\end{align}
	where $N_1=\int_\Omega  \frac{\fa}{\sqrt{n}}\big[F_{1:\fa}(x){-}F_1(x)\big]\frac{n{-}\fa}{\sqrt{n}}\big[F_{\fa+1:n}(x){-}F_1(x)\big]dx\stackrel{\D}{\to} \int_\Omega H(\alpha,x)[H(1,x)-H(\alpha,x)]dx$. Since 
	\begin{align}
		|N_2| &= \Bigg|\int_\Omega  \frac{\fa}{\sqrt{n}}\big[F_{1:\fa}(x){-}F_1(x)\big]\frac{n{-}\fa}{\sqrt{n}}\big[F_1(x){-}F_0(x)\big]dx       \Bigg| \nonumber \\
		&\leq \Big\{ \int_\Omega  \frac{\fa^2}{n}\big[F_{1:\fa}(x){-}F_1(x)\big]^2dx    \Big\}^{1/2} \frac{n{-}\fa}{\sqrt{n}}\|F_1-F_0\|_2 \nonumber
	\end{align}
	and $ \int_\Omega  \frac{\fa^2}{n}\big[F_{1:\fa}(x){-}F_1(x)\big]^2dx \stackrel{\D}{\to} \int_\Omega\big[H(\alpha,x)\big]^2dx$, we have $N_2=O_p(\sqrt{n}\|F_1-F_0\|_2)$. For the same reason, $N_3=\int_\Omega  \frac{\fa}{\sqrt{n}}\big[F_{1}(x){-}F_0(x)\big]\frac{n{-}\fa}{\sqrt{n}}\big[F_{\fa+1:n}(x){-}F_1(x)\big]dx=O_p(\sqrt{n}\|F_1-F_0\|_2)$. So the numerator is of the same order as $n\|F_1-F_0\|_2^2$.
	
	For the denominator of $T_n$, let $b_k(x)=F_{\fa+1:\fa+k}(x){-}F_{\fa+1:n}(x)$ and we have for $k=\fa+1,\dots,n$
	\begin{align}
		(S_k-S_n)^2 &=\Big\{\int_\Omega \frac{\fa}{\sqrt{n}}\big[F_{1:\fa}(x){-}F_1(x)\big]b_k(x)dx {+}\frac{\fa}{\sqrt{n}}\int_\Omega\big[F_1(x){-}F_0(x)\big]b_k(x)dx         \Big\}^2      \nonumber \\
		&\leq 2\Big\{\int_\Omega \frac{\fa}{\sqrt{n}}\big[F_{1:\fa}(x){-}F_1(x)\big]b_k(x)dx   \Big\}^2 {+}2\Big\{\frac{\fa}{\sqrt{n}}\int_\Omega\big[F_1(x){-}F_0(x)\big]b_k(x)dx         \Big\}^2  \nonumber \\
		&\leq 2\Big\{\int_\Omega \frac{\fa}{\sqrt{n}}\big[F_{1:\fa}(x){-}F_1(x)\big]b_k(x)dx   \Big\}^2 {+}2\frac{\fa^2}{n}\|F_1-F_0\|_2^2\int_\Omega b_k(x)^2dx. \label{eq_app2_2}
	\end{align}
	Since 
	\begin{align}
		&\frac{1}{(n{-}\fa)^2}\sum_{k=\fa{{+}}1}^n(k{-}\fa)^2\Big\{\int_\Omega \frac{\fa}{\sqrt{n}}\big[F_{1:\fa}(x){-}F_1(x)\big]b_k(x)dx   \Big\}^2 \nonumber \\
		\stackrel{\D}{\to}& \frac{1}{(1{-}\alpha)^2} \int_{\alpha}^{1} \Big\{ \int_\Omega H(\alpha,x)\big[H(r,x)-H(\alpha,x){-}\frac{r{-}\alpha}{1{-}\alpha}(H(1,x){-}H(\alpha,x))\big]  dx                     \Big\}^2 dr \nonumber
	\end{align}
	and $\frac{1}{(n{-}\fa)^2}\sum_{k=\fa{{+}}1}^n(k{-}\fa)^2\int_\Omega b_k(x)^2dx\stackrel{\D}{\to} \frac{1}{(1{-}\alpha)^2} \int_{\alpha}^{1} \Big\{\int_\Omega \big[H(r,x)-H(\alpha,x){-}\\\frac{r{-}\alpha}{1{-}\alpha}(H(1,x){-}H(\alpha,x))\big]^2  dx     \Big\}^2 dr $, the denominator of $T_n$ is of the order $O_p(\sqrt{n}\|F_1-F_0\|_2)$ and part (ii).1 of Theorem \ref{th_1null2} is proved.
	
	To prove part (ii).2 and (ii).3 of Theorem \ref{th_1null2}, by continuous mapping theorem, 
	\begin{align*}
		P_\alpha(x) = \frac{\fa}{\sqrt{n}}[F_{1:\fa}-F_0(x)] \rightsquigarrow H(\alpha,x)+\alpha c(x){=}\wt c(x)
	\end{align*}
	\begin{align}
		T_n &\stackrel{\D}{\to} \frac{\int_\Omega \wt c(x)[H(1,x){-}H(\alpha,x)]dx{+} (1{-}\alpha)\int_\Omega\wt c(x)c(x)dx}{\Big\{\frac{1}{1-\alpha}\int_\alpha^1\big\{\int_\Omega \wt c(x)[H(r,x)-H(\alpha,x)]dx-\frac{r-\alpha}{1-\alpha}\int_\Omega \wt c(x)[H(1,x)-H(\alpha,x)]dx\big\}^2 dr\Big\}^{1/2}} = \wt U_1.\nonumber
	\end{align}
	Since conditioning on $\{H(\alpha,x)\}_{x\in\Omega}$, $\{\int_\Omega \wt c(x)[H(r,x)-H(\alpha,x)]dx\}_{r\in (\alpha,1]} \coloneqq \{\wt B^A(r)\}_{r\in (\alpha,1]}$ is a Gaussian process with covariance function $\cov[\wt B^A(r_1),\wt B^A(r_2)]=\min\{r_1-\alpha,r_2-\alpha\} C^A$ where $C^A= \int_\Omega\int_\Omega \wt c(x)C(x,y)\wt c(y)dxdy$, we have 
	\begin{align}
		\wt U_1\Big|\{H(\alpha,x)\}_{x\in\Omega} =& \frac{(C^A)^{-1/2}\wt B^A(1)+(C^A)^{-1/2}(1-\alpha)\wt c( x)}{\Big\{\frac{1}{1-\alpha}\int_\alpha^1\big\{(C^A)^{-1/2}\wt B^A(r)-(C^A)^{-1/2}\frac{r-\alpha}{1-\alpha}\wt B^A(1)\big\}^2 dr\Big\}^{1/2}} \Big|\{H(\alpha,x)\}_{x\in\Omega} \nonumber  \\
		\stackrel{d}{=}& \frac{ B(1)- B(\alpha)+(C^A)^{-1/2}(1-\alpha)\wt c( x)}{\Big\{\frac{1}{1-\alpha}\int_\alpha^1\big\{ B(r)-B(\alpha)-  \frac{r-\alpha}{1-\alpha}[B(1)-B(\alpha)]\big\}^2 dr\Big\}^{1/2}}  \label{app_alt2}    \\
		\stackrel{d}{=}& \frac{B(1-\alpha)+(C^A)^{-1/2}(1-\alpha)\wt c( x)}{\Big\{\frac{1}{1-\alpha}\int_\alpha^{1}\big\{ B(r-\alpha)- \frac{r-\alpha}{1-\alpha}B(1-\alpha)\big\}^2 dr\Big\}^{1/2}}    \label{app_alt31}      \\	
		\stackrel{d}{=}& \frac{ B(1-\alpha)+(C^A)^{-1/2}(1-\alpha)\wt c( x)}{\Big\{\frac{1}{1-\alpha}\int_0^{1-\alpha}\big\{ B(r)- \frac{r}{1-\alpha}B(1-\alpha)\big\}^2 dr\Big\}^{1/2}}    \label{app_alt3}      \\
		\stackrel{d}{=}&  \frac{ B(1)+\sqrt{\frac{1-\alpha}{C^A}}\wt c( x)}{\Big\{\int_0^{1}\big\{ B(r)- rB(1)\big\}^2 dr\Big\}^{1/2}},  \label{app_alt4}
	\end{align}
	Equation (\ref{app_alt2}) follows from the fact that conditioning on $\{H(\alpha,x)\}_{x\in\Omega}$, $\{\wt B^A(r)/\sqrt{C^A}\}_{r\in (\alpha,1]}\stackrel{d}{=}\{B(r)-B(\alpha)\}_{r\in (\alpha,1]}$, and in Equation (\ref{app_alt3}), we used the change of variable property for integration. For Equation (\ref{app_alt31}), we used the property $\{B(r+\alpha)-B(\alpha):r\in[0,1-\alpha]\}\stackrel{d}{=}\{B(r):r\in[0,1-\alpha]\}$. For Equation (\ref{app_alt4}), we used the property $\{B((1-\alpha)r):r\in[0,1]\} \stackrel{d}{=}\{\sqrt{1-\alpha}B(r):r\in[0,1]\}$. So part (ii).2 and (ii).3 of Theorem \ref{th_1null2} is proved.

	\subsection{Proofs of Theorem \ref{th_nnull}}\label{appen_th_nnull}
	
	First we prove part (i) of Theorem \ref{th_nnull}. By continuous mapping theorem, we have 
	\begin{align}
		P'_\alpha(x) = &\frac{\fa}{\sqrt{n}}\big[F'_{1:\fa}(x)-F'_0(x;\boldsymbol{\theta}_0)-R_{1:\fa}(x)\big] \nonumber \\
		=&  \frac{\fa}{\sqrt{n}}\big[F'_{1:\fa}(x)-F'_0(x;\boldsymbol{\theta}_0)+o_p(n^{-1/2})\big] \nonumber \\
		\rightsquigarrow &H'(\alpha,x) \mbox{ on }\ell_\infty[\Omega]. 
	\end{align}
	For the numerator of $T'_n$, we have
	\begin{align}
		\sqrt{n{-}\fa}S_n' & =\sqrt{n{-}\fa}\int_\Omega P_\alpha'(x)[F_{\fa{+}1:n}(x){-}F_0(x;\boldsymbol{\hat\theta}_{\fa{+}1:n})] dx  \nonumber \\
		& = \sqrt{n{-}\fa}\int_\Omega\frac{\fa}{\sqrt{n}}\big[F'_{1:\fa}(x)-F'_0(x;\boldsymbol{\theta}_0)][F'_{\fa{+}1:n}(x){-}F'_0(x;\boldsymbol{\theta}_{0})] dx  +o_p(1) \nonumber \\
		& \stackrel{\D}{\to}(1{-}\alpha)^{-1/2}\int_\Omega H'(\alpha,x)[H'(1,x)-H'(\alpha,x)] dx . \nonumber
	\end{align} 
	Let the square of the denominator of $T'_n$ be $V'_n$, and denote $A_k = (k{-}\fa)\int_\Omega P'_\alpha(x)[F'_{\fa{+}1:k}(x){-}\\F'_{\fa{+}1:n}(x)]dx$, $B_k= (k{-}\fa)\int_\Omega P'_\alpha(x)R_{\fa+1:k}(x)dx$ and $C_k=(k{-}\fa)\int_\Omega P'_\alpha(x)R_{\fa+1:n}(x)dx$, we have
	\footnotesize 
	\begin{align}
		V'_{n}&=(n{-}\fa)^{{-}2}\sum_{k=\fa{+}1}^{n} (k{-}\fa)^2\Big\{\int_\Omega P'_\alpha(x)\big[F_{\fa{+}1:k}(x){-}F_0(x;\boldsymbol{\hat\theta}_{\fa{+}1:k}){-}F_{\fa{+}1:n}(x){+}F_0(x;\boldsymbol{\hat\theta}_{\fa{+}1:n})\big]dx\Big\}^2 \nonumber \\
		& = (n{-}\fa)^{{-}2}\sum_{k=\fa{+}1}^{n}\Big\{ A_k^2+B_k^2+C_k^2-2A_kB_k+2A_kC_k-2B_kC_k         \Big\}.
	\end{align}
	\normalsize
	By Assumption \ref{assump_nnull}, $(n{-}\fa)^{{-}2}\sum_{k=\fa{+}1}^{n}A_k^2 \stackrel{\D}{\to} \frac{1}{(1-\alpha)^2}\int_\alpha^1\big\{\int_\Omega H'(\alpha,x)[H'(r,x)-H'(\alpha,x)-\frac{r-\alpha}{1-\alpha}(H'(1,x)-H'(\alpha,x))]dx\big\}^2 dr$, $(n{-}\fa)^{{-}2}\sum_{k=\fa{+}1}^{n}B_k^2=o_p(1)$ and $(n{-}\fa)^{{-}2}\sum_{k=\fa{{+}}1}^{n}C_k^2=o_p(1)$. From this and the continuous mapping theorem, we have 
	$$		T'_n \stackrel{\D}{\to} \frac{\int_\Omega H'(\alpha,x)[H'(1,x)-H'(\alpha,x)]dx}{\Big\{\frac{1}{1-\alpha}\int_\alpha^1\big\{\int_\Omega H'(\alpha,x)[H'(r,x)-H'(\alpha,x)]dx-\frac{r-\alpha}{1-\alpha}\int_\Omega H'(\alpha,x)[H'(1,x)-H'(\alpha,x)]dx\big\}^2 dr\Big\}^{1/2}}. $$
	Following the conditioning idea used in the proof for part (i) of Theorem \ref{th_1null2}, we can finish the proof for part (i) of Theorem \ref{th_nnull} in an analogous manner. 
	
	The proof for part (ii) of Theorem \ref{th_nnull} is also similar to that of Theorem \ref{th_1null2} and we omit the details.

	\subsection{Proofs of Theorem \ref{th_cpn}}\label{appen_th_cpn}
	
	\subsubsection{Proof of part (i)}
	
	By simple calculation we can see that
	\begin{align}
		P_b(x) \rightsquigarrow   H(b,x)-\big[ H(1,x)- H(1-b,x)\big] 
		\stackrel{d}{=}&  \wt P_b(x).\nonumber
	\end{align}
	Also, let $k= \lfloor nr  \rfloor $, by continuous mapping theorem, we have
	\footnotesize 
	\begin{align}
		T_n(k)& = \frac{n^{3/2}}{(n-2\fb)^{3/2}} \Bigg\{\frac{(n-\fb-\fr)}{n}\frac{(\fr -\fb) }{\sqrt{n}}\int_\Omega P_b(x)\Big[{F}_{\fb +1:\fr }( x)-{F_0}(x)\Big]dx \nonumber\\
		& \qquad\qquad- \frac{(\fr-\fb)}{n}\frac{(n-\fb-\fr)}{\sqrt{n}}\int_\Omega P_b(x)\Big[{F}_{\fr +1:n-\fb }(x)-{F_0}(x)\Big]dx  \Bigg\} \nonumber\\
		& \rightsquigarrow \frac{1}{(1-2b)^{3/2}} \Big\{(1-b-r)\int_\Omega \wt P_b(x)\big[H(r,x)-H(b,x)\big]dx-(r-b)\int_\Omega \wt P_b(x)\big[H(1-b,x)-H(r,x)\big]dx  \Big\}  \nonumber\\
		& = \frac{1}{\sqrt{1-2b}} \Big\{\int_\Omega \wt P_b(x)[H(r,x)-H(b,x)]dx-\frac{r-b}{1-2b}\int_\Omega \wt P_b(x)\big[H(1-b,x)-H(b,x)\big]dx \Big\} \nonumber \\
		& :=T(r)\label{eqap1}
	\end{align}
	\begin{align}
		V_n(k) \rightsquigarrow \,& \frac{1}{1-2b} \Bigg\{\frac{1}{(r-b)^2}\int_b^{r} \Big\{(r {-} s)\int_\Omega \wt P_b(x)\big[H(s,x)-H(b,x)\big]dx {-} (s {-}b)\int_\Omega \wt P_b(x)\big[H(r,x)-H(s,x)\big]dx\Big\}^2 ds  \nonumber\\
		& + \frac{1}{(1-b-r)^2}\int_{r}^{1-b} \Big\{(s{-}r)\int_\Omega \wt P_b(x)\big[H(1-b,x)-H(s,x)\big]dx {-} (1{-}b{-}s)\int_\Omega \wt P_b(x) \big[H(s,x)-H(r,x)\big]dx \Big\}^2 ds\Bigg\}^{1/2}  \nonumber\\
		= &\frac{1}{1-2b} \Bigg\{\int_b^{r} \Big\{\int_\Omega \wt P_b(x)\big[H(s,x)-H(b,x)\big]dx {-} \frac{s-b}{r-b}\int_\Omega \wt P_b(x)\big[H(r,x)-H(b,x)\big]dx\Big\}^2 ds  \nonumber\\
		& + \int_{r}^{1-b} \Big\{\int_\Omega \wt P_b(x)\big[H(1-b,x)-H(s,x)]dx{-} \frac{1-b-s}{1-b-r} \int_\Omega \wt P_b(x)\big[H(1-b,x)-H(r,x)\big]dx \Big\}^2 ds\Bigg\}^{1/2} \nonumber \\
		:=&V(r),\nonumber
	\end{align}
	\normalsize
	so the limiting distribution of $G_n$ is $\sup_{r \in [b,1-b]}\frac{T(r)}{V(r)}$. By the orthogonal increment property of Brownian motion, we have $\{H(t,x)\}_{t\in[0,b],x\in\Omega}$, $\{H(s,y)-H(b,y)\}_{s\in[b,r],y\in\Omega}$, $\{H(1-b,z)-H(u,z)\}_{u\in[r,1-b],z\in \Omega}$ and $\{H(v,w)-H(1-b,w)\}_{v\in[1-b,1],w\in\Omega}$ are independent for any $r \in [b,1-b]$. Denote $C_b = \int_\Omega\int_\Omega \wt P_b(x)C(x,y)\wt P_b(y)dxdy$, then the conditional distribution of $\sup_{r \in [b,1-b]}\frac{T(r)}{V(r)}$ given $\wt P_b(x)$ is 
	\footnotesize 
	\begin{align}
		&\sup_{r \in [b,1{-}b]}\frac{T(r)}{V(r)} \Big| \wt P_b(x) \stackrel{d}{=} \sup_{r \in [b,1{-}b]}\frac{\frac{T(r)}{\sqrt{ C_b}}}{\frac{V(r)}{\sqrt{ C_b}}} \Big| \wt P_b(x) \nonumber\\
		\stackrel{d}{=} &\sup_{r \in [b,1{-}b]}  \frac{\sqrt{1{-}2b}\big(B(r){-}B(b){-} \frac{r{-}b}{1{-}2b}(B(1{-}b){-}B(b))  \big)}  {\Big\{\int_{b}^{r} \big[B(s){-}B(b){-}\frac{s{-}b}{r{-}b}(B(r){-}B(b))\big]^2 ds {+} \int_{r}^{1{-}b}\big[ B(1{-}b){-}B(s){-}\frac{1{-}b{-}s}{1{-}b{-}r}(B(1{-}b){-}B(r))    \big]^2 ds  \Big\}^{1/2}     }  \label{cpeq11}\\
		\stackrel{d}{=}& \sup_{r \in [b,1{-}b]}  \frac{\sqrt{1{-}2b}\big(B(r){-}B(b){-} \frac{r{-}b}{1{-}2b}(B(1{-}b){-}B(b)) }  {\Big\{\int_{0}^{r{-}b} \big[B(s{+}b){-}B(b){-}\frac{s}{r{-}b}(B(r){-}B(b))\big]^2 ds {+} \int_{r{-}b}^{1{-}2b}\big[ B(1{-}b){-}B(s{+}b){-}\frac{1{-}2b{-}s}{1{-}b{-}r}(B(1{-}b){-}B(r))    \big]^2 ds    \Big\}^{1/2}   }  \label{cpeq21}\\
		\stackrel{d}{=}& \sup_{r \in [b,1{-}b]}  \frac{\sqrt{1{-}2b}\big(B(r{-}b){-} \frac{r{-}b}{1{-}2b}(B(1{-}2b)) \big)}  {\Big\{\int_{0}^{r{-}b} \big[B(s){-}\frac{s}{r{-}b}B(r{-}b)\big]^2 ds {+} \int_{r{-}b}^{1{-}2b}\big[ B(1{-}2b){-}B(s){-}\frac{1{-}2b{-}s}{1{-}b{-}r}(B(1{-}2b){-}B(r{-}b)) \big]^2 ds\Big\}^{1/2} } \label{cpeq31}\\
		\stackrel{d}{=}& \sup_{r \in [0,1]}  \frac{\sqrt{1{-}2b}\big(B((1{-}2b)r){-} rB(1{-}2b) \big)}  {\Big\{\int_{0}^{(1{-}2b)r} \big[B(s){-}\frac{s}{(1{-}2b)r}B((1{-}2b)r)\big]^2 ds {+} \int_{(1{-}2b)r}^{1{-}2b}\big[ B(1{-}2b){-}B(s){-}\frac{1{-}2b{-}s}{(1{-}2b)(1{-}r)}(B(1{-}2b){-}B((1{-}2b)r)) \big]^2 ds\Big\}^{1/2} } \label{cpeq41} \\
		\stackrel{d}{=}& \sup_{r \in [0,1]}  \frac{\sqrt{1{-}2b}\big(B((1{-}2b)r){-} rB(1{-}2b) \big)}  {\Big\{\int_{0}^{r} \big[B((1-2b)s){-}\frac{s}{r}B((1{-}2b)r)\big]^2 ds {+} \int_{r}^{1}\big[ B(1{-}2b){-}B((1-2b)s){-}\frac{1{-}s}{1{-}r}(B(1{-}2b){-}B((1{-}2b)r)) \big]^2 ds \Big\}^{1/2}} \label{cpeq411} \\
		\stackrel{d}{=}& \sup_{r \in [0,1]}  \frac{B(r){-} rB(1) }  {\Big\{\int_{0}^{r} \big[B(s){-}\frac{s}{r}B(r)\big]^2 ds {+} \int_{r}^{1}\big[ B(1){-}B(s){-}\frac{1{-}s}{1{-}r}(B(1){-}B(r)) \big]^2 ds \Big\}^{1/2}} \label{cpeq51} 
	\end{align}
	\normalsize
	For Equation (\ref{cpeq11}), we used the fact that for any fixed $\wt P_b(x)$, $\{\int_\Omega\wt P_b(x) H(r,x)dx/\sqrt{C_b}\}_{r\in[0,1]}\stackrel{d}{=}\{B(r)\}_{r\in[0,1]}$ and $\frac{T(r)}{V(r)}$ is independent of $\wt P_b(x)$. For Equation (\ref{cpeq21}) we used the change of variable property for integration. For Equation (\ref{cpeq31}) we used the property $\{B(r{+}b){-}B(b):r\in[0,1{-}2b]\}\stackrel{d}{=}\{B(r):r\in[0,1{-}2b]\}$. For Equation (\ref{cpeq41}) we substitute $r-b$ with $(1-2b)r$. For Equation (\ref{cpeq411}) we used the change of variable property for integration and for Equation (\ref{cpeq51}) we used the fact $\{B((1-2b)r):r\in[0,1]\} \stackrel{d}{=}\{\sqrt{1-2b}B(r):r\in[0,1]\}$.

	\subsubsection{Proofs of part (ii)}

	To prove part (ii).1 of Theorem \ref{th_cpn}, denote $P'_b(x)=P_b(x)+\frac{\fb}{\sqrt{n}}\Delta(x)$, $Q'_{a:d}=\int_\Omega P'_b(x)[F_{a:d}(x)-\frac{1}{d-a+1}\sum_{t=a}^dF_t(x)]dx$ and $a_n = \frac{(k^\ast-\fb)(n-\fb-k^\ast)}{n^{3/2}}$, we have 
	\footnotesize
	\begin{align}
		T_n( k^\ast)=&  \frac{n^{3/2}}{(n-2\fb)^{3/2}} \Big\{a_nQ_{\fb+1,k^\ast}-a_nQ_{k^\ast+1,n-\fb}\Big\} \nonumber\\
		=&  \frac{n^{3/2}}{(n-2\fb)^{3/2}} \Big\{ a_nQ'_{\fb+1,k^\ast}-a_nQ'_{k^\ast+1,n-\fb} +a_n\frac{\fb}{\sqrt{n}}\|\Delta\|^2_2           \Big\}.
	\end{align}
	\normalsize
	Similar to Equation (\ref{eqap1}), by part (\ref{assump_cp_c}) of Assumption \ref{assump_cp}, we have $$\frac{n^{3/2}}{(n-2\fb)^{3/2}} \Big\{ a_nQ'_{\fb+1,k^\ast}{-}a_nQ'_{k^\ast+1,n-\fb} \Big\}=O_p(1),$$
	so $T_n( k^\ast)$ is of the same order as $n\|\Delta\|_2^2$. For $V^2_n( k^\ast)$, note that for $t=k^\ast+2,\dots,n-\fb$,
	\begin{align}
		& \frac{(n{-}\fb{-}t{+}1)^2(t{-}k^\ast{-}1)^2}{(n{-}\fb{-}k^\ast)^2}\big\{ Q_{t:n-\fb}{-} Q_{k^\ast+1:t-1} \big\}^2 \nonumber \\
		\leq &  \frac{(n{-}\fb{-}t{+}1)^2(t{-}k^\ast{-}1)^2}{(n{-}\fb{-}k^\ast)^2}\Big\{ \big(Q'_{t:n-\fb}{-} Q'_{k^\ast+1:t-1}\big)^2 {+}    \frac{\fb^2}{n}\|\Delta\|^2_2\int_\Omega\big[F_{t:n-\fb}(x){-} F_{k^\ast+1:t-1}(x)\big]^2dx\Big\}\nonumber \\
		=&\D_1(t)+\D_2(t). \nonumber
	\end{align}
	Since $(n-2\fb)^{-2}\sum_{t=k^\ast+2}^{n-\fb}\D_1(t)=O_p(1)$ and
	\begin{align}
		&(n-2\fb)^{-2}\sum_{t=k^\ast+2}^{n-\fb}\D_2(t) \nonumber \\
		=& \frac{\fb^2\|\Delta\|^2_2}{n}(n-2\fb)^{-2}\sum_{t=k^\ast+2}^{n-\fb} \frac{(n{-}\fb{-}t{+}1)^2(t{-}k^\ast{-}1)^2}{(n{-}\fb{-}k^\ast)^2}\int_\Omega\big[F_{t:n-\fb}(x){-} F_{k^\ast+1:t-1}(x)\big]^2dx    \nonumber \\
		=&n\|\Delta\|^2_2 O_p(1), \nonumber
	\end{align}
	we have 
	\begin{align}
		(n-2\fb)^{-2}\sum_{t=k^\ast+2}^{n-\fb}\frac{(n{-}\fb{-}t{+}1)^2(t{-}k^\ast{-}1)^2}{(n{-}\fb{-}k^\ast)^2}\{ Q_{t:n-\fb}{-} Q_{k^\ast+1:t-1} \}^2 = O_p(n\|\Delta\|^2_2). \label{eqap2}
	\end{align}
	For the same reason, we can show that 
	\begin{align}
		(n-2\fb)^{-2}\sum_{t=\fb+1}^{k^\ast-1}\frac{(t-\fb)^2(k^\ast-t)^2}{(k^\ast-\fb)^2}\{ Q_{\fb+1:t}- Q_{t+1:k^\ast}\}^2 = O_p(n\|\Delta\|^2_2).\label{eqap3}
	\end{align}
	Combining Equations (\ref{eqap2}) and (\ref{eqap3}), we have $V^2_n(k^\ast)=O_p(n\|\Delta\|^2_2)$, so $G_n\geq \frac{T_n(k^\ast)}{V_n(k^\ast)}\stackrel{p}{\to}\infty$ and part (ii).1 of Theorem \ref{th_cpn} is proved.

	To prove part (ii).2, for $r\in[0,1]$ and $x\in \Omega$ define $J(r,x)=(r-\tau_0)\delta(x)\id(r\geq \tau_0)$ and 
	\begin{align}
		Q(r,x) = H_1(r,x)\id(r< \tau_0)+H_1(\tau_0,x)\id(r\geq \tau_0)+\big[H_2(r,x)-H_2(\tau_0,x)\big]\id(r\geq \tau_0) \nonumber
	\end{align} 
	then we have 
	$$
	\frac{1}{\sqrt{n}}\sum_{s=\fr}^{\ft}\Delta(x)\id(s\geq \lfloor n\tau \rfloor  +1)\rightsquigarrow  J(t,x)-J(r,x) \mbox{ on } \ell_\infty[\Gamma\times \Omega]\nonumber$$
	and 
	$$		\frac{\ft{-}\fr  {+}1}{\sqrt{n}}\Big\{{F}_{\fr :\ft}(x){-}     \frac{1}{\ft{-}\fr  {+}1}         \sum_{s=\fr}^{\ft}   {F}_s(x)\Big\}  \rightsquigarrow   {Q}(t,x){-}{Q}(r,x) \mbox{ on } \ell_\infty[\Gamma\times \Omega]. \nonumber
	$$
	Denote $U(r,x) = Q(r,x)+J(r,x)-Q(b,x)$ and $U'(r,x) = Q(r,x)+J(r,x)$, similar to the proof of part (i) of Theorem \ref{th_cpn}, we can get $\wt G'' = \sup_{r \in [b,1-b]}\frac{\wt T(r)}{\wt V(r)}$, where 
	\footnotesize 
	\begin{align} \label{eqg1}
		\wt T(r)& = \frac{1}{\sqrt{1-2b}} \Big\{\int_\Omega \wt c''(x)U(r,x)dx-\frac{r-b}{1-2b}\int_\Omega\wt c''(x)U(1-b,x)dx \Big\}
	\end{align}
	\normalsize
	and 
	\footnotesize 
	\begin{align} \label{eqg2}
		\wt V(r)	= &\frac{1}{1-2b} \Bigg\{\int_b^{r} \Big\{\int_\Omega  \wt c''(x)U(s,x)dx {-} \frac{s-b}{r-b}\int_\Omega  \wt c''(x)U(r,x)dx\Big\}^2 ds  \nonumber\\
		& + \int_{r}^{1-b} \Big\{\int_\Omega  \wt c''(x)\big[U'(1-b,x)-U'(s,x)]dx{-} \frac{1-b-s}{1-b-r} \int_\Omega \wt c''(x)\big[U'(1-b,x)-U'(r,x)\big]dx \Big\}^2 ds\Bigg\}^{1/2}.
	\end{align}
	\normalsize
	
	For part (ii).3 and (ii).4 of Theorem \ref{th_cpn}, note that under part (\ref{assump_cp_d}) and (\ref{assump_cp_e}) of Assumption \ref{assump_cp}, we have $\cov[Q(r_1,x_1),Q(r_2,x_2)] = \min\{r_1,r_2\}C_1(x_1,x_2)$. Denote $J(r)=\frac{1}{\sqrt{\int_\Omega\int_\Omega \wt c''(x) C_1(x,y)\wt c''(y)dxdy}}\int_\Omega\wt c''(x){J}(r,x)dx$, $C(r)=B(r)+J(r)-B(b)$, $C'(r)=B(r)+J(r)$, $D(r)=B(r-b)+J(r)$, $a_s=\frac{s}{(1-2b)r}$ and $b_s = \frac{1-2b-s}{(1-2b)(1-r)}$. We can use the continuous mapping theorem and the orthogonal increment property of Brownian motion as in the proof of part (i) of Theorem \ref{th_cpn} to derive the conditional distribution of $\wt G''$ given $\wt c''(x)$ as follows:
	\footnotesize
	\begin{align}
		&\wt G''\Big|\wt c''(x) \nonumber\\
		\stackrel{d}{=} &\sup_{r \in [b,1{-}b]}  \frac{\sqrt{1{{-}}2b}\big(C(r){-} \frac{r{-}b}{1{-}2b}(C(1-b))  \big)}  {\Big\{\int_{b}^{r} \big[C(s){-}\frac{s{-}b}{r{-}b}(C(r))\big]^2 ds {+} \int_{r}^{1{-}b}\big[ C'(1-b){-}C'(s){-}\frac{1{-}b{-}s}{1{-}b{-}r}(C'(1-b){-}C'(r))    \big]^2 ds  \Big\}^{1/2}  }  \nonumber \\
		\stackrel{d}{=} &\sup_{r \in [b,1{-}b]}  \frac{\sqrt{1{-}2b}\big(C(r){-} \frac{r{-}b}{1{-}2b}C(1-b) \big)}  {\Big\{\int_{0}^{r{-}b} \big[C(s+b){-}\frac{s}{r{-}b}C(r)\big]^2 ds {+} \int_{r{-}b}^{1{-}2b}\big[ C'(1-b){-}C'(s+b){-}\frac{1{-}2b{-}s}{1{-}b{-}r}(C'(1-b){-}C'(r))    \big]^2 ds     \Big\}^{1/2}  }  \nonumber \\
		\stackrel{d}{=} &\sup_{r \in [b,1{-}b]}  \frac{\sqrt{1{-}2b}\big(D(r){-} \frac{r{-}b}{1{-}2b}D(1-b) \big)}  {\Big\{\int_{0}^{r{-}b} \big[D(s+b){-}\frac{s}{r{-}b}D(r)\big]^2 ds {+} \int_{r{-}b}^{1{-}2b}\big[ D(1-b){-}D(s+b){-}\frac{1{-}2b{-}s}{1{-}b{-}r}(D(1-b){-}D(r))    \big]^2 ds    \Big\}^{1/2}   }  \nonumber \\
		\stackrel{d}{=} &\sup_{r \in [b,1{-}b]}  \frac{\sqrt{1{-}2b}\big(D((1{-}2b)r{+}b){-} rD(1{-}b) \big)}  {\Big\{\int\limits_{0}^{(1{-}2b)r} \big[D(s{+}b){-}a_sD((1{-}2b)r{+}b)\big]^2 ds {+} \int\limits_{(1{-}2b)r}^{1{-}2b}\big[ D(1{-}b){-}D(s{+}b){-}b_s(D(1{-}b){-}D((1{-}2b)r{+}b))    \big]^2 ds \Big\}^{1/2}  }  \nonumber\\		
		\stackrel{d}{=} &\sup_{r \in [b,1{-}b]}  \frac{D((1{-}2b)r{+}b){-} rD(1{-}b) }  {\Big\{\int_{0}^{r} \big[D((1{-}2b)s{+}b){-}\frac{s}{r}D((1{-}2b)r{+}b)\big]^2 ds {+} \int_{r}^{1}\big[ D(1{-}b){-}D((1{-}2b)s{+}b){-}\frac{1{-}s}{1{-}r}(D(1{-}b){-}D((1{-}2b)r{+}b))    \big]^2 ds \Big\}^{1/2}  }  \nonumber\\		
		\stackrel{d}{=} &\sup_{r \in [0,1]}  \frac{B''(r)- rB''(1) }  {\Big\{\int_{0}^{r} \big[B''(s)-\frac{s}{r}B''(r)\big]^2 ds + \int_{r}^{1}\big[ B''(1)-B''(s)-\frac{1-s}{1-r}(B''(1)-B''(r)) \big]^2 ds \Big\}^{1/2} }\nonumber.
	\end{align}
	\normalsize
	So part (ii).3 and (ii).4  of Theorem \ref{th_cpn} are proved.

\end{document}